\documentclass[a4paper,twocolumn,floatfix]{article}

\pdfoutput=1

\usepackage{times}
\usepackage{amsfonts}
\usepackage{amssymb}
\usepackage{bm}
\usepackage{amsmath}
\usepackage{graphicx}
\usepackage{epsfig}
\usepackage{mathrsfs}
\usepackage{color}
\usepackage{soul}
\usepackage{physics}
\usepackage{xcolor}
\usepackage{mathtools}
\usepackage{lipsum}
\usepackage{titling}
\usepackage{hyperref}
\usepackage[switch]{lineno} 
\usepackage{caption}
\usepackage{amsthm}

\usepackage{tikz}
\usetikzlibrary{
    positioning,
    arrows,
    decorations.markings,
    hobby,
    decorations.pathreplacing
}

\definecolor{tenne}{rgb}{0.8, 0.34, 0.0}

\tikzstyle{beamsplitter} = [
    fill=gray,
    fill opacity=0.3,
    minimum height=1cm,
    minimum width=1cm
]

\tikzset{
  on each segment/.style={
    decorate,
    decoration={
      show path construction,
      moveto code={},
      lineto code={
        \path [#1]
        (\tikzinputsegmentfirst) -- (\tikzinputsegmentlast);
      },
      curveto code={
        \path [#1] (\tikzinputsegmentfirst)
        .. controls
        (\tikzinputsegmentsupporta) and (\tikzinputsegmentsupportb)
        ..
        (\tikzinputsegmentlast);
      },
      closepath code={
        \path [#1]
        (\tikzinputsegmentfirst) -- (\tikzinputsegmentlast);
      },
    },
  },
  mid arrow/.style={postaction={decorate,decoration={
        markings,
        mark=at position .5 with {\arrow[#1]{stealth}}
      }}},
}

\newtheorem{lemma}{Lemma}

\newcommand{\red}{\color{red}}
\newcommand{\mg}{\color{magenta}}
\newcommand{\blk}{\color{black}}
\newcommand{\blu}{\color{blue}}

\definecolor{maroon}{rgb}{0.7,0,0}

\definecolor{ngreen}{rgb}{0.2,0.6,0.2}

\definecolor{golden}{rgb}{0.8,0.6,0.1}

\newcommand{\beq}{\begin{equation}}
\newcommand{\eeq}{\end{equation}}
\newcommand{\bqa}{\begin{eqnarray}}\usepackage{lineno}
\linenumbers
\newcommand{\eqa}{\end{eqnarray}}

\newcommand{\dg}{^\dagger}

\newcommand{\smallfrac}[2]{\mbox{$\frac{#1}{#2}$}}

\newcommand{\half}{\smallfrac{1}{2}}

\newcommand{\cu}[1]{\left\{ {#1} \right\}}
\newcommand{\ro}[1]{\left( {#1} \right)}

\newcommand{\trc}[1]{{\rm Tr}[{#1}]}

\newcommand{\s}[1]{\hat{\sigma}_{#1}}

\newcommand{\xfrac}[2]{{#1}/{#2}}

\newcommand{\la}{\langle} 
\newcommand{\ra}{\rangle}

\newcommand{\be}{\begin{equation*}} 
\newcommand{\ee}{\end{equation*}}
\newcommand{\bea}{\begin{eqnarray*}}
\newcommand{\eea}{\end{eqnarray*}} 
\renewcommand{\bra}[1]{\langle #1|}
\renewcommand{\ket}[1]{|#1\rangle} 

\newcommand{\eref}[1]{Eq.~(\ref{#1})}
\newcommand{\fref}[1]{Fig.~\ref{#1}}
\newcommand{\sref}[1]{Sec.~\ref{#1}}

\newcommand{\etal}{ \emph{et al.}}
\newcommand{\coh}{\mathfrak{C}}

\graphicspath{{../figures/}}
        
\newtheorem{theorem}{Theorem}

\renewcommand{\mg}{\blk}

\renewcommand{\blu}{\blk}

\renewcommand{\red}{\blk}

\DeclareCaptionLabelSeparator{vert}{ $\vert$ }
\newcounter{lastnote}

\pretitle{\begin{flushleft}\LARGE\bfseries}
\posttitle{\par\end{flushleft}}
\preauthor{\begin{flushleft}\large}
\postauthor{\par\end{flushleft}}
\predate{\begin{flushleft}}
\postdate{\par\end{flushleft}\vskip 0.25em}
\title{\huge \bf{The Heisenberg limit for laser coherence}}
\author{
\bf{
Travis J. Baker\thanks{Centre for Quantum Dynamics, Griffith University, Brisbane, QLD 4111, Australia}
               \thanks{These authors contributed equally to this work.} ,
S.~N.~Saadatmand$^{*\dagger}$,
Dominic W. Berry\thanks{Department of Physics and Astronomy, Macquarie University, Sydney, NSW 2109, Australia} ,
Howard  M.  Wiseman$^{*}$\thanks{E-mail: h.wiseman@griffith.edu.au}
} }
\date{\em This paper is now published [Nature Physics  DOI:10.1038/s41567-020-01049-3 (26 October 2020)]. For copyright reasons, this arxiv paper is based on a version of the paper prior to the accepted (21 August 2020) version.}

\begin{document} 

\nolinenumbers

\maketitle 



\noindent
\textbf{To quantify quantum optical coherence requires both the particle- and wave-natures of light. For an ideal laser beam~\cite{McClung62,Gla63a,Carmichael99_book}, 
it can be thought of roughly as the number of photons emitted consecutively into the beam with the same phase. This number, $\coh$, can be much larger than $\mu$, the number of photons in the laser itself. The limit on $\coh$ for an ideal laser was thought to be of order $\mu^2$~\cite{SchTow58,Wis99}. Here, assuming nothing about the laser operation, 
only that it produces a beam with  certain \blk properties close to those \blk  
of an ideal laser beam, and that it does not have external sources 
of coherence, we derive an upper bound: $\coh = O(\mu^4)$. Moreover, using \blk 
the matrix product states (MPSs) method~\cite{Perez-Garcia07,McCulloch08,Orus14,Schollwock11}, we find a model that achieves this scaling, \blu and show that it could in principle be realised using circuit quantum electrodynamics (QED) \cite{Har20}. \blk Thus $\coh = O(\mu^2)$ is only a standard quantum limit (SQL); the ultimate quantum limit, or Heisenberg limit, is quadratically better.
}



Quantum theory underpins much modern technological development, and
sets the ultimate limits to the performance of devices --- the best conceivable performance towards which scientists and engineers can work under the constraint of a given resource (such as energy or power). In this context, a {\em quantum enhancement} exists when the ultimate limit, also known as a {\em Heisenberg limit}, scales better in terms of the resource than the SQL~\cite{GLM04}. The latter is also derived employing quantum theory,  but  using a set of \blk `standard' assumptions on how the device must work. 
 The quadratic quantum enhancement found in static phase estimation~\cite{Caves,Yurke} is well known,  
and there are many other metrological examples~\cite{GLM04,BerHalWis13}.  
Here, by contrast, we prove that there can be a quadratic quantum  enhancement  in the {\em production} of a physical property of great importance for both classical and quantum technology: optical coherence. 

A laser beam epitomizes optical coherence in all its aspects, including a long coherence time. 
 This time can be converted  to a dimensionless measure of coherence, 
$\coh$,  by multiplying by ${\cal N}$, the number of photons emitted per unit time. This gives, 
\blk loosely speaking, the number of photons emitted consecutively that are mutually coherent; \blu 
for its relation to other measures of coherence, see Supplementary Sec.~1F. \blk
The quantum limit to  the coherence time was famously studied by Schawlow and Townes over 60 years ago~\cite{SchTow58}. 
However, even more rigorous subsequent work~\cite{Wis99} 
made assumptions 
not entailed by fundamental requirements such as local conservation of energy. In the 21st century, our ability to engineer and control quantum systems~\cite{Say11,Chou18} 
has  changed our  conception  of what is practical. At the same time, our understanding of quantum 
processes has been deepened 
through theoretical  and numerical \blk techniques such as 
operational super-selection rules (SSRs)~\cite{BarWis03,BRS07}, and 
\blk MPSs. Hence it is plausible that the Schawlow--Townes limit to laser coherence is only a SQL, and that a Heisenberg limit can be proven to lie beyond it.  

Lasers are extremely widely used in science and technology, and improvement in 
coherence time 
is a  perennial  
goal for national laboratories~\cite{Mei09}.  
Essentially, a laser can be thought of as a device that creates  an optical output field with high coherence, 
from inputs that have  negligible coherence. 
 Here our measure of coherence, $\coh$, can be defined  for general fields as  the mean number of photons 
 in the maximally populated spatial mode  
 (within some frequency band, if required). 
Despite  its importance,  the fundamental physics of 
coherence creation 
 receives surprisingly little attention. 
 For example, we know 
 no textbook that directly states the cardinal fact that, for a typical laser, 
 $\frak{C}$ is far larger than $\mu$, 
 the mean number of photons (or other optical-frequency excitations) 
 in the laser itself. The bound $\frak{C} \leq \mu$ does apply for a $Q$-switched laser~\cite{McClung62}, or the ``atom lasers''~\cite{Ketterle02}, which 
 dump a Bose condensate from a trap, as the best outcome is if all the bosons inside the laser are in the same mode and stay in the same mode as they are dumped. By contrast, 
 for a laser that is pumped as it is damped, giving rise to a time-stationary continuous beam, there is no such simple argument. 
 
 For the continuous-output laser models developed in the 1960s and 70s~\cite{Lou73,SarScuLam74}, 
 the output  beam is a one-parameter (time) field, \blk with annihilation operator $\hat{b}(t)$, 
normalized so that 
$\hat b\dg(t) \hat b(t)$ is the photon flux operator.  \blk 
 In the right parameter regime and in the absence of technical noise~\cite{Lou73,SarScuLam74,Wis99}, 
 the state of the beam 
 \blk is well described by 
  an eigenstate of $\hat{b}(t)$ with eigenvalue $\beta(t) = \sqrt{{\cal N}}e^{i\phi(t)}$, \blk 
where $\phi(t)$ is a stochastic phase, undergoing pure diffusion~\cite{Carmichael99_book,Lou73}. 
 As detailed in \blu Supplementary Sec.~1E, \blk the coherence 
 for such lasers 
 evaluates to be $\frak{C}^{\rm ideal}_{\rm SQL} = \Theta(\mu^2)$.
 (This $\Theta$-notation means that $\frak{C}^{\rm ideal}_{\rm SQL}$ scales as $\mu^2$.) 
 Moreover, performance close to this ideal has been observed experimentally~\cite{Vit12}.
 One might therefore assume that the ultimate limit to laser coherence, in terms of the resource $\mu$, the number of stored energy quanta, would be $\Theta(\mu^2)$. 
 
 The central result of this paper is that the true ultimate limit, the Heisenberg limit, for a laser beam  having properties akin to those of the ideal models, \blk is $\frak{C}^{\rm ideal}_{\rm HL} = \Theta(\mu^4)$. This quadratic improvement implies vastly better performance in the limit $\mu \gg 1$ which characterizes most lasers. 
We prove the existence of a quadratic quantum enhancement in laser coherence creation by the following steps. 
First, we state our conditions on the laser and its beam. Second, we show analytically from these  the upper bound of $\frak{C} = O(\mu^4)$. Third, we introduce a \blk $\mu$-parameterized family of laser models and show numerically that $\coh = \Theta(\mu^4)$. Fourth, we show that all conditions are satisfied in our model, either exactly or numerically. \blu 
\blu In addition, we propose an implementation of the model in a system based on current architecture used in the field of circuit QED, and we \blk
conclude with a discussion of open questions. 

\emph{First}, we state our conditions on the laser and its beam. We use the term ``laser'' and ``laser cavity''  interchangeably, but we actually make no assumptions about the physical medium storing excitations in the laser.
 Similarly, we refer to the 
energy quanta in the frequency band of interest as ``photons'', regardless of how they may be stored.  The laser cavity need not comprise a single harmonic oscillator; the laser cavity photon number operator $\hat{n}_{\rm c}$ can be degenerate. 
\blk 
The names (in bold) for the conditions below are for shorthand reference and should not be taken to capture everything important. \blu For more details on each, see Supplementary Sections 1A--1D, respectively. \blk 
\begin{enumerate}
\item[1.] {\bf One-dimensional beam.}  The beam propagates away from the laser in a particular direction, at a constant speed, and has a single transverse mode and a single polarization. 
\end{enumerate} 
Mathematically, 
at any time $T\in \mathbb{R}$, \mg the beam can be described as a scalar quantum field with a single argument $t \in (-\infty,T)$, which is the time at which that infinitesimal part of the beam, b$_t$, was created by the laser. Thus the quantum state of b$_t$ is independent of $T$ (as long as $T>t$). The annihilation operator $\hat b(t)$ for the field satisfies 
$[\hat b(t),\hat b\dg(s)] = \delta(t-s)$, so that $\hat b\dg(t)\hat b(t)$ is the operator for photon flux (photons per unit time). \blk 
\begin{enumerate}
\item[2.] {\bf Endogenous phase.} The coherence of the beam proceeds only from the laser.  
That is, a phase shift imposed on the laser state at some time $T$ will lead, in the future, to the same phase shift on the beam emitted after time $T$, as well as on the laser state. 
\end{enumerate} 
 
 The phase shift at time $T$ on the laser (which may have been prepared by measurement 
 on the beam generated prior to $T$) is described by the unitary transformation $\hat{U}^{\zeta}_{\rm c} = \exp(i\zeta \hat{n}_{\rm c})$. 
 The effect of this, \red at any time $T'>T$, on the state of the cavity plus the beam segment 
 generated in the interval $(T,T']$, is 
 described by the unitary transformation $\hat{U}^{\zeta}_{\rm cb} = \exp(i\zeta (\hat{n}_{\rm c} + \hat{n}_{\rm b}))$, 
 where  $\hat{n}_{\rm b} = 
\int_T^{T'} \hat{b}\dg(s)\hat{b}(s) ds$ is the photon number operator for the generated beam segment.
\begin{enumerate}
\item[3.] {\bf Stationarity.} The statistics of the laser and beam \mg have a long-time limit that is unique and invariant under time translation. 
In particular, $\langle \hat n_{\rm c}\rangle$ has a unique limit, $\mu$. \blk
\end{enumerate}
Note that this Condition rules out a laser based on the ``catalytic coherence'' 
model of 
Ref.~\cite{Aberg14}, as its mean photon number would grow linearly in time.
\begin{enumerate}
\item[4.] {\bf Ideal Glauber$^{(1),(2)}$-coherence.}  The stationary beam is close to the ideal laser beam as described in the introduction --- an eigenstate of $\hat b(t)$ of eigenvalue 
 $\beta(t) = \sqrt{{\cal N}}e^{i\sqrt{\ell}W(t)}$, with $W(t)$ a Wiener process~\cite{Carmichael99_book} --- in the sense that the beam's first- and second-order Glauber coherence functions~\cite{Gla63a} 
 well approximate those of the ideal beam. 
\end{enumerate} 

The first- and second-order Glauber coherence functions are defined as 
\begin{linenomath}
\begin{align} 
G^{(1)}(s,t) &= \langle \hat b\dg(s) \hat b(t) \rangle, \notag \\  
G^{(2)}(s,s',t',t) &= \langle \hat b\dg(s) \hat b\dg(s') \hat b(t') \hat b(t) \rangle.
\end{align}
\end{linenomath}
$G^{(1)}$ yields the photon flux ${\cal N} = G^{(1)}(t,t)$, and 
the coherence $\coh = \max_\omega |\int_{-\infty}^\infty G^{(1)}(s,t)e^{-i\omega s}ds|$. 
The latter relation is derived in \blu Supplementary Sec.~1A. \blk 
The ideal beam coherence functions are evaluated by letting $\hat b(t) \to \beta(t)$, and 
what it means to 
``well approximate'' these is formulated \blu quantitatively \blk 
below. 
\blk The requirement on $G^{(1)}$ in this Condition is that the laser power spectrum is Lorentzian (or close to it) with linewidth  
$\ell \equiv 4 {\cal N}/\coh$, while the requirement on $G^{(2)}$ implies that the photon statistics are Poissonian (or close to it). 
\blu

\emph{Second}, we show analytically 
that $\frak{C} = O(\mu^4)$:
\begin{theorem}[Informal; see Theorem~2.2~ in Supplementary Sec. 2]\label{theo:main}
For an ideal laser beam satisfying Conditions 1--4, the coherence $\coh^{\rm ideal}$ in the asymptotic limit $\mu\rightarrow\infty$ is bounded from above by
\begin{equation}
	\frak{C}^{\rm ideal} \lesssim  2.9748 \mu^4,
	\label{ubCmu4}
\end{equation}
where $\mu$ is the mean number of stored excitations, as above. 
\end{theorem}\blk 
The detailed proof of this theorem is given in \blu Supplementary Sec.~2; \blk
here we sketch it. \blk
We assume \blu steady-state operation (after a time $\gg \ell^{-1}$) \blk and $\coh, \mu \gg 1$.  
 The proof involves consideration of measurements of optical phase. Such measurements require an independent phase reference, 
 and all phases mentioned are relative to this ``local oscillator''~\cite{Carmichael99_book,WisMil10}. 
 It plays no role in the dynamics of the laser and in reality would simply be another laser with much greater coherence $\coh$. 
  
  Consider a heterodyne measurement~\cite{WisMil10} of the laser beam in the interval 
  $[T-\tau, T)$, where $\tau = \sqrt{3/(2\mathcal{N}\ell)}$ (this value is chosen to give the tightest bound, below). From the result, 
  the observer  can form an estimate, $\phi_{\rm F}$, of the phase of the laser beam at  time $T$.  
Consider two methods by which a second, newly arrived, observer can estimate 
$\phi_{\rm F}$.   
The first method is by heterodyne measurement of the 
laser beam in the interval $ ( T,T+\tau]$. From  Condition 4,  we can show that 
the mean square error (MSE) for this estimate can attain $4\sqrt{2/(3\coh)}$.  
 The second method is by direct phase measurement on the laser cavity at time $T$, which is optimal, because 
 Condition 2 ensures that $\phi_{\rm F}$ is encoded in the state at time $T$ via the 
generator $\hat{n}_{\rm c}$.  
 Now the MSE  of 
any such estimate is bounded below by  $k/\mu^{2}$, where 
$k\!\approx\!1.8936$~\cite{Ban91}, and 
 $\mu=\langle \hat{n}_{\rm c} \rangle$ from  Condition 3.  
Thus, the MSE  from the first method  
cannot be smaller than that of the  second,  
which cannot be smaller than  $k/\mu^{2}$. 
Theorem~\ref{theo:main} follows.

\emph{Third}, we introduce a family of laser models,  parameterized by $\mu$, satisfying Conditions 1--3,  and show numerically that they give 
$\coh = \Theta(\mu^4)$. We derived our model and results in the framework of infinite  MPSs  (iMPSs)~\cite{McCulloch08,Schollwock11}. MPS methods have been applied previously in quantum optics~\cite{Schon05,Schon07,Jarzyna13,Man17},  
but not to coherence creation in lasers.  The details, including physical motivations, 
are given in Methods below (and further discussed in \blu Supplementary Sec.~3). \blk
Here we just present the final results, in the continuous-time limit.  
We make no claim that the model below is the most general one 
satisfying  Conditions 1--4; we do not require that to show achievability of the scaling in \eref{ubCmu4}.  

\blk 
The laser model comprises a $D$-level `cavity' with non-degenerate number operator $\hat n_{\rm c} = \sum_{n=0}^{D-1} \blu n \blk \ket{n}\bra{n}$. 
Working in a phase-rotating frame at the \blu cavity frequency $\omega_{\rm c}$, \blk we describe the laser cavity 
dynamics by the master equation $\dot{\rho} = {\cal L}\rho$ where 
${\cal L} = {\cal D}[\hat G]  + {\cal D}[\hat L]$. 
Here, in standard notation~\cite{WisMil10}, ${\cal D}[\hat c]\bullet \coloneqq \hat{c}\bullet\hat{c}\dg - \half\cu{\hat c\dg \hat c,\bullet}$, while $\hat G$ and $\hat L$ are one-photon gain and loss operators.  That is, their only nonzero elements 
are $G_n = \bra{n}\hat G \ket{n-1}$ and $L_n = \bra{n-1}\hat L \ket{n}$, for $0<n<D$. 
If all $G_n$ and $L_n$ are nonzero,  
then there is a unique stationary state: ${\cal L}\rho_{\rm ss}=0$, with nonzero elements $\rho_n = 
\bra{n}\rho_{\rm ss}\ket{n}$,  for $0<n<D-1$, \blk determined by 
$\rho_{n} = | \xfrac{G_n}{L_n} |^2 \rho_{n-1}$. To obtain the ultimate limit to coherence, we do not assume linear damping ($\blu \hat L \propto \hat a \implies \blk L_n \propto \sqrt{n}$). 
Rather, guided by iMPS optimization of $\coh$, we define our family of models by the choices 
\begin{linenomath}
\begin{align}
\rho_n \propto \sin^4 \ro{\pi \frac{n+1}{D+1}},\ G_n = 1\,,  \ L_n \in \mathbb{R}_+,
\label{eq:model_choices}
\end{align}
\end{linenomath}
for which $\mu$ (\blu see \blk Condition 3) equals $(D-1)/2$, and ${\cal N}$ (\blu see \blk Condition 4) equals $1$ (which simply sets a convenient time unit). 

\begin{figure}[bth]
\begin{center}
\includegraphics[width=1.00\columnwidth]{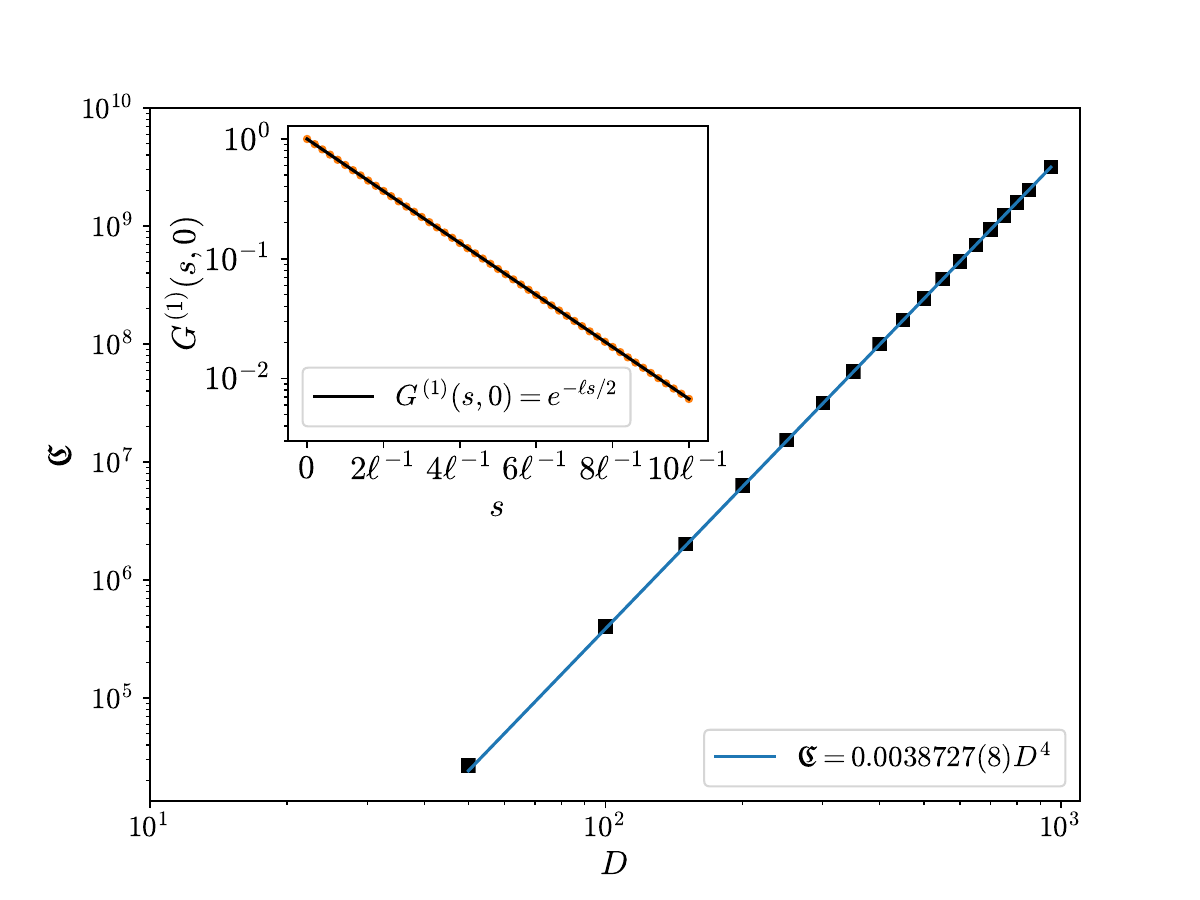}
\caption{\label{fig:numerics} \blk iMPS calculations (squares)  
of the beam coherence $\coh$ for our laser model as a function of dimension $D= 2\mu+1$. The line is a fit 
$\mathfrak{C} \propto D^4$. Insert: 
\mg iMPS \blk  calculations (discs) of first-order Glauber coherence function 
$G^{(1)}(s,0)$  for $D=300$, \blk which is indistinguishable from the ideal exponential decay of a phase-diffusing coherent state (solid line). All error-bars (estimated or actual) are smaller than symbol sizes.}
\end{center}
\end{figure}

We can derive the properties of the beam from the above master equation if the ${\cal D}[\hat L]$ term arises from 
a quantum \red vacuum white noise coupling~\cite{WisMil10}. This means that \mg ${\rm b}_t$, the 
infinitesimal segment of beam emitted in the interval $[t,t+dt)$, as introduced in Condition 1, 
can be described by a truncated Fock space ${\rm span}\cu{\ket{0},\ket{1}}$, \red  
with $\hat b(t) \sqrt{dt}= \ket{0}\bra{1}$. Moreover, the state of ${\rm b}_t$ 
is generated from (and thus remains correlated with) the cavity state \mg via \red 
\begin{linenomath}
\begin{align} \label{genfield}
\rho_{{\rm cb}_t}(t+dt) = (1 &+ {\cal D}[\hat G\otimes \hat{I}]dt  + {\cal D}[\hat L\otimes \hat b\dg(t) dt] 
 \notag \\
 &+ \bar{\cal H}[\hat L\otimes \hat b\dg(t) dt])(\rho_{\rm c}\mg(t)\red\otimes\ketbra{0}),
\end{align}
\end{linenomath}
where $\bar{\cal H}[\hat c]\bullet  \coloneqq \red  \hat{c}\bullet+ \bullet\hat{c}\dg$~\cite{WisMil10}.
From this it follows that $G^{(1)}(s,t) = \trc{ \hat L\dg e^{{\cal L}(s-t)} (\hat L \rho_{\rm ss})}$,  
with an analogous expression for $G^{(2)}$.  
For our model, $G^{(1)}(s,t)$ is real and positive, so the coherence can be evaluated as 
$\coh = \int_{-\infty}^\infty G^{(1)}(s,t) ds = -2\trc{ \hat L\dg {\cal L}_+^{-1} (\hat L \rho_{\rm ss})}$, 
 where ${\cal L}_+^{-1}$ is the inverse of ${\cal L}$ on its row space. \blk 
 We evaluate this for the above family of models, with $D$  up to $1000$. \blk 
The data fit a power law  $\coh \sim 0.06196(1) \mu^{4}$, \blk as shown in Fig.~\ref{fig:numerics}. That is, the ultimate bound on the scaling derived earlier is achieved, albeit with a far smaller coefficient than that in \eref{ubCmu4}, \blu which we do not expect to be achievable. \blk  

\begin{figure}[bth]
\begin{center}
\includegraphics[width=1.00\columnwidth]{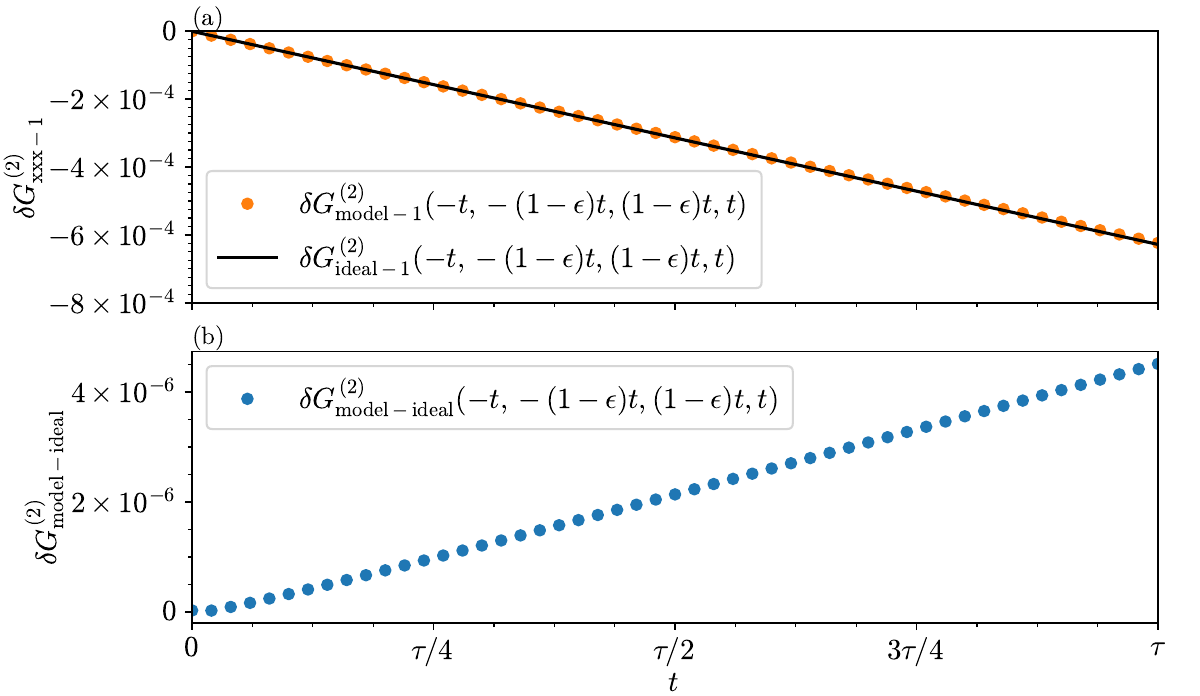}
\caption{
iMPS calculations \blu (discs) \blk of the 
four-time 
Glauber coherence function 
for our model \mg with $D=500$, \blk for representative arguments of order the time scale of interest,  $\tau =  (3\coh/8)^{1/2}$. \blk 
(a) The deviation from unity, \red $\delta G_{\rm model-1}^{(2)} \coloneqq  G_{\rm model}^{(2)} - 1$, \blk is \mg almost \blk indistinguishable 
from the \red ideal, $\delta G_{\rm ideal-1}^{(2)} \coloneqq \blk  G_{\rm ideal}^{(2)} - 1$, 
\mg corresponding \red to a phase-diffusing coherent state (solid line). 
(b) The deviation of \red the model from the ideal,  
$\delta G_{\rm model-ideal}^{(2)} \coloneqq \blk  G_{\rm model}^{(2)} - G_{\rm ideal}^{(2)}$, is \mg about $10^{-2}$ times \mg $|\delta G_{\rm ideal-1}^{(2)}|$. 
 Here, $\epsilon$ is an arbitrarily small positive number \blk and all error-bars (estimated or actual) are smaller than symbol sizes.}
\label{fig:numerics2}
\end{center}
\end{figure}

{\em Fourth}, we show that \red our model satisfies Conditions 1--4. Conditions 1 and 3 have already been covered above. 
We show in \blu Supplementary Sec.~4A \blk
 that Condition 2 follows from 
 \eref{genfield} and 
the facts that, under the transformation 
$\hat{U}_{\rm c}^\zeta$, 
${\cal L}$ is U(1)-invariant~\cite{BRS07} and $\hat L$ is U(1)-covariant ($\hat U_{\rm c}^{-\zeta} \hat L \hat U_{\rm c}^{\zeta} = e^{i\zeta}\hat L$)~\cite{BRS07}. 
Condition 4 \red requires \blk that $G^{(1)}(s,t)$ be close to the ideal value of $e^{-\ell |s-t|/2}$, and we show this exponential decay over two decades in Fig.~\ref{fig:numerics}, where $\ell = 4/\coh$. For Theorem~\ref{theo:main} 
in fact \blu it is more critical \blk that $G^{(2)}(s,s',t',t) \approx G^{(2)}_{\rm ideal}(s,s',t',t) \blu \equiv\blk\!\! e^{-\ell( |s-t|+|s'-t'|+|s-t'|+|t-s'|-|s-s'|-|t-t'|)/2}$, 
for its arguments in the range $[-\tau,\tau]$, where  $\tau = \sqrt{3\coh/8}$  as above. 
Fig.~\ref{fig:numerics2} 
shows that this \blu approximation is indeed very good for the 
$(s, s', t, t')$ 
that maximize the deviation 
$\delta G_{\rm model-ideal}^{(2)} \coloneqq G^{(2)}_{\rm model} - G^{(2)}_{\rm ideal}$.  
Already with $D=500$, \mg $\delta G_{\rm model-ideal}^{(2)}$ is orders of magnitudes smaller than $\delta G_{\rm ideal-1}^{(2)} \coloneqq G^{(2)}_{\rm ideal} - 1$. (It is $\delta G_{\rm ideal-1}^{(2)}$ \blk which \red sets the scale of \blk the result in \eref{ubCmu4}.) The requirement that our model well-approximate Glauber$^{(2)}$-ideality is precisely formulated in Methods below, 
and we show numerically, in \blu Supplementary Sec.~4B, \blk 
that it is fulfilled. 

\begin{figure}[bth]
\begin{center}
\includegraphics[width=1.00\columnwidth]{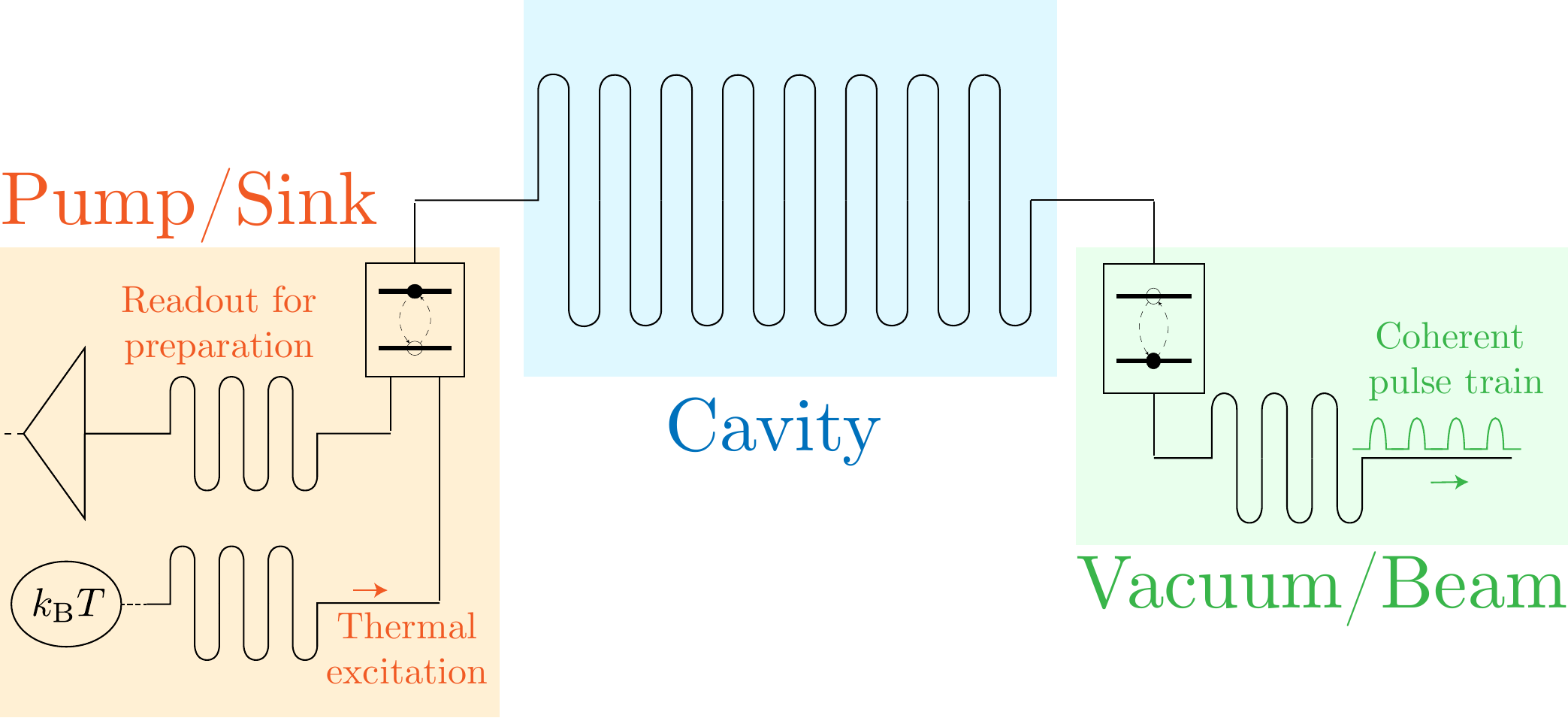}
\caption{\label{fig:CirQED}
\blu Proposal for a circuit QED device to realise a Heisenberg-limited maser. 
The left qubit is controllably coupled to a thermal source and a non-demolition energy detector to prepare it in the excited state. 
The qubit is then controllably coupled to the cavity mode so that as it loses its excitation the cavity energy is raised via the operator 
$\hat G$. The right qubit is then controllably coupled to the cavity so that as it gains an excitation the cavity energy is lowered via the operator 
$\hat L$. The qubit is then controllably coupled to a microwave transmission line so as to release its energy as a pulse, repreparing it 
in the ground state. All controls are achieved by tuning the qubits' frequencies.}
\end{center}
\end{figure}

\blu 
{\em Fifth}, we show that the model introduced above could be realised with physical couplings in circuit QED \cite{Har20}, 
yielding a microwave output field with a pulsed (rather than flat) spatial structure, with coherence scaling as the Heisenberg limit. 
We consider a waveguide cavity, capacitively coupled weakly to two superconducting qubits, as shown in Fig.~\ref{fig:CirQED}.  
Making the rotating-wave approximation, in a frame rotating at the maser cavity frequency $\omega_{\rm c}$, this coupling is described by the Hamiltonian
\begin{linenomath}
\begin{align} \label{maser_couple}
\hat H(t) &= \sum_{s={\rm l,r}} \hbar \left[ g_s\left( \hat a\dg \hat  \sigma_s^- + \hat \sigma_s^+ \hat a  \right)  + \Delta_s(t) \hat \sigma^z_s \right] ,
\end{align}
\end{linenomath} 
where the $\sigma$-operators for the left (l) and right (r) qubit have their usual meanings. In Supplementary Sec.~5, we show using Lie algebraic techniques that the ability 
to control the qubit frequencies $\omega_s(t) = \omega_{\rm c} + \Delta_s(t)$, with $|\Delta_s| \ll \omega_{\rm c}$, is sufficient to realise the 
discrete-time (see Methods) version of our model. 


\blk

In conclusion, the Schawlow--Townes limit to laser coherence $\coh$ (the number of mutually coherent photons emitted in the beam) 
is only a standard quantum limit. Beyond it lies a Heisenberg limit which scales quadratically better in terms of $\mu$ (the number of excitations in the laser itself). For $\mu$ large this represents a vast improvement in laser coherence properties. 
We constructed a model that achieves this quantum enhancement \blk in scaling,  
while retaining the same first- and second-order coherence properties as an ideal laser beam 
(a constant-intensity coherent state with a diffusing phase). \blk It is the assumption of these coherence properties that allowed us to prove the Heisenberg limit $\coh^{\rm ideal}_{\rm HL} = O(\mu^4)$. It is thus natural to ask whether relaxing this assumption would enable an even higher scaling to be achieved. Preliminary results suggest that this is indeed the case. 
However, this remains to be investigated, along with other 
 fundamental issues, as well as the question of 
what technology \blu is most suitable to \blk enable optical coherence beyond the SQL to be observed experimentally. 



\section*{Acknowledgments} 
\small 
We thank Ignacio Cirac, Howard Carmichael, \blu Mazyar Mirrahimi, Ian McCulloch, \blk Michael Hall, and Antoine Tilloy for useful discussions. 
This work was supported by ARC Discovery Projects DP170101734, 
DP160102426, and DP190102633,  
and an Australian Government RTP Scholarship.
\normalsize

\section*{Author contributions}
\small 
HMW conceived, acquired funding for, and \blu directed \blk the project. DWB had the key idea for theorem~\ref{theo:main}. All authors contributed to the analytics. SNS did the numerics. 
TJB did the figures. HMW drafted the manuscript, while TJB and SNS drafted Supplementary Information. All authors contributed to revisions.
\normalsize

\section*{Competing interests}
\small 
The authors declare no competing interests.
\normalsize

\section*{Additional information}
\small

\textbf{Correspondence and requests for materials} should be addressed to H.M.W.

\section*{Note Added in arxiv v2}
\small 
After the final acceptance of this paper, a related paper appeared on the arxiv~\cite{Liu20}, 
reporting independent theoretical work on achieving a laser linewidth beyond the 
SQL (though not at the Heisenberg limit) in circuit QED.

\normalsize


\section*{Methods}

\small

\begin{figure}[htb]
\begin{center}
\includegraphics[width=1.00\columnwidth]{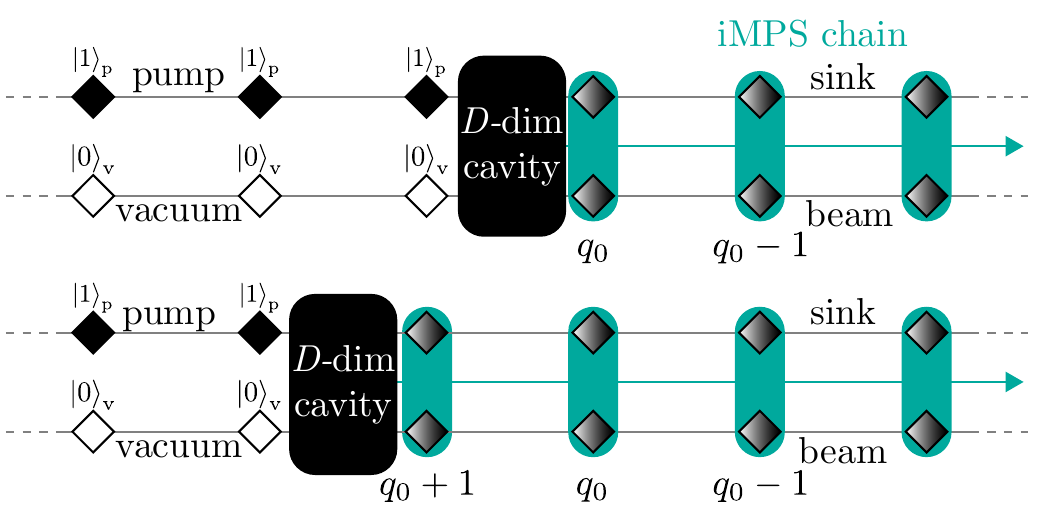}
\caption{\label{fig:toy_model_mini}
\blu \small  Conceptual diagram of our laser model. 
    From the upper figure to the lower, one time step has passed, converting 
    one pair of input qubits (pump and vacuum) into 
    a new pair of output qubits (beam and sink), with position label  
    $q_0+1$. The indefinite length string of pairs of output 
    qubits is described by an iMPS of bond-dimension $D$,     
    equal to the Hilbert space dimension of the laser cavity.}
\end{center}
\end{figure}

\noindent
\textbf{The \lowercase{i}MPS description of a laser beam.} The discretized 
(assuming a time interval of $\delta t$) laser system we described in the main text \blu is illustrated in Fig.~\ref{fig:toy_model_mini}.
It \blk contains five elements: 
the cavity, a pump, a vacuum input, the beam output, and a sink.
All of these are essential for laser operation, and we use the simplest possible model for all.
We can consider the sink (s) and beam (b) as a joint four-level system in order to maintain consistency with the requirements of an MPS sequential generation scheme~\cite{Schon05,Schon07}, possessing a single output at a time.
In doing so, we may consider the beam alone by simply tracing over the sink. 
The time evolution of the cavity (c)  and its outputs \blk is governed by the generative interaction 
\begin{equation}
	\hat{V}_{ q} = \sum_{j_{ q+1},m,n} A^{[j_{q+1}]}_{mn} \ket{m}_{ \rm c}\bra{n} \blu\otimes\blk \ket{j_{q+1}}_{ \rm o},
	\label{eq:interaction_isometry}
\end{equation}
which maps a $D$-dimensional vector space into a $4 \times D$-dimensional one, \blu and for the output space we have defined $\ket{j_{q+1}}_{ \rm o} \coloneqq \ket{\lfloor j/2\rfloor_{q+1}}_{\rm b}\otimes \ket{(j\mod 2)_{q+1}}_{\rm s}$. \blk
This results in a completely-positive trace-preserving map which evolves the laser cavity one time step. 
The isometry $\hat V$  can be related to the generative unitary interaction, $\hat{U}_{\rm int}$\blk, according to $\hat{V} \ket{\psi}_{\rm c} \equiv \hat{U}_{\rm int} (\ket{\psi}_{ \rm c}\ket{1}_{\rm p} \ket{0}_{ \rm v} )$, where the subscripts p and v correspond to pump and vacuum states.
The isometry condition,  $\hat{V}^\dagger \hat{V} = I_{D}$, \blk where $I_m$ is the $m \times m$ identity matrix, translates to
a completeness/orthonormality relation
\begin{equation}
  \sum_{j=0}^3 A^{[j]}{}^\dagger A^{[j]} = I_D~.
\label{eq:IsometryCond}
\end{equation}
Each $A^{[j]}$ here is treated as an operator whose elements in the cavity photon number basis, $\ket{m}_{\rm c}$, 
are those of the $A$-matrices introduced above. \blu See Supplementary Sec.~3A. for details.\blk 

We are interested in the \emph{one-site unit-cell} iMPS that $\hat{V}$ eventually creates (for arbitrary consecutive times $q_0$ and $q_0+1$). In terms of the $A$-operators used above, the
iMPS is given by  %
\begin{linenomath}
\begin{align}
  &\ket{\Psi_{\rm iMPS}}\!=\!\sum_{...,j_{q_0},j_{q_0-1},j_{q_0-2},...} \bra{\Phi(q\!=\!+\infty)}_{ \rm c} \, 
                           \cdots A^{[j_{q_0}]}_{(q_0)} \notag \\
                           &A^{[j_{q_0-1}]}_{(q_0-1)} A^{[j_{q_0-2}]}_{(q_0-2)} \cdots
                           \ket{\Phi(q\!=\!-\infty)}_{ \rm c}  
                           \ket{...,j_{q_0},j_{q_0-1},j_{q_0-2},...}_{ \rm o}~,
\end{align}
\end{linenomath}
%
If we impose the condition that 
the largest-magnitude eigenvalue of its \emph{transfer matrix}~\cite{Schollwock11,Orus14}, 
\blu $\mathcal{T}=\sum_{j=0}^3 A^{[j]}{}^* \otimes A^{[j]}$\blk, 
be non-degenerate then 
the iMPS is translationally invariant and independent of the boundary states $\ket{\Phi(q\!=\!+\infty)}_{\rm c}$ and $\ket{\Phi(q\!=\!-\infty)}_{\rm c}$.  \blk

\mg 
The relationship between the iMPS model above and the laser model in the main text is as follows. 
First we set $A^{[2]}=0$, $A^{[0]} = \sqrt{\delta t} B^{[0]}$, $A^{[3]} = \sqrt{\delta t} B^{[3]}$, and $A^{[1]}=A^{[1]}{}\dg$ 
determined by \eref{eq:IsometryCond}. Then we can identify  $B^{[0]} = \hat G$ and $B^{[3]} = \hat L$, \blk
if we consider the $B$-matrices as operators acting on the Hilbert space of the cavity
\blu (see Supplementary Sec.~3C for details). \blk 

Now we want to calculate $\frak{C}$ and Glauber$^{(1),(2)}$ correlators in the iMPS formalism. \blk  
In \blu Supplementary Sec.~3B, \blk we find that 
\begin{linenomath}
\begin{align}
   \frak{C} & =
    - 2\blk (1| ~ B^{[3]*} \otimes I_D \cdot \text{inv}(\mathbb{Q}\mathcal{L}\mathbb{Q}) \cdot I_D \otimes B^{[3]} ~ |1)~,
\label{eq:C-final-kappa0_indep}
\end{align}
\end{linenomath}
\blk
where we have reshaped any $D \times D$-size operator into flattened $D^2 \times 1$-dimensional vector 
forms as in $\hat{E}_{m,n}\rightarrow ( E |_{(m,n)}$ and $1 \times D^2$-dimensional vector forms as in $\hat{F}_{m,n}\rightarrow | F )_{(m,n)}$ for collective $(m,n)$ indices. Using this notation, $(1| \leftrightarrow \hat{I}_D$, is the identity operator of 
the Hilbert space of the cavity and left leading eigenvector of $\mathcal{T}$;  it should be clear from \eref{eq:IsometryCond} that \blk $(1| \mathcal{T} = (1| \lambda_1 = (1|$, where eigenvalues of $\mathcal{T}$ are arranged as 
$1=|\lambda_1| \geq  |\lambda_2| \geq \cdots \geq |\lambda_{D^2}|$. In addition, 
the familiar reduced density matrix, $|1) \leftrightarrow { \rho}^{\rm ss}$,
satisfies the fixed-point (steady-state) equation 
$\sum_j \hat{A}^{[j]} \rho^{\rm ss} \hat{A}^{[j]}{}^\dagger = \rho^{\rm ss}$ 
(in other words, it is the right leading eigenvector as $\mathcal{T} |1) = |1)$). 
Furthermore,  in \eref{eq:C-final-kappa0_indep}, \blk we have set $\mathbb{Q} = I_{D^2} - |1)(1|$
and $\mathcal{T} = I_{D^2} + \delta t \mathcal{L}$  (note $\mathcal{L}$ is the flattened-space
version of the same operator introduced in the main text)\blk. Notice 
\eref{eq:C-final-kappa0_indep} is equivalent to the coherence form expressed above in the main text and 
is how we evaluated \blk $\frak{C}$ in practice for \fref{fig:numerics}. 

In the iMPS language, $G_{\rm model}^{(1)}(s,0) = (\sigma^+| \mathbb{L}_{\rm exp}(s) |\sigma^-)$, 
where we have defined $\mathbb{L}_{\rm exp}(t) \coloneqq  \exp(\mathcal{N}t\mathcal{L}) = \exp(t\mathcal{L})$ 
and used the transformation $\sqrt{\delta t}~\hat{b} \rightarrow \hat{\sigma}^-_{ \rm b}$. 
This is how we produced the inset in 
\fref{fig:numerics}. 
\blu We calculate $\max_{s,s',t',t \in \blu[-\tau,\tau]\blk}[\delta G_{\rm ideal}^{(2)}(s,s',t',t)]$ \blu in a similar manner. 
Taking \blk the ordering $\blu s\blk<s'<t'<t$ for example (again with $\delta t \rightarrow 0^+$), 
the flattened-space form is 
\begin{linenomath}
\begin{align}
   G^{(2)}(\blu s\blk,s',t',t) =\, & (\sigma^+| \mathbb{L}_{\rm exp}(s'-\blu s\blk) (B^{[3]*}\otimes I_D) \times \notag \\
     &\mathbb{L}_{\rm exp}(t'-s') (I_D\otimes B^{[3]}) \mathbb{L}_{\rm exp}(t-t') |\sigma^-),
\label{eq:exampleG2s}
\end{align} 
\end{linenomath}\blk
where we again used 
the transformation $\sqrt{\delta t}~\hat{b} \rightarrow \hat{\sigma}^-_{ \rm b}$ 
and commutator relations $[\hat{b}^\dagger(t),\hat{b}(t'\neq t)]=[\hat{b}(t),\hat{b}(t')]=0$  for $t\neq t'$.
\blu Other time orders can be calculated similarly. \blk 

\noindent
\textbf{Efficient numerical calculations of Glauber$^{(2)}$ coherence functions.} The main numerical challenge for calculating such second-order (and even first-order) Glauber coherence functions is to exponentiate \blu the \blk$D^2 \times D^2$ matrix $\mathcal{L}$ for large bond dimensions; this is equivalent to precisely finding a large enough number of its eigenvalues. 
Direct scaling and squaring algorithms (directly estimating \blu a \blk matrix exponential based on Pad\'{e} approximants) provide high precision, but are memory extensive with computational time scaling as $O(D^6)$. 
Luckily, the $\mathcal{L}$ superoperator is highly sparse (in fact, we manipulate all the operators in our codes as sparse inputs), 
and Krylov subspace  projection techniques can be used to efficiently 
find the result of the action of $\mathbb{L}_{\rm exp}(t)$
on a vector. Here, we employed the `matrix-free' Krylov method introduced in \cite{Sid98}. As such,
we constructed the iMPS forms for all required permutations of arguments in 
$\delta G_{\rm ideal}^{(2)}(\blu s\blk,s',t',t) = G^{(2)}(\blu s\blk,s',t',t) - G_{\rm ideal}^{(2)}(\blu s\blk,s',t',t)$,
which was used to produce the results in \fref{fig:numerics2}. Furthermore, we employed the highly scalable \blu interior-point optimization method\blk, discussed in \blu Supplementary Sec.~3B, \blk to find the global maximum of all 
such functions for $\{\blu s\blk,s',t',t\} \in \blu[-\tau, \tau]$ for bond dimensions up to $250$ to numerically prove that Condition 4 of the main text is indeed satisfied for our laser model. 

\noindent
\textbf{Well-approximating Glauber$^{(2)}$-ideality.} In \blu Theorem~\ref{theo:main}, we proved (see Supplementary Sec.~4B) that $\frak{C}=O(\mu^4)$ 
by taking $G^{(2)}$ to equal $G^{(2)}_{\rm ideal}$, where the latter is that for an ideal beam. \blk 
The calculation involved four point correlations $G^{(2)}(s,s',t',t)$, with time arguments spanning an interval of twice the filtering (retrofiltering) time: $s,s',t',t \in \blu[-\tau,\tau]$. Crucially, the \blu the upper bound (\ref{ubCmu4}) 
arises because, for some such values of its arguments, $\delta G^{(2)}_{\rm ideal - 1} 
= \Theta \blk (\coh^{-1/2})$.\blk

Therefore, for  \blu Theorem~\ref{theo:main} to be relevant \blk for our laser model, it would be sufficient to demand that differences between $G^{(2)}$ for the ideal laser and our model, $\delta G^{(2)}_{\rm model - ideal}$, for \emph{all} its time arguments, over a time interval of length $2\tau = \sqrt{3\coh/2}$, be dominated by $\coh^{-1/2}$.
That is, 
\begin{equation}
	\max\limits_{s,s',t',t \, \in \blu\left[-\tau,\tau\right]} \delta G_{\rm model - ideal}^{(2)}(s,s',t',t) =  o \blk\left( \frak{C}^{-1/2} \right)
	\label{eq:this_is_wellapproximate}
\end{equation}
\blu is a precise and sufficient condition \blk for the requirement that $G^{(2)}_{\rm model}$ well approximates $G^{(2)}_{\rm ideal}$ in Condition 4.


\normalsize

\section*{Data availability}
\small
All data files, including those presented in the main text and supplementary information's figures, are available from the corresponding author on request.
\normalsize

\section*{Code availability}
\small
The 
iMPS codes used in this study are available from the
corresponding author upon request.
\normalsize

\vspace{60ex}

\pagebreak
\onecolumn
\setcounter{equation}{0}
\setcounter{figure}{0}
\setcounter{table}{0}
\setcounter{page}{1}
\makeatletter
\renewcommand{\theequation}{S\arabic{equation}}
\renewcommand{\thefigure}{S\arabic{figure}}

\begin{center}
{\Large  \bf Supplementary Information for } \vspace{1ex} \\ {\Large \bf \emph{The Heisenberg limit for laser coherence}}
\end{center}
\vspace{2ex}

In this Supplementary Information, we present further details of the analytic and numerical methods that have been employed to derive the results presented in the main text, ``The Heisenberg limit for laser coherence''. 
Following the systematic steps discussed in the main text, 
this document is organized accordingly.
\emph{First}, in \sref{sec:1lasers}, we elaborate on the four conditions we have assumed to derive the Heisenberg limit for coherence.
This entails an evaluation of the coherence $\coh$---the maximal occupation number over all field modes--for a laser.
This section also contains a number of mathematical results which follow from the four conditions, and will later allow us to rigorously derive the upper bound on $\mathfrak{C}$.
Moreover, we discuss Glauber coherence functions and how these relate to the standard ideal laser model, which gives the standard quantum limit to laser coherence.
\emph{Second}, we derive the upper bound on the coherence in \sref{sec:UpperBound}, making no assumptions on the laser besides the four conditions stated in the main text.
\emph{Third}, we introduce a family of laser models in \sref{sec:MPS-methods}, described using tensor networks.
The details of our tensor network simulations, showing that a Heisenberg-limited scaling for $\coh$ is achievable by our model, are also presented in that section. 
\emph{Fourth}, we show that our family of models satisfies Conditions 1--4 in \sref{sec:proving_all_conditions_hold}.
\blu \emph{Fifth}, we introduce a circuit QED system which can realise our model in discrete time, producing a sequence of microwave pulses achieving the Heisenberg limit in \sref{sec:pathway}.
\blk

\tableofcontents

\section{Lasers and Coherence}
\label{sec:1lasers}

In this first section, we supplement the presentation of the four conditions we have placed on our laser model with an in-depth discussion of each.
One subsection below is dedicated to each condition.
We derive a number of mathematical results which stem from these conditions, which will be essential for proving the upper bound on $\coh$ in the next section.
\blu In \sref{sec:ideallaser}, \blk a derivation of the coherence for a conventional laser model \blu is provided, \blk which we have referred to as the {\it standard quantum limit}.
\blu To conclude, in \sref{sec:other_coherence_measures} we derive a quantitative relation between our measure of coherence and a standard \blu measure of coherence used in quantum information theory. \blk

\subsection{Description of a 1-D field}
\label{sec:1Dfield}

In the main text, we defined the coherence $\mathfrak{C}$ of a bosonic field very generally  to be the mean number of photons in a single spatial mode, maximized over all modes $\mathfrak{u}$ within some frequency band.
This measure of coherence can be written as
\begin{equation}
	\coh \coloneqq \max_{u \in \mathfrak{u}} \langle \hat b_{u}\dg \hat{b}_u \rangle,
\end{equation}
where $\hat{b}_u$ is the annihilation operator for mode $u$,
the maximization is over the set of modes $\mathfrak{u}$ which are normalised, $\int\! |u({\bf r})|d^3{\bf r}\,=\,1$, and are such that the support of the Fourier transform  $\tilde{u}$ is contained within a spherical shell in the space of wavevectors.
The thickness of the shell corresponds to the frequency band of interest. 
The consideration of the frequency band is important in order to avoid the infrared divergence, whereby a thermal state with an arbitrarily low frequency has a diverging mean photon number $\mu = (\exp(\hbar \omega/k_B T)-1)^{-1}$.
By our measure of coherence, the radiation from such a thermal beam may have a large coherence, but it is not at a useful frequency. 
Note that a field with high coherence need not have an absolute phase or mean field; it would be nonsensical to impose such a requirement because all phases that we can observe are relative phases~\cite{sBRS07}.

\begin{figure}[htbp]
   \centering
   \includegraphics{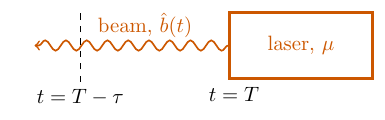}
   \caption{A laser cavity with stored energy $\mu$ which produces one-dimensional beam, described by the field operator $\hat{b}(t)$.}
   \label{fig:cavity_and_a_beam}
\end{figure}

Here, we  apply this definition to the case of a one-dimensional beam travelling at a fixed speed, with translationally invariant statistics. 
Due to the fixed speed, we can describe this bosonic beam by the one-parameter field operator $\hat{b}(t)$ satisfying $[\hat{b}(t),\hat{b}^\dagger(s)]=\delta(t-s)$, and having statistics independent of translation in $t$ as depicted in Fig.~\ref{fig:cavity_and_a_beam}.
Similarly, we can define the mode over which we maximize the occupation by the complex waveform $u(t)$. 
 The  annihilation operator for this mode is 
\begin{equation}
  \hat{b}_u = \frac{1}{{\sqrt{I_u}}} \int_{-\infty}^{\infty} dt~ 
              u(t) \hat{b}(t)~,
\label{eq:b_u}
\end{equation}
 where 
${ I_u} = \int_{-\infty}^{\infty} dt~ |u(t)|^2$.
The mean occupation number of  this mode is  $\la \hat{N}_u \ra =
\la \hat{b}_u^\dagger \hat{b}_u \ra$.  
Now, we are interested in the maximum of this over all modes $u$,
%
\begin{equation}
  \frak{C} = \max_u \big[ \la \hat{N}_u \ra \big]~.
\label{eq:C-continuum-def}
\end{equation}

 Here we show that for a beam with 
translationally invariant statistics, the maximum is attained with a 
\emph{flat} waveform, with $|u(t)|^2=\text{const.}$
(Precisely speaking, such waveforms are  unnormalisable,  but 
one can consider a regularized waveform such as  $u(t)=\text{const.}~e^{-|t|/l}$ and eventually  take the limit   
$l\rightarrow\infty$.) 
Define a coherence correlation function as $\frak{c}(t') \coloneqq \la \hat{b}^\dagger(t+t') \hat{b}(t) \ra$, which is independent of choice of $t$. The occupation number can now be written as
\begin{equation}
  \la \hat{N}_u \ra = \frac{1}{{ I_u}} \iint_{t,t'=-\infty}^{\infty} dt \, dt'~
              u^*(t') u(t) \frak{c}(t'-t)~,
\label{eq:N_u}
\end{equation}
The above can be rewritten as
\begin{align}
\la \hat{N}_u \ra  
              &= \frac{1}{I_u} \iiint_{t,t',t''=-\infty}^{\infty} dt \, dt' \, dt''~
              u^*(t') u(t) \frak{c}(t'') \delta(t''-(t'-t)) \nonumber \\
              &= \frac{1}{2\pi I_u} \iiint_{t,t',t''=-\infty}^{\infty} dt \, dt' \, dt''~
              u^*(t') u(t) \frak{c}(t'') \int_{-\infty}^{\infty} d\omega \, e^{-i\omega(t''-(t'-t))} \nonumber \\
              &= \frac{1}{2\pi I_u} \int_{-\infty}^{\infty} d\omega~|\tilde{u}(\omega)|^2 
                 \tilde{\frak{c}}(\omega),
\label{eq:Nu-sorting}
\end{align}
where $\tilde{u}$ and $\tilde{\frak{c}}$ are the Fourier transforms of the waveform and coherence 
correlation functions, respectively. 
This final form of $\la \hat{N}_u \ra$, \eref{eq:Nu-sorting}, clearly indicates that the choice of 
the waveform that leads to the maximum occupation number is
$ |\tilde{u}(\omega)|^2 = \delta(\omega - \omega_{\rm max}) $, where 
$\omega_{\rm max} \coloneqq  {\rm argmax}_\omega\big(\tilde{\frak{c}}(\omega)\big)$. 
 
 Since the power spectrum is given by $P(\omega) \coloneqq  (1/2\pi)\tilde{\frak{c}}(\omega)$ \cite{sWisMil10},  the coherence can  also be written  as 
\begin{equation}
  \frak{C} = 2\pi \max_\omega \big[ {P}(\omega) \big]~,
\label{eq:C-P_omega-def}
\end{equation}
as claimed in the main text. 
 In fact, by moving to a suitable rotating frame, it is always possible to define $\hat{b}(t)$ such that the peak of the power spectrum is at $\omega=0$ (leading to $u(t)=\text{const.}$) and so 
 \begin{equation}
	\frak{C} = \int_{-\infty}^\infty dt' \la \hat{b}^\dagger(t+t') \hat{b}(t) \ra. 
\end{equation}

As mentioned in the main text, the state of a laser beam for the models developed in the 1960s and 70s~\cite{sLou73,sSarScuLam74} is equivalent to a fixed-intensity coherent state that undergoes phase diffusion. 
For such a beam, the diffusion rate $\ell$ is the full-width at half-maximum-height of the beam's (Lorentzian) dimensionless power spectrum $P(\omega)$, and reciprocally related to its coherence time, while the beam's photon statistics are Poissonian with a flux rate ${\cal N} = \int  d\omega P(\omega)$,  so that the peak of $P(\omega)$ is $(2/\pi){\cal N}/\ell$. 
It is notable that the coherence for such lasers  evaluates  to $\frak{C} = 4{\cal N}/\ell$ --- this is easy to interpret as the number of photons emitted in an interval over which the phase stays roughly the same. 
In the ideal limit of standard laser models~\cite{sLou73,sSarScuLam74,sWis99}, $\frak{C}$  turns out  to be $\frak{C}^{\rm ideal}_{\rm SQL} = \Theta(\mu^2)$.
We  take  the above description  to define  an ideal laser  beam. 

\subsection{Endogenous phase}
\label{sec:phase_autoch}

Condition 2 states that the phase of the laser beam is endogenous 
to the laser cavity.
In other words, all phase information imprinted onto the output beam proceeds {\it only} from the cavity.
This ensures there are no sources driving the laser cavity---if there were, such sources would have to be considered as part of the laser cavity, and contribute to the mean photon number $\mu$.

Since a laser is a device which transforms resources with a negligible amount of coherence into a beam with a high degree of optical coherence, it is natural to consider the laser cavity to be coupled to an environment  (in addition to the output beam). 
Denote by $\mathcal{B}(\mathcal{H})$ the set of bounded linear operators on a Hilbert space  $\mathcal{H}$.
At time $T$, the global state of the cavity plus environment is given by $\rho_{\rm c e}\in  \mathcal{B}(\mathcal{H}_c \otimes \mathcal{H}_{\rm e})$, where $\mathcal{H}_c$ and $\mathcal{H}_e$ are the Hilbert spaces associated with cavity and environment states, respectively.
At a later time $T' > T$, a new part of the beam will have been emitted in the interval $t\in[T,T']$.
For this part of the beam, we introduce a third Hilbert space, $\mathcal{H}_{\rm b}$. 
From time $T$ to $T'$, the evolution of the system must be unitary, which can be described generally by defining the superoperator ~$\mathcal{U}^{T\rightarrow T'}_{\rm ce}: \mathcal{B}(\mathcal{H}_c \otimes \mathcal{H}_{\rm e}) \rightarrow \mathcal{B}(\mathcal{H}_{c} \otimes \mathcal{H}_{\rm b} \otimes \mathcal{H}_{\rm e'})$.
The initial (unprimed) environment contains the incoherent inputs into the laser cavity, and the primed environment contains everything not counted as part of the cavity or beam---for example, cavity photons which are lost but do not end up in the beam.
Note that the dimension of $\mathcal{H}_{\rm e'}$ will be strictly less than that of $\mathcal{H}_{\rm e}$, for the evolution of the entire system to be unitary.
In this description, the state of the cavity and its beam is obtained by tracing over the environment. \blu
Throughout this Supplementary Information, we adopt the notational convention $\mathcal{U} (\bullet) \coloneqq \hat{U} \bullet \hat{U}\dg$, relating unitary superoperators to their corresponding unitary operators.
\blk

To express the endogenous phase condition in a convenient manner, we define unitary superoperators for the laser cavity \blu $\mathcal{U}^\theta_{\rm c} (\rho_{\rm c}) \equiv  \hat{U}^\theta_{\rm c} \rho_{\rm c}  \hat{U}^{\theta\dagger}_{\rm c}$, where $\hat{U}^\theta_{\rm c}\coloneqq e^{i\theta\hat{n}_{\rm c}}$ \blk is the unitary that describes an optical phase shift of $\theta$. \blk
 That is, the generator of the phase shift is  the cavity photon number $\hat{n}_{\rm c}$.
A superoperator of this kind generalizes to those which may act on joint systems. 
We do this by placing additional subscripts, such as $\mathcal{U}^\theta_{\rm cb} (\rho_{\rm cb}) \coloneqq \blu e^{i\theta(\hat{n}_{\rm c} + \hat{n}_{\rm b})} \blk \rho_{\rm cb}\blu e^{-i\theta(\hat{n}_{\rm c} + \hat{n}_{\rm b})}$, which shifts the phase of both the cavity and beam by $\theta$.
We also adopt $\mathcal{I}$ to represent the identity superoperator.
In terms of these superoperators, Condition 2 (endogenous phase) requires that the equivalence 
\begin{equation}
	\Tr_{\rm e'}[\mathcal{U}^{T\rightarrow T'}_{\rm ce}((\mathcal{U}^\theta_{\rm c} \otimes \mathcal{I}_{\rm e}) \rho_{\rm ce}(T))] \equiv \Tr_{\rm e'}[(\mathcal{U}^\theta_{\rm cb} \otimes \mathcal{I}_{\rm e'}) (\mathcal{U}^{T\rightarrow T'}_{\rm ce}(\rho_{\rm ce}(T)))]
	\label{eq:endogenous_phase}
\end{equation}
holds, for all times $T' > T$, and all $\theta\in[0,2\pi)$.

The endogenous phase condition plays a crucial role in deriving the upper bound for coherence in Sec.~\ref{sec:UpperBound}. 
We conclude this subsection by deriving  a  lemma  showing  that, in the long-time limit, a phase covariant measurement on the beam prepares the cavity in a state which has the outcome encoded as an optical phase.
This will allow us to treat the encoded phase as a parameter which another observer can estimate (see \cite{sWisMil10}, Chapter 2).
The lemma involves the act of \emph{filtering}, which is a method of measuring the phase of the laser at time $T'$, say, by performing a measurement on a portion of the beam emitted \emph{before} $T'$.
Following sHolevo \cite{sHolevo}, we use the term \emph{covariant} measurement to refer to any positive operator valued measure (POVM) element $\hat{E}^\phi$, with $\phi$ being its outcome, that obeys $\mathcal{U}^{\theta}(\hat{E}^\phi) = \hat{E}^{\phi + \theta}$. 
 

\begin{lemma}[Cavity encoded phase from filtering]\label{lem:cavenc}
Suppose at time $T$ the cavity is in  a phase invariant  state  $\rho^{\rm inv}_{\rm c}$,  and the cavity, beam and environment are evolved up to time $T'$ by the unitary map ~$\mathcal{U}^{ T\rightarrow T'}_{\rm ce}: \mathcal{B}(\mathcal{H}_c \otimes \mathcal{H}_{\rm e}) \rightarrow \mathcal{B}(\mathcal{H}_{c} \otimes \mathcal{H}_{\rm b} \otimes \mathcal{H}_{\rm e'})$.
If a covariant phase measurement is performed on the beam emitted over the interval $[T, T')$, and outcome $\phi_F$ is obtained at time $T'$, the conditioned state of the cavity is equivalent to a fiducial state $\rho^{\rm fid}_{\rm c\blu}$ with an optical phase $\phi_F$ encoded by the generator $\hat{n}_{\rm c}$.
That is,
\begin{equation}
	\rho_{ {\blu\rm c} | \phi_F}(T') ={} \mathcal{U}^{\phi_F}_{\rm c}(\rho^{\rm fid}_{\blu\rm c}), \label{eq:stationarity_invariance}
\end{equation} \blk
where  $\mathcal{U}^{\phi_F}_{\rm c} (\rho_{\rm c}) \coloneqq \blu e^{i\phi_F\hat{n}_{\rm c}} \blk\rho_{\rm c} \blu e^{-i\phi_F\hat{n}_{\rm c}}$, \blk and the fiducial state $\rho^{\rm fid}_{\blu\rm c}$ is independent of $\phi_F$.
\begin{proof}
At time $T$, the state of the cavity is in the state  $\rho^{\rm inv}_{\rm c}(T)$, which is  invariant under optical phase shifts,  $\rho^{\rm inv}_{\rm c}(T) = \mathcal{U}^\theta_{\rm c} (\rho^{\rm inv}_{\rm c}(T))~\forall \theta$. 
Suppose the filtering measurement, represented by the phase covariant beam operator $\hat{E}^{\phi_F}_{\rm b}$ with outcome $\phi_F$, is performed on the beam emitted over the interval $[T, T')$.
The  unnormalised state of the cavity and environment  at time $T'$,  conditioned on performing covariant phase measurement with result $\phi_F$ on the beam, can be written as  
\begin{align}
	\rho_{ {\rm \blu c} | \phi_F} &\propto{} \Tr_{\rm b\blu e' \blk}\left[ \mathcal{U}^{T\rightarrow T'}_{\rm ce}((\mathcal{U}^\theta_{\rm c}\otimes \mathcal{I}_{\rm e})(\rho^{\rm inv}_{\rm ce}(T))) (I_{\rm c} \otimes \hat{E}^{\phi_F}_{\rm b} \otimes I_{\rm e'}) \right] \\
	&={} \Tr_{\rm b\blu e' \blk}\left[ \mathcal{U}^\theta_{\rm cb} (\mathcal{U}^{T\rightarrow T'}_{\rm ce}(\rho^{\rm inv}_{\rm ce}(T))) (I_{\rm c} \otimes \mathcal{U}^{\phi_F}_{\rm b}(\hat{E}^{0}_{\rm b}) \otimes I_{\rm e'}) \right] \\ 
	&={} \Tr_{\rm b\blu e' \blk}\left[ ( \mathcal{U}^{\theta}_{\rm c} \otimes \mathcal{U}^{\theta - \phi_F}_{\rm b}  \otimes \mathcal{I}_{\rm e'}) \mathcal{U}^{T\rightarrow T'}_{\rm ce}(\rho^{\rm inv}_{\rm ce}(T)) (I_{\rm c} \otimes \hat{E}^{0}_{\rm b}\otimes I_{\rm e'}) \right],
\end{align} \blk
where we have used Condition 2 (endogenous phase) from the first to the second line, and the fact that $\hat{E}^{\phi_F}_{\rm b}$ is phase covariant. 
The norm of this state is the probability of its realization.
Since the above expression must be true for all $\theta$, we can choose  $\theta=\phi_F$. 
This means the cavity and environment are in the normalized state
\begin{align}
	\rho_{ {\rm \blu c} | \phi_F} &={} \frac{\Tr_{\rm b\blu e'\blk}\left[ (\mathcal{U}^{\phi_F}_{\rm c}\otimes \mathcal{I}_{\rm b}\otimes \mathcal{I}_{\rm e'})(\rho_{\rm cbe'}(T')) (I_{\rm c} \otimes \hat{E}^{0}_{\rm b}\otimes I_{\rm e'}) \right]}{\Tr \small[ \rho_{\rm b}\hat{E}^{0}_{\rm b} \small]} \\
	&={} \mathcal{U}^{\phi_F}_{\rm c}(\rho^{\rm fid}_{\blu\rm c}),
\end{align} \blk
where we have introduced a fiducial state $\rho^{\rm fid}_{\blu\rm c} \coloneqq \Tr_{\rm b\blu e'} [ \rho_{\rm cbe'} (I \otimes \hat{E}^{0}_{\rm b}\otimes I_{\rm e'})] / \Tr \small[ \rho_{\rm cb}(I \otimes \hat{E}^{0}_{\rm b}) \small]$, \blk which does not depend on $\phi_F$.
\end{proof}
\end{lemma}

\subsection{Stationarity}
\label{sec:stationarity}

Condition 3 in the main text places the restriction of stationarity on the laser cavity and its beam.
Restated, this condition requires that the statistics of the laser and beam are invariant under time translation, in the long-time limit.
\blu When the laser cavity is in this unique steady state, \blk we denote it by $\rho^{\rm ss}_{\rm c}$, for which $\langle \hat{n}_{\rm c} \rangle = \mu$.
An important consequence of this condition is that the steady state of the cavity must be invariant under optical phase shifts, which is shown by the following lemma.
\begin{lemma}[Steady state phase shift invariance]\label{lem:stationary_shift_invariance}
\blu If there is a unique steady state of the cavity, it \blk is invariant under all optical phase shifts,
\begin{equation}
	\mathcal{U}^\theta_{\rm c} (\rho^{\rm ss}_{\rm c})=\rho^{\rm ss}_{\rm c}~~\forall~\theta.
\end{equation}
\begin{proof}
Suppose at time $T$, the laser is in its steady state, \blu which is assumed to be unique. \blk
From Condition 2 (endogenous phase), applying an optical phase shift of $\theta$ at this time  and waiting until time $T'$  is equivalent to  waiting  until time $T'$ and  then  shifting the phase of both the cavity and  newly created beam segment  by $\theta$,
\begin{equation}
	\Tr_{\rm e'}[\mathcal{U}^{T\rightarrow T'}_{\rm ce}((\mathcal{U}^\theta_{\rm c} \otimes \mathcal{I}_{\rm e}) \rho_{\rm ce}(T))] \blu = \blk \Tr_{\rm e'}[(\mathcal{U}^\theta_{\rm cb} \otimes \mathcal{I}_{\rm e'}) (\mathcal{U}^{T\rightarrow T'}_{\rm ce}(\rho_{\rm ce}(T)))].
	\label{eq:endogenous_phase_trb}
\end{equation}
To prove the lemma, we take the limit $T'-T \rightarrow \infty$ and trace over the beam.
The left-hand side becomes
\begin{equation}
	\lim\limits_{T'-T\rightarrow\infty}\Tr_{\rm be'}[\mathcal{U}^{T\rightarrow T'}_{\rm ce}((\mathcal{U}^\theta_{\rm c} \otimes \mathcal{I}_{\rm e}) \rho_{\rm ce}(T))] = \rho^{\rm ss}_{\rm c},
\end{equation}
and the right-hand side is
\begin{equation}
	\lim\limits_{T'-T\rightarrow\infty} \Tr_{\rm be'}[(\mathcal{U}^\theta_{\rm cb} \otimes \mathcal{I}_{\rm e'}) (\mathcal{U}^{T\rightarrow T'}_{\rm ce}(\rho_{\rm ce}(T)))] = \mathcal{U}^\theta_{\rm c}(\rho^{\rm ss}_{\rm c}).
\end{equation}
Since Eq.~\eqref{eq:endogenous_phase_trb} must hold $\forall \theta$, the lemma follows.
\end{proof}
\end{lemma}
From this lemma, it is straightforward to verify that $\rho^{\rm ss}_{\rm c}$ must be diagonal when expressed in the photon number basis.

Using the stationarity assumption, we are now in a position to prove  a lemma  which  arises  from Conditions 2 and 3, that will be crucial in proving the upper bound on $\coh$ in the next section.
It ensures that all states of the cavity which are conditioned upon covariant phase measurements of the beam have  the same  mean photon number.

\begin{lemma}[Phase encoding preserves photon number statistics]\label{lem:mu_invariant}
For arbitrary covariant phase measurements on the beam, the fiducial state of the cavity $\rho^{\rm fid}_{\rm\blu c}$ defined in Lemma~\ref{lem:cavenc}, has the same photon number statistics as the  phase-invariant  steady state, $\bra{n}\rho^{\rm fid}_{\rm c} \ket{n} = \bra{n}\rho^{\rm ss}_{\rm c} \ket{n}$. 
\begin{proof}
Begin with the laser cavity at time  $T$  in its steady state $\rho^{\rm ss}_{\rm c}$.
At a later time  $T'$,  the joint state of the cavity and the beam it emitted since  $T$  will be  $\rho_{\rm cb}(T')=\Tr_{\rm e'}[\mathcal{U}^{T\rightarrow T'}_{\rm ce}(\rho^{\rm ss}_{\rm ce}(T))]$.
In the absence of measurements on the beam which may steer the state of the cavity, the cavity remains invariant,  $\Tr_{\rm b}[\rho_{\rm cb}(T')] = \rho^{\rm ss}_{\rm c}$. 
To prove the lemma, we need to relate $\rho^{\rm ss}_{\rm c}$ to the fiducial state $\rho^{\rm fid}_{\rm c}$.
To this end, for phase covariant POVM elements $\hat{E}^\phi_{\rm b}$ we use the identity  $\int d\phi (2\pi)^{-1} \hat{E}^\phi_{\rm b} = I_{\rm b}$.
From the endogenous phase and stationarity assumptions, the state of the cavity is given by
\begin{align}
	\rho^{\rm ss}_{\rm c} &={} \Tr_{\rm be'}\left[ \mathcal{U}^{T\rightarrow T'}_{\rm ce}(\mathcal{U}^\theta_{\rm c}\otimes \mathcal{I}_{\rm e}(\rho^{\rm ss}_{\rm ce}(T)))\left( I_{\rm c} \otimes \int \frac{d\phi}{2\pi} \hat{E}^\phi_{\rm b}\otimes I_{\rm e'} \right)\right] \\
	&={} \int \frac{d\phi}{2\pi} \Tr_{\rm be'}\left[ \mathcal{U}^\theta_{\rm cb}\otimes \mathcal{I}_{\rm e'}(\rho_{\rm cbe'}(T') )\left(I_{\rm c} \otimes \mathcal{U}^{\phi}_{\rm b} (\hat{E}^0_{\rm b})\otimes I_{\rm e'} \right) \right] \\
	&={} \int \frac{d\phi}{2\pi} \Tr_{\rm b}\left[ \mathcal{U}^\theta_{\rm c} \otimes \mathcal{U}^{\theta - \phi }_{\rm b}(\rho_{\rm cb}(T')) \left(I_{\rm c} \otimes  \hat{E}^0_{\rm b}\right) \right].
\end{align}
The second line follows from the covariance of $\hat{E}^\phi_{\rm b}$.
Choosing $\theta = \phi$  here shows that
\begin{align}
	\rho^{\rm ss}_{\rm c} &={} \int \frac{d\phi}{2\pi} \Tr_{\rm b}\left[ (\mathcal{U}^{\phi}_{\rm c} \otimes \mathcal{I}_{\rm b}) \rho_{\rm cb}(T') \left(I_{\rm c} \otimes  \hat{E}^0_{\rm b}\right) \right] \\
	&={} \int \frac{d\phi}{2\pi} \mathcal{U}^{\phi}_{\rm c} (\rho^{\rm fid}_{\rm c}).
\end{align}
Since $\rho^{\rm fid}_{\rm c}$ is independent of $\phi$, the photon number  probability distribution  is
\begin{align}
	\bra{n} \rho^{\rm ss}_{\rm c} \ket{n} &={} \int \frac{d\phi}{2\pi} \bra{n} \mathcal{U}^{\phi}_{\rm c} (\rho^{\rm fid}_{\rm c}) \ket{n} \\
	&={} \bra{n} \rho^{\rm fid}_{\rm c} \ket{n}.
\end{align}
\end{proof}
\end{lemma}

\subsection{Glauber-$^{(1),(2)}$ ideality}
\label{sec:glauber_ideality}

The fourth Condition we have placed on the laser involves the family of correlation functions used by Glauber \cite{sGla63a} to describe the optical coherence properties of a quantum field.
This family of $n^{\rm th}$ order correlation functions are defined most generally in terms of $2n$ field operators being evaluated at different spatial coordinates $s_i$ of a field,
\begin{equation}
	G^{(n)}(s_1,...,s_{2n}) \coloneqq  \langle \hat{b}^\dagger(s_1) \cdots \hat{b}^\dagger(s_n) \hat{b}(s_{n+1})\cdots \hat{b}(s_{2n}) \rangle.
\end{equation}
The phrase \emph{$n^{\rm th}$ order coherence} was used to refer to a radiation field for which the normalized correlation function \cite{sGla63a} 
\begin{equation}
	g^{(m)}\coloneqq \frac{G^{(m)}(s_1,...,s_{2n})}{\Pi_{i=1}^{2n} [G^{(1)}(s_i, s_i)]^{1/2}}
\end{equation}
is equal to unity, for all $m\leq n$.  
A perfectly coherent beam was considered by Glauber to be one that is coherent to all orders; for example, a classical plane wave which has a well defined intensity and phase.
 Note that for generic time arguments, $g^{(1)}$ cannot be equal to unity due to finite linewidth of any realistic laser, but must decay over the timescale of $\ell^{-1}$. 
This is true for generic arguments of the higher-order normalized Glauber functions too.

In the context of deriving the Heisenberg limit for laser coherence, we are not concerned with arbitrarily high orders of the Glauber coherence functions.
Rather, we restrict the discussion to the first and second order functions, since they are sufficient for proving the upper bound on the coherence.
Condition 4 is the constraint that these two functions for a laser model,
\begin{align}
	g^{(1)}(s,t) &={} \mathcal{N}^{-1}\langle \hat b\dg(s) \hat b(t) \rangle, \\
	g^{(2)}(s,s',t',t) &={} \mathcal{N}^{-2}\langle \hat b\dg(s) \hat b\dg(s') \hat b(t') \hat b(t) \rangle, \label{eq:G2definition}
\end{align}
\emph{well approximate} those of a stochastic coherent state.
\blu Note that these are the normalized versions of the Glauber functions defined in Eq.~(1) of the main text. \blk
As we saw in Sec.~\ref{sec:1Dfield}, the two-point correlations $G^{(1)}(s,t)$ directly connect to the coherence of the beam, via $\frak{C} = \int_{-\infty}^\infty ds~ G^{(1)}(s,t)$, and thus relates to the relative phase properties of the beam.
The second order correlations provide information about the beam's intensity and phase correlations, and will play a crucial role in deriving the upper bound on $\coh$ in the next section.

Later, we will introduce a laser model that achieves the Heisenberg limit for laser coherence.
In order to show this model satisfies the condition of Glauber-$^{(1),(2)}$ ideality, we define two functions which quantify the difference between the functions $g^{(1)}(s,t)$ and $g^{(2)}(s,s',t',t)$ from the  ideal laser model. 
These are
\begin{align}
	\delta g_{\rm model - ideal}^{(1)}(s,t) &\coloneqq {} g^{(1)}_{\rm model}(s,t) - g_{\rm ideal}^{(1)}(s,t) \label{eq:g1_ideality}\\
	\delta g_{\rm model - ideal}^{(2)}(s,s',t',t) &\coloneqq {} g^{(2)}_{\rm model}(s,s',t',t) - g_{\rm ideal}^{(2)}(s,s',t',t). \label{eq:g2_ideality} 
\end{align}
Here, ``ideal'' refers to the standard laser model, in which the quantities of interest are derived from assuming the beam is in a  constant intensity  coherent state with its phase undergoing pure diffusion.
The label ``model'' refers to the model we introduced in the main text,  which  is described in  detail  in Sec.~\ref{sec:MPS-methods}.
Note that in the main text, we adopted  $\mathcal{N}=1$ for all our calculations, and hence wrote $\delta G_{\rm model - ideal}^{(n)}$ in place of $\delta g_{\rm model - ideal}^{(n)}$. 
These difference functions allow us to recast the phrase \emph{well approximate}  from Condition 4 in terms of mathematical statements about their behaviour in the limit of large $\mu$.
The details of this analysis are given in \sref{sec:proving_all_conditions_hold}.

We now derive expressions for the first- and second-order coherence functions in the ``ideal'' case analytically, by assuming the laser beam is described by a phase-diffusing coherent state.
 This  stochastic coherent state satisfies the eigenvalue equation $\hat{b}(t) \ket{\beta(t)}  = \beta(t) \ket{\beta(t)}$, where $\beta(t) = \sqrt{\cal N} e^{i\sqrt{\ell}W(t)}$ and $W(t)$ is a Wiener process~\cite{sCarmichael99_book}.
From these relations, the first order  coherence  function is given by
\begin{equation}
	g^{(1)}_{\rm ideal}(s,t) ={} \langle e^{i\sqrt{\ell}[W(t) - W(s)]} \rangle.
\end{equation}
Now, for a Gaussian distribution of a variable $X$ with variance $\sigma^2$ and mean zero, it can be shown that $\langle e^{iX} \rangle = e^{-\sigma^2/2}.$
This gives 
\begin{equation}
	g^{(1)}_{\rm ideal}(s,t) ={} \exp\left[ -\frac{\ell}{2}\langle [W(t) - W(s)]^2\rangle\right].
	\label{eq:g1_ideal_wiener}
\end{equation}
The variance term in the exponential above is  $\langle [W(t)-W(s)]^2 \rangle = |t-s|$. 
Hence, for a coherent beam with its phase undergoing pure diffusion, 
\begin{equation}
	g^{(1)}_{\rm ideal}(s,t) ={} \exp\left[ -\frac{\ell}{2}|s-t|\right].
	\label{eq:g1_ideal}
\end{equation}

The derivation of the second order correlation function is similar.
 The  four-time correlation function in Eq.~\eqref{eq:G2definition} is 
\begin{equation}
	g^{(2)}_{\rm ideal}(s,s',t',t) ={}\langle e^{i\sqrt{\ell}[W(t)+W(t')-W(s)-W(s')]} \rangle.
\end{equation}
 It is not difficult to show that 
\begin{equation}
\langle [W(s)+W(s')-W(t)-W(t')]^2 \rangle = |s-t|+|s'-t'|+|s-t'|+|t-s'|-|s-s'|-|t-t'|,
\end{equation}
and so
\begin{equation}
g^{(2)}_{\rm ideal}(s,s',t',t) = \exp \left[
-\frac{\ell}2 \left( |s-t|+|s'-t'|+|s-t'|+|t-s'|-|s-s'|-|t-t'|\right)\right].
\label{eq:g2_idealsts't'}
\end{equation}

In the main text, we introduced one additional function, $\delta g_{\rm model - 1}^{(2)}\coloneqq  g_{\rm model }^{(2)} - 1$,  which quantifies the difference between the normalized second-order correlation function in our laser model, and unity.
In fact, this leads to a sufficient (as we shall later show) quantitative condition, that ensures $g_{\rm model }^{(2)}$  \emph{well approximates} $g_{\rm ideal }^{(2)}$,  given by
\begin{equation} \label{firstwellapp}
	\delta g_{\rm model - ideal}^{(2)} = o(\delta g_{\rm ideal - 1}^{(2)}).
\end{equation}
\blu For the first-order coherence function the analogous quantity can also be introduced, but a far weaker constraint, 
\begin{equation} \label{secondwellapp}
	\delta g_{\rm model - ideal}^{(1)} = o(1), 
\end{equation}
which is implied by any reasonable interpretation of ``well approximate'', will serve as a sufficient condition for this. 
These conditions arise \blk from demanding that the upper bound on $\coh$, derived in \sref{sec:UpperBound}, applies for our laser model.
We refer the reader to \sref{sec:G1ideality} for further discussion and evaluation of \blu the relevant quantities.\blk

\subsection{The ideal laser model and the standard quantum limit}
\label{sec:ideallaser}

For completeness, we will now present a derivation of the standard quantum limit to the scaling of $\coh$, which can be achieved using an argument based on the uncertainty relation \cite{sWis99}.
Consider an ideal single-mode laser which can be considered to be in a coherent state $\ket{\alpha}$ with fixed amplitude $|\alpha| = \sqrt{\mu} \gg 1$, and a phase which will vary randomly as we shall see. 
The phase/number uncertainty relation $V (\phi) V (n) \gtrsim 1/4$ (which is rigorous in terms of the sHolevo phase variance~\cite{sHol84}) is---to a good approximation---saturated for large-amplitude coherent states, $V (\phi) V (n) \sim 1/4$. 
Since for coherent states $V(n) = \langle \hat{n} \rangle = \mu$, we have initially $V (\phi) = 1/4\mu$.

Let the cavity damping rate due to the output coupling creating the beam be $\kappa$, and assume a linear coupling between the laser cavity and the beam.
Consider the infinitesimal time $dt$.
The effect of damping over this time will induce a change in the mean photon number $\mu \rightarrow \mu (1 - \kappa dt)$. 
Since linear damping takes coherent states to coherent states, the change in phase variance after this short time is
\begin{equation}
	V(\phi) + dV (\phi)=  \frac{1}{4\mu(1 -\kappa dt)} \implies dV(\phi) \sim  \frac{\kappa dt}{4\mu}.
	\label{eq:tinyphasevariance}
\end{equation}
Now consider the process of gain into the laser cavity.
If there is no coherent input to the laser, then $dV(\phi)$ cannot decrease from the value in Eq.~\eqref{eq:tinyphasevariance}.
In the best case scenario where no phase noise is added, the amplitude of the coherent state is restored to its original value of $\mu$. 
Therefore, for a cavity which is pumped as it is damped, the phase variance must increase by at least $\kappa dt/(4\mu)$. 
Since each such time increment is independent, we can think of this as a coherent state $\ket{\sqrt{\mu}e^{i\theta(t)}}$ with an unknown phase $\theta(t)$ undergoing a classical random walk with variance growing as $\kappa t/(4\mu)$.
Now the coherence for such a beam is given by 
\begin{align}
	\mathfrak{C}^{\rm ideal}_{\rm SQL} ={} \int\limits_{-\infty}^{\infty} ds  ~G^{(1)}_{}(s,t) ={} \kappa \mu \int\limits_{-\infty}^{\infty} ds ~e^{-V[\phi(s) - \phi(t)]/2} 
\end{align} 
where we have used the result again that $\langle e^X \rangle = e^{-V(X)/2}$ for any zero-mean Gaussian variable $X$.
Using the arguments above to replace the variance in the exponential by $\kappa|s-t|/(4\mu)$, we can compute the integral to find 
\begin{equation}
	\mathfrak{C}^{\rm ideal}_{\rm SQL} \leq{} \kappa \mu \int\limits_{-\infty}^{\infty} ds ~e^{-\kappa |s - t|/(8\mu)} ={} 16 \mu^2.
\end{equation} 
This is the scaling for the standard quantum limit for laser coherence, $\mathfrak{C}^{\rm ideal}_{\rm SQL} = \Theta(\mu^2)$. 

\subsection{Relation to other coherence measures}
\label{sec:other_coherence_measures}
\blu
Although we have been primarily concerned with describing \emph{optical} coherence, there are many definitions of coherence in quantum information theory (see \cite{sStr17} for a review).
In this section, we discuss how our definition of optical coherence in Eq.~\eqref{eq:C-P_omega-def} relates to one such coherence measure.
In particular, we will derive an explicit relationship between our measure, and the measure of coherence introduced in Ref.~\cite{sVac08}.
Given a group $G$, with unitary representation $U(g)$ for $g\in G$, the $G$-asymmetry of a quantum state $\rho$ is defined as
\begin{equation}
    A_G(\rho) \coloneqq S(\mathcal{G}_G[\rho]) - S(\rho),
    \label{eq:def_asymmetry}
\end{equation}
where $\mathcal{G}_G[\rho] = \int dg \, U(g) \rho U\dg(g)$ is the group average with respect to the Haar measure $dg$, and $S(\rho)\coloneqq -\Tr \rho\ln\rho$ is the von-Neumann entropy.
States which are invariant under the action of $U(g) \forall g\in G$ are said to be symmetric with respect to $G$.
We adopt this quantity as an entropic measure of the asymmetry of the state $\rho$ with respect to $G$.

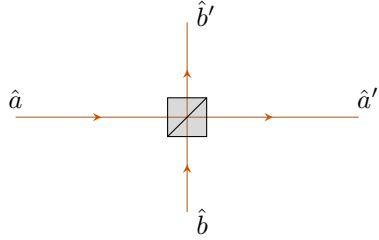
\begin{figure}[!h]
\centering
\begin{tikzpicture}
    \node (source) {};
    \node[right=2 of source, minimum size=0.5cm] (splitter) {}; 
    \node[above=1 of splitter] (top) {};
    \node[below=1 of splitter] (bottom) {};
    \node[right=2 of splitter] (right) {};
    \draw[color=tenne, postaction={on each segment={mid arrow=tenne}}] (source.east) node[above] {\textcolor{black}{$\hat{a}$}}  -- (splitter.center) ;

    \draw[color=tenne, postaction={on each segment={mid arrow=tenne}}] (splitter.center) -- (right) node[above] {\textcolor{black}{$\hat{a}'$}};
    \draw[color=tenne, postaction={on each segment={mid arrow=tenne}}] (splitter.center) -- (top) node[right] {\textcolor{black}{$\hat{b}'$}};
    \draw[color=tenne, postaction={on each segment={mid arrow=tenne}}] (bottom) node[right] {\textcolor{black}{$\hat{b}$}} -- (splitter.center);
    \draw[beamsplitter] (splitter.north west) rectangle (splitter.south east);
    \draw (splitter.north east) -- (splitter.south west);
\end{tikzpicture}
\caption{Beam splitter and its input/output modes, which we consider in order to measure the relative coherence of a beam.}
\label{fig:beamsplitter}
\end{figure}

Recall that all measurements of optical phase require an independent phase reference.
This means that---in the context of analyzing the coherence (optical or otherwise) of any laser beam---one can only make meaningful statements about \emph{relative} coherence.
In the absence of a phase reference, we can obtain a measure of a beam's relative coherence by imposing it on a 50:50 beam splitter, with a vacuum state incident in the other input port.
To this end, we consider the four optical field modes shown in Fig.~\ref{fig:beamsplitter}, where $\hat{a}, \hat{b} ~(\hat{a}', \hat{b}')$ are annihilation operators for the input (output) modes.
In the Heisenberg picture, the action of the beamsplitter transforms the input mode operators according to
\begin{align}
	\hat{a} &= \frac{1}{\sqrt{2}} \left( \hat{a}' + \hat{b}' \right), \label{eq:heisenberg_a_transformation}\\
	\hat{b} &= \frac{1}{\sqrt{2}} \left( \hat{a}' - \hat{b}' \right).
\end{align}
\blk

\blu
Suppose the input state to the beamsplitter has $N$ photons in mode $a$. 
Explicitly, $\ket{\psi_{\text{in}}} = \ket{N}_a\ket{0}_b = (N!)^{-1/2}(\hat{a}\dg)^N \ket{0}_a\ket{0}_b$.
Using the Heisenberg picture operator transformation in \eref{eq:heisenberg_a_transformation}, it is straightforward to calculate the output state of the beamsplitter as
\begin{align}
	\ket{\psi_{\text{out}}} &= \frac{1}{\sqrt{2^N N!}} \left( \hat{a}^{\prime\dagger} + \hat{b}^{\prime\dagger} \right)^N \ket{0}_{a'}\ket{0}_{b'} \\
	&= \frac{1}{\sqrt{2^N}} \sum\limits_{m=0}^N \sqrt{N \choose m} \ket{m}_{a'} \ket{N-m}_{b'}.
\end{align}
This means that, for a fixed number of input photons, the probability distribution of finding $n_{a'}$ and $n_{b'}$ photons in modes $a'$ and $b'$ respectively is binomial,
\begin{equation}
	P(n_{a'}, n_{b'}| N) = w_{n_{a}'|N} \, \delta_{n_{b'}, N-n_{a'}}, \ \textrm{ for } \ w_{i|j} \coloneqq \frac{1}{2^j}{j \choose i} . 
\end{equation}

To connect the above result to the measure of coherence $\coh$ used in the main text, we note that it was shown in the Appendix of Ref.~\cite{sWis16a} that any
monochromatic mode of any statistically stationary optical field that can be described by a stochastically varying coherent state, is populated thermally. 
That is to say, if the mode has mean occupation $\bar{n}$ then its state is  
\begin{equation}
	\rho_{\text{th}} (\bar{n}) =  \sum\limits_{j=0}^\infty \wp_j \ketbra{j}, \ \textrm{ for } \ \wp_j = \frac{1}{1+\bar{n}}\left( \frac{\bar{n}}{1+\bar{n}} \right)^j. 
\end{equation}
Thus we can write the output state of the beamsplitter, defined over modes $a'$, $b'$, as
\begin{equation}
	\rho' = \sum\limits_{j=0}^\infty  \sum\limits_{i,h=0}^j \wp_j \sqrt{w_{i|j}w_{h|j}} \ket{i,j-i} \bra{h,j-h}.
	\label{eq:beamsplitter_output}
\end{equation}

We now calculate the $G$-asymmetry of the state $\rho'$, defined by Eq. \eqref{eq:def_asymmetry},
\begin{equation}
    A_G(\rho') = S(\mathcal{G}_G[\rho']) - S(\rho'),
    \label{eq:beamsplitter_asymmetry}
\end{equation}
for which we take the group $G$ to be that represented by the group of unitary phase shift operators, $\{U(\theta): \theta\in[0,2\pi), U(\theta)=\exp(i\theta\hat{n})\}$, 
where $\hat{n}$ is the number operator for {\em one} of the output modes ($a'$ or $b'$)---it does not matter which one. That is, $A_G(\rho')$ will measure the 
{\em relative} coherence of the two output modes, which is the only sort of coherence that is operationally meaningful.   
The density operators which are symmetric with respect to this group are those diagonal in the photon number basis.
This implies that taking the average of a state $\rho$ with respect to this group is equivalent to applying the dephasing map, $\mathcal{G}_G[\rho]\equiv \sum_k (\ketbra{k}\otimes I)\rho (\ketbra{k}\otimes I)$.
Using Eq.~\eqref{eq:beamsplitter_output}, we obtain 
\begin{equation}
	\mathcal{G}_G[\rho'] = \sum\limits_{j=0}^\infty  \sum\limits_{k=0}^j \wp_j w_{k|j} \ketbra{k,j-k}. 
\end{equation}
Thus its von-Neumann entropy simplifies to the Shannon entropy $H$ of the distribution defined by the diagonal elements of this matrix,
\begin{align}
	S(\mathcal{G}_G[\rho']) &= H(\{ \wp_j w_{k|j} \}_{j,k}) \\
	&= H(\{\wp_j \}_j) + \sum_j \wp_j H(\{w_{k|j}\}_k),
\end{align}
where $H(\{\wp_j \}_j)\coloneqq -\sum_j \wp_j \ln \wp_j$, and the second line follows from the strong additivity of the Shannon entropy.
Now, since the action of the beamsplitter is a unitary, the second term in Eq.~\eqref{eq:beamsplitter_asymmetry} is 
\begin{align}
	S(\rho') &= S(\rho_{\text{th}} (\bar{n}) \otimes \ketbra{0}) \\
	&= H(\{ \wp_j \}_{j}).
\end{align}
From the above expressions, the $G$-asymmetry simplifies to
\begin{equation}
	S(\mathcal{G}_G[\rho']) = \sum\limits_j \wp_j H(\{w_{k|j} \}_{k}).
	\label{eq:last_sum_asymmetry}
\end{equation}
We calculate an explicit expression for this quantity in the limit of a large number of photons in the input port of the beamsplitter, $\bar{n}\rightarrow\infty$.
First, note that in this limit we can approximate $\bar{n}/(1+\bar{n})$ as $e^{-1/\bar{n}}$, so that $\wp_j \rightarrow \bar{n}^{-1} e^{-j/\bar{n}}$. 
We can also use the asymptotic formula for the entropy of a binomial distribution (see e.g.~Ref.~\cite{sHug96}), 
\begin{equation}
	-\sum_{k=0}^j \frac{1}{2^j}{j \choose k} \ln \left[\frac{1}{2^j}{j \choose k}\right] \sim \frac{1}{2} \ln \left[\frac{\pi e j}{2}\right]
\end{equation}
as $j\rightarrow\infty$, \blu because the greatest contribution of the sum in \eqref{eq:last_sum_asymmetry} will come from large $j$. Finally, 
we can convert this sum into \blu the integral
\begin{equation}
	A_G(\rho') \sim \frac{1}{2\bar{n}} \int\limits_{0}^\infty dj ~e^{- j/\bar{n}} \ln j,
\end{equation} 
where we have dropped additive factors of order unity. 
Using known integrals \cite[p.~571]{sGra14}, and again dropping additive factors of order unity, this evaluates to
\begin{align}
	A_G(\rho') &= \frac{1}{2} \ln(\bar{n}). 
\end{align} 
Maximizing the coherence by choosing the maximally populated mode of the laser beam, 
so that $\bar{n} = \coh$, we get finally
\begin{align}
	A_G(\rho') &=  \frac{1}{2} \ln\coh .  	
\end{align}
Thus, we see a direct correspondence between our measure of laser coherence, and the relative $G$-asymmetry of coherence.
This means that the Heisenberg limit scaling for $\coh$ has implications for the fundamental limits on other types of coherence used in quantum information.
\blk

\section{Analytical upper bound on $\mathfrak{C}$}
\label{sec:UpperBound}

In this section we prove the claim stated in the main text that $\frak{C} = O(\mu^4)$, for any laser model which satisfies our four conditions.  
We begin with an illustration which outlines the proof, before presenting the full derivation.

\subsection{Sketch of the proof}
\label{sec:proofsketch}

The proof involves consideration of two observers, and three methods by which the optical phase of the laser at a time $T$ could be estimated. 
The first method we refer to as {\it filtering}, wherein an observer performs heterodyne measurements over the beam emitted before $T$ in a finite interval  $[T-\tau, T)$ of length $\tau$. 
The second method we call {\it retrofiltering}, which---like filtering---relies on measuring the phase of the beam proceeding from the cavity, with the difference that the finite segment of the measured beam is emitted \emph{after} $T$.
That is, retrofiltering involves heterodyning the beam over the interval $(T,T+\tau]$.
Finally, the most direct way that the phase of the cavity could be estimated at $T$ is by performing a direct measurement on the cavity at this time.

\begin{figure}[htbp]
   \centering
   \includegraphics{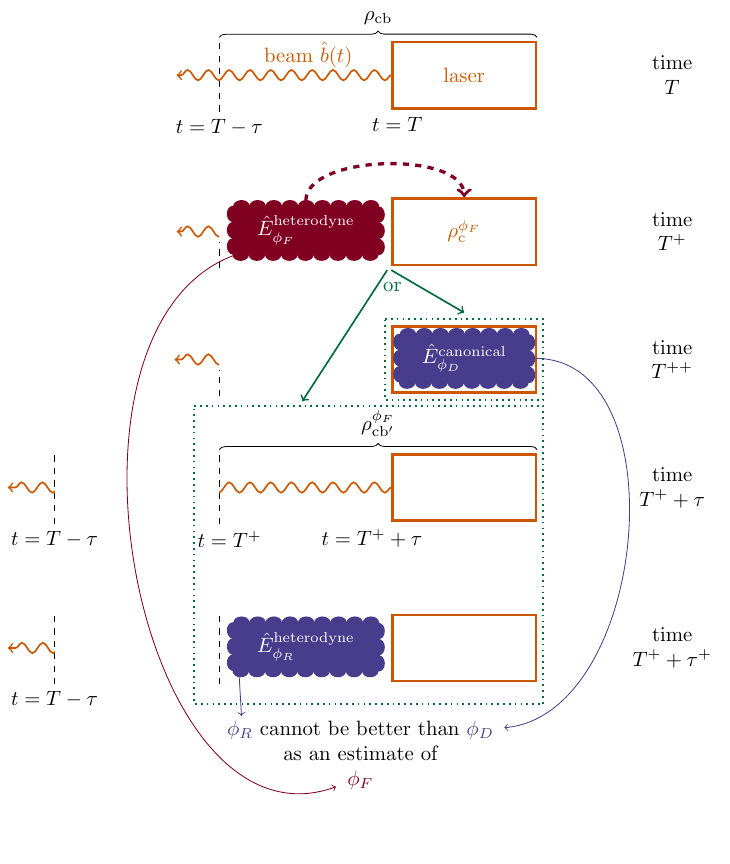}
   \caption{Illustration of the method for deriving the upper bound $\coh = \Theta(\mu^4)$.
   Time increases from the top to the bottom but is branched. 
   Initially, we consider the cavity at time $T$ in steady state, and the segment of the beam emitted since time $T-\tau$.
   At an immediately following time $T^+>T$, we suppose an observer, Effie, can perform heterodyne filtering over the beam emitted in the interval $[T-\tau,T)$, to obtain a phase estimate $\phi_F$ of the cavity at time $T$.
   The state of the cavity, conditioned on her measurement is $\rho^{\phi_F}_{\rm c}$.
   Now, there are two ways a second observer, Rod, could measure the phase of the cavity (green arrows).
   The first method allowing the cavity to emit the beam up until time $T^++\tau$, upon which heterodyne retrofiltering could be performed over the interval $(T^+,T^+ +\tau]$, yielding an estimate $\phi_R$ at time $T^+ + \tau^+$. 
   The second would consist of a canonical phase measurement performed directly on the cavity at time $T^{++}$ immediately following $T^+$, with outcome $\phi_D$.
   Since the result $\phi_R$ cannot be better than $\phi_D$ as a phase estimate of $\phi_F$ , the upper bound on $\coh$ follows from known results on optimal covariant phase estimation.}
   \label{fig:upperboundvisualisation}
\end{figure}

The methodology behind the proof requires considering the first observer, Effie, to  undertake a  filtering measurement  which encodes her result as an optical phase in the state of the cavity.
This creates a phase estimation problem for a second observer, Rod, who can  perform either  retrofiltering or a direct cavity measurement to infer Effie's result.
Since, as we show, Rod's retrofiltering cannot outperform his direct cavity measurement as an estimate of the encoded phase, we can use known bounds on covariant phase estimation to prove the upper bound on the coherence.
These ideas are illustrated in Fig.~\ref{fig:upperboundvisualisation}.

\subsection{Main theorem}

To rigorously prove the upper bound, we first begin by defining unitary operators $e^{i\hat{\phi}_{F}}$ and  $e^{i\hat{\phi}_{R}}$ which will allow us to describe filtering ($F$) and retrofiltering ($R$) of heterodyne measurements on the beam.
These operators have unit-modulus complex eigenvalues, the arguments (phases) of which  equal  the results (phase estimates) $\phi_{F}$ and $\phi_{R}$ for filtering and retrofiltering, respectively.
From the theory of heterodyne detection (see e.g.~\cite{sWisMil10}), these exponentials can be written as
\begin{align}
	e^{i\hat{\phi}_{F}} &={} \frac{\hat{F}}{|\hat{F}|} \label{defeiphiF} \\
	e^{i\hat{\phi}_{R}} &={} \frac{\hat{R}}{|\hat{R}|}
\end{align}
with $|\hat{A}| \coloneqq  \sqrt{\hat{A}^{\dagger} \hat{A}}$, and 
\begin{align}
	\hat{F} &\coloneqq {} \blu \int_{T-\tau}^T\blk dt\, u_F(t) \hat{b}(t) + \hat{a}^\dagger_F \label{eq:phasecovF}\\
	\hat{R} &\coloneqq  \blu\int_T^{T+\tau}\blk dt\, u_R(t) \hat{b}(t) + \hat{a}^\dagger_R. \label{eq:phasecovR}
\end{align}
Here, $\hat{b}(t)$ is the one-dimensional field operator satisfying $[\hat b(t),\hat b\dg(s)] = \delta(t-s)$  as in the main text, while $\hat{a}_{F(R)}$  is the  annihilation operator for the  ancillary  vacuum mode which enters into heterodyne detection, satisfying  $[\hat{a}_{F(R)},\hat{a}^\dagger_{F(R)}] = 1$.
Both $u_F(t)$ and  $u_R(t)$ are normalized ($\int_{-\infty}^{\infty}|u(t)|^{2}dt=1$) filter functions which have support on the intervals $[T-\tau,T)$ and $(T,T+\tau]$, respectively.  
From these properties,  all commutators of $\hat{F}, \hat{F}\dg, \hat{R}, \hat{R}\dg$ vanish. 

We first prove these operators are phase covariant.
This allows the observers to obtain a phase estimate of the laser at a given time, while ensuring Lemmas \ref{lem:cavenc} and \ref{lem:mu_invariant} from the previous section are relevant. 
The lemma below is proved explicitly for the filtering operator $\hat{F}$, from which the proof for $\hat{R}$ follows trivially.
\begin{lemma}[Phase covariance] \label{lem:phase_covariance}
The filtering observable $e^{i\hat{\phi}_{F}}$ changes covariantly when the beam undergoes an optical phase shift by arbitrary angle $\theta$. That is, the probability to obtain the result $e^{i{\phi}_{F}}$ changes as 
\begin{equation}
	P(e^{i{\phi}_{F}}|\blu\mathcal{U}^{\theta}_{\rm b}\blk(\rho_{\rm b})) = P(e^{i(\blu{\phi}_{F}-\theta\blk)}|\rho_{\rm b})
\end{equation} 
\begin{proof}
For reasons that will become apparent, consider  
$\mathcal{U}^\theta_{\rm b-a}(\bullet) = \hat{U}^\theta_{\rm b-a} \bullet \hat{U}^{\theta\dagger}_{\rm b-a}\coloneqq \blu e^{i\theta(\hat{n}_{\rm b} - \hat{n}_{\rm a})} \bullet e^{-i\theta(\hat{n}_{\rm b} - \hat{n}_{\rm a})}$, \blk where $\hat{n}_b$ and $\hat{n}_a$ are photon number operators for the beam and ancillary vacuum modes, respectively. Then we can show  
\begin{align}
	\mathcal{U}^\theta_{\rm  b-a}(\hat{F}) &= \hat{U}^{\theta}_{\rm  b-a}\left( \blu\int_{ T-\tau}^T\blk dt\, u_F(t) \hat{b}(t) + \hat{a}^\dagger_F \right)\hat{U}^{\theta\dagger}_{\rm  b-a} \\
	&=  \hat{U}^{\theta}_{\rm b}\left[\blu\int_{ T-\tau}^T\blk dt\, u_F(t) \hat{b}(t) \right] \hat{U}^{\theta\dagger}_{\rm b}  + \hat{U}_{\rm a}^{ -\theta}\hat{a}^\dagger_F\hat{U}_{\rm a}^{- \theta}{}^{\dagger} \\
	&={} \blu e^{-i\theta}\blk\hat{F}. 
\end{align}
Note that $\hat{F}$ is a normal operator, since $[\hat{F},\hat{F}\dg]=0$.
Therefore, by the spectral theorem \cite{sWisMil10}, there exists an orthonormal basis in the joint Hilbert space of the beam and vacuum mode ${\rm span} \{ \ket{F} : F\blu\in\mathbb{C}\blk\}$ such that $\hat{F}\ket{F} = F \ket{F}$.
Thus,
\begin{equation}
	\hat{U}^{\theta\dagger}_{\rm b-a}\ket{F} = \ket{F\blu e^{-i\theta}\blk}, 
	\label{eq:phase_shifted_eigenket}
\end{equation}
where $\ket{F\blu e^{-i\theta}\blk}$ is an eigenstate of $\hat{F}$, with eigenvalue $F\blu e^{-i\theta}$. \blk
This can be seen from observing that $\hat{F} \hat{U}^{\theta\dagger}_{\rm b-a}\ket{F} = \blu e^{-i\theta}\blk\hat{U}^{\theta\dagger}_{\rm b-a}\hat{F} \ket{F}= F \blu e^{-i\theta}\blk\hat{U}^{\theta\dagger}_{\rm b-a}\ket{F}$.

Now, to prove the lemma, \blu the \blk probability of obtaining outcome $F$ when measuring the beam in state $\rho_{\rm b}$ is 
\begin{align}
	P(F\blu|\rho_{\rm b}\blk) = \bra{F}(\rho_{\rm b}\otimes \ket{0}_{\rm a}\bra{0})\ket{F},
\end{align}
where the subscript a labels the vacuum state in the ancillary mode which enters into heterodyne detection.
This means that, if the state of the beam undergoes a rotation in phase by amount $\theta$, the probability of obtaining outcome $F$ becomes
\begin{align}
	P(F| \blu \mathcal{U}^{\theta}_{\rm b}\blk (\rho_{\rm b}))   &={} \bra{F}\mathcal{U}^\theta_{\rm b}(\rho_{\rm b})\otimes \ket{0}_{\rm a}\bra{0}\ket{F} \\
	&\equiv{} \bra{F}(\mathcal{U}^\theta_{\rm b}\otimes \mathcal{U}^{-\theta}_{\rm a})(\rho_{\rm b}\otimes \ket{0}_{\rm a}\bra{0})\ket{F} \\
	&={} (\bra{F}\hat{U}^{\theta}_{\rm b-a}) \rho_{\rm b}\otimes \ket{0}_{\rm a}\bra{0}(\hat{U}^{\theta\dagger}_{\rm b-a}\ket{F}) \\
	&={}\bra{ F\blu e^{-i\theta}\blk}(\rho_{\rm b}\otimes \ket{0}_{\rm a}\bra{0})\ket{F\blu e^{-i\theta}\blk} \\
	&={} P(F\blu e^{-i\theta}\blk| \rho_{\rm b}). 
\end{align}
The second line follows from the first due to the phase-shift invariance of the ancillary vacuum state $\ket{0}_{\rm a}\bra{0}$, and the fourth follows from  Eq.~\eqref{eq:phase_shifted_eigenket}.
Thus, from \eref{defeiphiF},  the result of measuring $e^{i\hat{\phi}_{F}}$ on a state shifted by $\theta$ in phase, is equivalent to making the same measurement on the original state and shifting the result by $\theta$. 
\end{proof}
\end{lemma}

We now proceed to \blu the formal proof of Theorem 1 \blk in the main text, namely that $\coh=O(\mu^4)$ for any laser which satisfies our four assumptions.

\begin{theorem}[Upper bound on $\coh$]\label{theo:main}
For an ideal laser beam satisfying Conditions 1--4 stated in the main text, the coherence $\coh$ of the beam is bounded from above by 
\begin{equation}
	\coh \leq  \frac{2}{3} \left| \frac{3}{z_A} \right|^{6} \mu^4,
\end{equation} 
in the asymptotic limit $\mu\rightarrow\infty$, where $\mu$ is the mean number of excitations stored inside the laser cavity, and $z_A \approx -2.338$ is the first zero of the Airy function.
\end{theorem}

In order to prove the theorem, we require one additional result proved by Bandilla, Paul and Ritze \cite{sBan91}.
This involves consideration a convenient measure of phase  spread  \cite{sBan69}
\begin{equation}
	1-|\langle e^{i\theta} \rangle|^2,
	\label{eq:convenient_uncertainty}
\end{equation}
we referred to as the mean-square error (MSE) in the main text,  when $\theta$ is an error.
This is justified by observing that, for small $\theta$, $1-|\langle e^{i\theta} \rangle|^2\approx \langle \theta^2\rangle - \langle \theta\rangle^2$, which is equivalent to the MSE for an unbiased estimate of $\theta$. 
We will use this quantity to describe how well Rod's estimate of the phase agrees with the actual encoded phase in the cavity.
The following lemma gives the required lower bound on this quantity in terms of the cavity's mean excitation number $\mu$, when Rod performs a direct measurement on the cavity,  from Ref.~\cite{sBan91}. 
\begin{lemma}[Minimum MSE for a cavity measurement] \label{lem:min_msse}
An optical phase measurement on the state of a system with mean photon number $\mu$, and for which the $U(1)$-mean phase is $\bar{\phi}$, will  give an estimate $\hat{\phi}$ with MSE bounded from below by 
\begin{equation}
 1-|\langle e^{i(\hat{\phi} - \bar{\phi})} \rangle|^2 \geq   4\left| \frac{z_A}{3}\right|^3 \mu^{-2}
\end{equation}
in the asymptotic limit $\mu\gg 1$.
\end{lemma}

\noindent\textit{Proof of Theorem \ref{theo:main}.}
Consider the problem of measuring the phase of the cavity at time $T$, once the laser system is in its steady state.
Let  one  observer (Effie) obtain a phase estimate $\phi_F$ by performing heterodyne filtering of the beam segment emitted in the interval $t\in[T-\tau,T)$.
Effie's measurement is represented by the phase covariant operator $\hat{F}$ defined in Eq.~\eqref{eq:phasecovF}.
By Lemma \ref{lem:phase_covariance}, we know Lemma~\ref{lem:cavenc} applies for this measurement operator, and so the state of the cavity conditioned on Effie obtaining outcome $\phi_F$ is a fiducial cavity state $\rho$ with $\phi_F$ encoded by the generator $\hat{n}_{\rm c}$.
That is, the state of the cavity is \emph{steered} to $\rho_{ {\rm c} | \phi_F} = \blu e^{i\phi_F\hat{n}_{\rm c}} \blk\rho_0 \blu e^{-i\phi_F\hat{n}_{\rm c}}$ \blk by Effie's measurement, which by Lemma~\ref{lem:mu_invariant} will have mean photon number $\mu$ for all outcomes $\phi_F$.

Now, the task of  a  second observer (Rod) is to estimate $\phi_F$, which has been encoded into the cavity by Effie's measurement.
Consider two ways Rod could choose to do this---either by allowing the cavity to emit a beam for a further $\tau$ time and performing retrofiltering over the beam emitted in $(T,T+\tau]$, or by performing a direct measurement on the cavity at time $T$.
Rod's retrofiltering measurement is represented by the phase covariant operator $\hat{R}$ defined in Eq.~\eqref{eq:phasecovR}.
We denote the two estimates he can obtain by these methods by $\phi_R$ and $\phi_D$, respectively.
In order to compare how correlated  $\phi_R (\phi_D)$ is with $\phi_F$,  we consider the quantity $1-|\langle e^{i(\hat{\phi}_{R(D)}-\hat{\phi}_F)} \rangle|^2$.

For the finite filtering/retrofiltering intervals discussed above, the filter function $u_F(t), u_R(t)$ must have support only for $t\in[T-\tau,T)$.
 For simplicity we  choose this filter function to be uniform over this interval, $u_R(t) = u_F(-t) = \tau^{-1/2} [H(t) - H(t-\tau)]$, where $H$ is the Heaviside step function.
 This symmetry,  together with the fact that $\left\langle \sin(  \hat{\phi}_R - \hat{\phi}_F  )\right\rangle=0$,  allows us to calculate the MSE in the first case where Rod retrofilters the beam as
\begin{align}
1-\left\langle \cos (\hat{\phi}_R - \hat{\phi}_F) \right\rangle^2 &= 1-\frac{1}{4} \left\langle\frac{\hat{F}^\dagger\hat{R} + \hat{R}^\dagger\hat{F}}{|\hat{R}|\, |\hat{F}|} \right\rangle^2
\end{align}
Defining the operator $\hat{S} \coloneqq  \hat{R}^\dagger \hat{F}$ allows this to be rewritten as
\begin{equation}
	1-\left\langle \cos (\hat{\phi}_R - \hat{\phi}_F) \right\rangle^2 = 1 - \left\langle \frac{\hat{S}}{\sqrt{\hat{S}\dg\hat{S}}}\right\rangle^2,
	\label{eq:MSSE_s_operators}
\end{equation}
where we have used the fact that $\langle\hat{S}\dg\rangle = \langle\hat{S}\rangle  \eqqcolon\bar{S}$  due to symmetries between $\hat{R}$ and $\hat{F}$.
Let $\hat{S}\coloneqq \bar{S}+\delta\hat{S}$ and $\hat{S}\dg\coloneqq \bar{S}+\delta\hat{S}\dg$.
We have
\begin{align}
	1-\left\langle \cos (\hat{\phi}_R - \hat{\phi}_F) \right\rangle^2 &={} 1 - \left\langle \frac{\bar{S} + \delta\hat{S}}{\sqrt{(\bar{S} + \delta\hat{S}\dg)(\bar{S} + \delta\hat{S})}}\right\rangle^2 \\
	&={} 1 - \left\langle \frac{1 + \delta\hat{S}/\bar{S}}{\sqrt{(1 +  \delta\hat{S}/\bar{S} +  \delta\hat{S}\dg/\bar{S} + \delta\hat{S}\delta\hat{S}\dg/\bar{S}^2)}}\right\rangle^2.
\end{align}
Since $\delta\hat{S}\dg, \delta\hat{S} \ll \bar{S}$, we can expand the operators in the denominator by performing a Maclaurin series expansion for $(1+x)^{-1/2} = 1 - x/2 + 3x^2/8 + O(x^3)$. 
Thus,  to leading order, 
\begin{align}
	1-\left\langle \cos (\hat{\phi}_R - \hat{\phi}_F) \right\rangle^2 &={}  1 - \left\langle \left(1 + \frac{\delta\hat{S}}{\bar{S}}\right)\left( 1 -\frac{\delta\hat{S}}{2\bar{S}} - \frac{\delta\hat{S}\dg}{2\bar{S}} -  \frac{\delta\hat{S}\dg\delta\hat{S}}{2\bar{S}^2}  + \frac{3(\delta\hat{S}\dg)^2}{8\bar{S}^2} + \frac{3(\delta\hat{S})^2}{8\bar{S}^2} + \frac{3\delta\hat{S}\dg\delta\hat{S}}{4\bar{S}^2} \right)\right\rangle^2 \nonumber\\
	&={} 1 - \left\langle 1 + \frac{1}{4}\left( \frac{(\delta\hat{S})^2}{\bar{S}^2} - \frac{\delta\hat{S}\dg\delta\hat{S}}{\bar{S}^2} \right) \right\rangle^2 \nonumber\\
	&=\frac{1}{2}\left\langle \frac{ \delta\hat{S}\dg\delta\hat{S} - (\delta\hat{S})^2}{\bar{S}^2} \right\rangle.
\end{align}
The large term on the denominator evaluates to $\bar{S}=\mathcal{N}\tau + O(1)$  for $\ell\tau\ll 1$,  and so, in terms of the original operators $\hat{S}$ and $\hat{S}\dg$,
\begin{equation}
	1-\left\langle \cos (\hat{\phi}_R - \hat{\phi}_F) \right\rangle^2 \sim \frac{1}{2}\frac{\left\langle \hat{S}\dg\hat{S} - \hat{S}^2\right\rangle}{\mathcal{N}^2\tau^2}
	\label{eq:error_simplified_s}
\end{equation}

The expectation value in the numerator requires careful calculation.
We use the definitions of $\hat{F}$ and $\hat{R}$ from Eqs.~\eqref{eq:phasecovF} and \eqref{eq:phasecovR}, together with the first- and second-order Glauber coherence functions for an ideal laser beam in Eqs.~\eqref{eq:g1_ideal} and \eqref{eq:g2_idealsts't'}, to evaluate these.
The first term is
\begin{align}
\langle \hat{S}\dg\hat{S} \rangle &= \langle R^\dagger F^\dagger F R \rangle  \\
&= \langle a_R a_F a^\dagger_F a^\dagger_R \rangle  + \frac{1}{\tau} \int_{{\blu T-\tau}}^{\blu T} ds \int_{{\blu T-\tau}}^{\blu T} dt \, \langle a_R \hat{b}^\dagger(s) \hat{b}(t) a^\dagger_R \rangle + \frac{1}{\tau} \int_{{\blu T}}^{\blu T+\tau} dt \int_{{\blu T}}^{\blu T+\tau} ds \, \langle b^\dagger(t) a_F a^\dagger_F b(s) \rangle \nonumber \\
&\quad + \frac {1}{\tau^2} \int_{{\blu T}}^{\blu T+\tau} ds \int_{{\blu T-\tau}}^{\blu T} ds' \, \int_{{\blu T}}^{\blu T+\tau} dt' \int_{{\blu T-\tau}}^{\blu T} dt \, \langle b^\dagger (s) b^\dagger(s') b(t') b(t) \rangle \\
&= {\blu 1 + \frac{{2\cal N}}{\tau} \int_0^{\tau} dt \int_0^{\tau} ds \, g^{(1)}(s,t) + \frac {{\cal N}^2}{\tau^2} \int_0^{\tau} ds \int_{-\tau}^0 ds' \, \int_0^{\tau} dt' \int_{-\tau}^0 dt \, g^{(2)}(s,s',t',t)} \label{eq:SSG2} \\
&= 1 + \frac{{2\cal N}}{\tau} \int_0^{\tau} dt \int_0^{\tau} ds \, e^{-\ell |s-t|/2} \nonumber \\
& \quad + \frac {{\cal N}^2}{\tau^2} \int_0^{\tau} ds \int_{-\tau}^0 ds' \, \int_0^{\tau} dt' \int_{-\tau}^0 dt \,e^{-\ell (|s-t|+|s'-t'|+|s-t'|+|t-s'|-|s-s'|-|t-t'|)/2} \\
&=1 + \frac{8 \mathcal{N} \left(\ell \tau +2 e^{-\frac{\ell \tau }{2}}-2\right)}{\ell^2 \tau } + \frac{16 \mathcal{N}^2 e^{-\ell \tau } \left(e^{\frac{\ell \tau }{2}} (\ell \tau -2)+2\right)^2}{\ell^4 \tau ^2}.
\end{align} 
\blu Note that here (and below) we have used the stationarity of the process to change the limits of integration when we rewrite the correlation in 
terms of Glauber functions. \blk 
The other term is
\begin{align}
\langle \hat{S^2} \rangle &={} \langle (R^\dagger)^2 F^2 \rangle \\
&=\blu \frac {1}{\tau^2} \int_T^{T+\tau} ds \int_T^{T+\tau} ds' \,\int_{T-\tau}^T dt' \int_{T-\tau}^T dt \, \langle b^\dagger (s) b^\dagger(s') b(t') b(t) \rangle \\
& ={\blu \frac {{\cal N}^2}{\tau^2} \int_0^{\tau} ds \int_0^{\tau} ds' \,\int_{-\tau}^0 dt' \int_{-\tau}^0 dt \, \, g^{(2)}(s,s',t',t)} \label{eq:SSG2_oneterm} \\
& =\frac {{\cal N}^2}{\tau^2} \int_0^{\tau} ds \int_0^{\tau} ds' \,\int_{-\tau}^0 dt' \int_{-\tau}^0 dt \, \, e^{-\ell (|s-t|+|s'-t'|+|s-t'|+|t-s'|-|s-s'|-|t-t'|)/2} \\
&= \frac{4 \mathcal{N}^2 e^{-4 \ell \tau } \left(e^{\frac{\ell \tau }{2}}-1\right)^4 \left(2 e^{\frac{\ell \tau }{2}}+3 e^{\ell \tau }+1\right)^2}{9 \ell^4 \tau ^2}.
\end{align}
This time there are no two-time correlations because the vacuum modes only have annihilation or creation operators, so give expectation values of zero.
For the purpose of proving the theorem, we only require Eq.~\eqref{eq:error_simplified_s} to leading order in $\ell/\mathcal{N}$.
To this end, we introduce a real parameter $\sigma$ and set $\tau\coloneqq \sigma/\sqrt{\mathcal{N}\ell}$, from which we find
\begin{align}
	1-\left\langle \cos (\hat{\phi}_R - \hat{\phi}_F) \right\rangle^2 & \sim  \frac{1}{2}\frac{\left\langle \hat{S}\dg\hat{S} - \hat{S}^2\right\rangle}{\mathcal{N}^2\tau^2} \\
	&={\blu \frac 1{\mathcal{N}\tau} + \frac {2\ell\tau}3+
	O\left( \frac{\ell}{\mathcal{N}} \right)} \label{eq:g12var} \\
	&={} \left(\frac{ 1}{\sigma }+\frac{2\sigma}{3 }\right)\sqrt{\frac{\ell}{\mathcal{N}}} + O\left( \frac{\ell}{\mathcal{N}} \right).
	\label{eq:numerator_series}
\end{align}
It is of interest (see \sref{sec:proving_all_conditions_hold}) to know how this result relates to the Glauber coherence functions introduced in \sref{sec:glauber_ideality}.
Clearly from Eqs.~\eqref{eq:SSG2} and \eqref{eq:SSG2_oneterm}, the quantity we have just evaluated can be written as a constant plus linear functionals (integrals) of $\delta g^{(1)}_{\rm ideal - 1}$ and $\delta g^{(2)}_{\rm ideal - 1}$. 
It turns out that, the $\delta g^{(1)}_{\rm ideal - 1}$ terms do  not contribute to the leading order expression in Eq.~\eqref{eq:numerator_series}.
Specifically, we can say
\begin{align}
	1-\left\langle \cos (\hat{\phi}_R - \hat{\phi}_F) \right\rangle^2 \sim {} &\frac{1}{\cal N\tau } + \frac{1}{\tau^4}\left[\frac{1}{2} \int_0^{\tau} ds \int_{-\tau}^0 ds' \, \int_0^{\tau} dt' \int_{-\tau}^0 dt \,\delta g^{(2)}_{\rm ideal - 1}(s,s',t',t) \right.\nonumber\\
	&{} \qquad \qquad \ - \frac{1}{2}\left.\int_0^{\tau} ds \int_0^{\tau} ds' \,\int_{-\tau}^0 dt' \int_{-\tau}^0 dt \, { \delta} g^{(2)}_{\rm ideal - 1}(s,s',t',t) \right].\label{eq:dgvar}
\end{align}
The first term comes from the shot noise in the phase estimate due to the finite number of photons, and decreases 
as the reciprocal of the number of photons in the time interval $\tau$. The second term is a sort of time-average, over that interval, 
of the deviation from complete coherence. That is, it would vanish if the phase of the beam were constant in time and because of phase diffusion it increases with $\tau$ as we see explicitly in the evaluated expression, Eq.~\eqref{eq:numerator_series} (remembering that 
$\sigma$ is just a scaled version of $\tau$).  
 The optimal time $\tau$, which minimizes the MSE, balances those two competing contributions. 
This translates to the tightest (given the method used)  upper bound on $\coh$, as we shall see below.
Taking the derivative of Eq.~\eqref{eq:numerator_series} with respect to $\sigma$ and setting it to zero shows that the minimum MSE occurs when $\sigma=\sqrt{3/2}$, corresponding to retrofiltering/filtering intervals of  duration
\begin{equation}\label{eq:tauchoice}
\tau = \sqrt{\frac{3}{2\mathcal{N}\ell}}.
\end{equation}
Substituting this filtering time in Eq.~\eqref{eq:numerator_series}  gives, asymptotically  
\begin{equation}
	1-\left\langle \cos (\hat{\phi}_R - \hat{\phi}_F) \right\rangle^2 \sim  2\sqrt{\frac{2\ell}{3\cal N}}.
	\label{eq:asymptotic_msse}
\end{equation} 

The second method available to Rod involves performing a direct optical phase measurement on the state of the cavity. 
In this case, the lower bound on the MSE from Lemma~\ref{lem:min_msse} can be applied, since by Lemma~\ref{lem:mu_invariant} the state of the cavity maintains a constant mean photon number $\mu$ after Effie has performed filtering.
From Lemma~\ref{lem:min_msse}, we can identify $\hat{\phi}$  with $\hat{\phi}_D$  and $\bar{\phi}$ with  $\phi_F$  plus the average phase value of the fiducial state.
We thus know that  $1-|\langle e^{i(\hat{\phi}_D-\phi_F)} \rangle|^2 \gtrsim 4|z_A/3|^3 \mu^{-2}$. 
Clearly, the direct measurement which achieves this lower bound on the MSE must outperform any other measurement of the phase at time $T$, since Lemma~\ref{lem:cavenc} shows that the phase information from Effie's measurement is encoded in the fiducial state  purely by  the generator $\hat{n}_{\rm c}$. 
It therefore follows that $\phi_R$ cannot be better than $\phi_D$ as an estimate of $\phi_F$, since the direct cavity measurement which achieves equality in Lemma~\ref{lem:min_msse} is optimal.
In terms of the MSEs between the phases, this means 
\begin{equation}
	1-|\langle e^{i(\hat{\phi}_R-\hat{\phi}_F)} \rangle|^2 \gtrsim 1-|\langle e^{i(\hat{\phi}_D-\hat{\phi}_F)} \rangle|^2
\end{equation} 
for $\mu\gg1$.
This immediately gives the inequality 
\begin{equation}
2\sqrt{\frac{2\ell}{3\cal N}} \gtrsim \frac{4|z_A/3|^3}{\mu^{2}},
\label{eq:msse_inequality_nocoh}
\end{equation}
and rewriting in terms of the coherence $\coh=4{\cal N}/\ell$, we have that
\begin{align}
	\coh &\lesssim \frac{2}{3} \left| \frac{3}{z_A} \right|^{6} \mu^4 \\
&\approx  2.9748 \mu^4, \nonumber
\end{align}
 as claimed. \qed
\\ \\

\section{Details of our Tensor-Network Laser Model}
\label{sec:MPS-methods}

The discretized version of the laser process that was 
described in the main text  is illustrated in Fig.~\ref{fig:model}.
For any finite-size portion of the beam, this process corresponds to a matrix product state (MPS) sequential quantum factory, or sequential generation scheme, previously introduced in~\cite{sSchon05,sSchon07}. 
For such processes, Schon\etal~\cite{sSchon05} proved that the generative interaction corresponds to an isometry, and the output can be expressed as an 
MPS with \emph{left-handed} orthogonality 
(see~\cite{sSchollwock11,sMcCulloch07,McCulloch08,sSaadatmand17_thesis} for details of orthogonality conditions on MPSs).

MPS methods are  widely used in condensed matter physics and 
quantum information theory~\cite{sAKLT87_original,sSchollwock11,sOrus14}, and have had some applications in quantum optics~\cite{sSchon05,sSchon07,sJarzyna13,sMan17}, but have been never used to describe the  creation of laser coherence,  to the best of our knowledge.
In particular, while Ref.~\cite{sSchon05} also pointed out that a cavity quantum-electrodynamics (QED) system can be generally emulated using such a sequential process, the authors  were not concerned with describing coherence generation (in the sense defined above) by a laser.
Rather, they assumed external sources of coherence, {\it i.e.}~lasers,  driving their cavity QED system.

\subsection{The \lowercase{i}MPS description of a laser beam}
\label{sec:MPS-description}

\begin{figure}[h]
  \begin{center}
    \includegraphics[width=0.75\linewidth]{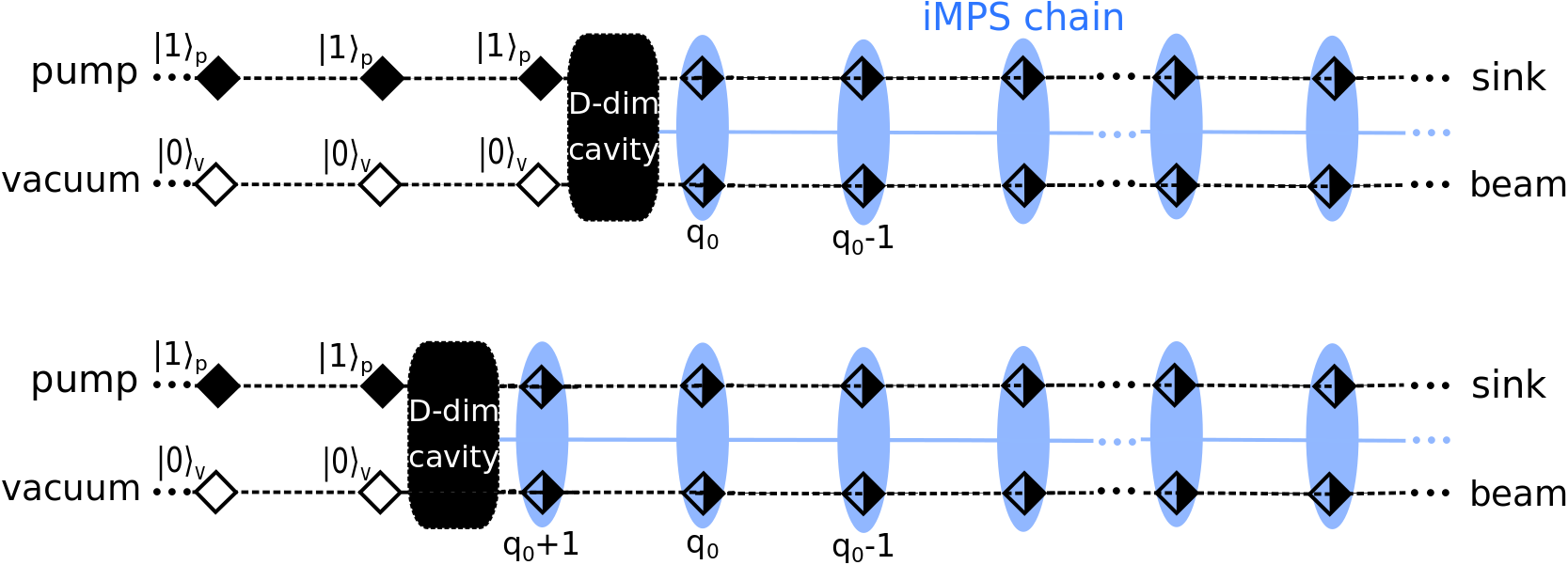}
    \caption{
    Conceptual diagram of our laser model. 
    From the upper figure to the lower, one time step has passed, converting 
    one pair of input qubits (pump and vacuum) into 
    a new pair of output qubits (beam and sink), with position label  
    $q_0+1$. The indefinite length string of pairs of output 
    qubits is described by an iMPS of bond-dimension $D$,     
    equal to the Hilbert space dimension of the laser cavity. 
    \label{fig:model}}
  \end{center}
\end{figure}

The descretized laser system we described in the main text (and is visualised in \fref{fig:model}) contains five elements:
the cavity, a pump, a vacuum input, the beam output, and a sink.
All of these are essential for laser operation, and we use the simplest possible model for all.
We can consider the sink (s) and beam (b) as a joint four-level system in order to maintain consistency with the requirements of an MPS sequential generation scheme, possessing a single output at a time.
In doing so, we may consider the beam alone by simply tracing over the sink. 
\blu From Eq.~(6) in the main text, \blk the time evolution of the cavity (c) and its outputs is governed by the generative interaction \blu
\begin{equation}
	\hat{V}_{ q} = \sum_{j_{ q+1},m,n} A^{[j_{q+1}]}_{mn} \ket{m}_{ \rm c}\bra{n} \otimes \ket{j_{q+1}}_{ \rm o},
	\label{eq:interaction_isometry}
\end{equation}
\blk
which maps a $D$-dimensional vector space into a $4 \times D$-dimensional one.
\blu The ket specifying the outputs (labelled o) acts on the composite beam and sink space $\mathcal{H}_{\rm b} \otimes \mathcal{H}_{\rm s}$, and is defined as $\ket{j_{q+1}}_{ \rm o} \coloneqq \ket{\lfloor j/2\rfloor_{q+1}}_{\rm b}\otimes \ket{(j\mod 2)_{q+1}}_{\rm s}$.
The action of the isometry $\hat V$ corresponds to a completely-positive trace-preserving map which evolves the laser cavity one time step. 
It can be related to the generative unitary interaction, $\hat{U}_{\rm int}$, according to  
\begin{equation}
	\hat{V} \ket{\psi}_{\rm c} \equiv \hat{U}_{\rm int} (\ket{\psi}_{ \rm c} \ket{1}_{\rm p} \ket{0}_{ \rm v})
	\label{eq:iMPS_interaction_unitary}
\end{equation}\blk
where the subscripts p and v correspond to pump and vacuum states.
The isometry condition,  $\hat{V}^\dagger \hat{V} = I_{D}$,  where $I_m$ is the $m \times m$ identity matrix, translates to
a completeness/orthonormality relation 
\begin{equation}
  \sum_{j=0}^3 A^{[j]}{}^\dagger A^{[j]} = I_D~.
\label{eq:IsometryCond}
\end{equation}
Each $A^{[j]}$ here is treated as an operator whose elements in the cavity photon number basis, $\ket{m}_{\rm c}$, 
are those of the $A$-matrices introduced above. 

We are interested in the \emph{one-site unit-cell} infinite MPS (iMPS) that $\hat{V}$ eventually creates, pictured in \fref{fig:model} for arbitrary consecutive times $q_0$ and $q_0+1$. In terms of the $A$-operators used above, the
iMPS is given by 
\begin{align}
  \ket{\Psi_{\rm iMPS}}\!=\!\sum_{...,j_{q_0},j_{q_0-1},j_{q_0-2},...} \bra{\Phi(q\!=\!+\infty)}_{ \rm c} \, 
                           \cdots A^{[j_{q_0}]}_{(q_0)} 
                           A^{[j_{q_0-1}]}_{(q_0-1)} A^{[j_{q_0-2}]}_{(q_0-2)} \cdots
                           \ket{\Phi(q\!=\!-\infty)}_{ \rm c} \notag \\ 
                           \ket{...,j_{q_0},j_{q_0-1},j_{q_0-2},...}_{ \rm o}~,
\label{eq:MPS-WF}
\end{align}
where $\ket{\Phi(q)}_{ \rm c}$ denotes the state of the cavity at integer time $q$. 
We suppose in the last step, $q=+\infty$, the cavity decouples from the output. 
Since there exists translational invariance in the outputs, we now drop the $(q)$-subscripts.
Later on, we shall impose an important condition on the iMPS (i.e.~the largest-magnitude eigenvalue of its \emph{transfer matrix} being non-degenerate) that
makes the boundary states $\ket{\Phi(q=+\infty)}_s$ and $\ket{\Phi(q=-\infty)}_s$ irrelevant---they would not appear in any expectation value calculations below (see also~\cite{sSchollwock11,sOrus14,sSaadatmand17_thesis}). 
Therefore, the iMPS in \eref{eq:MPS-WF} is equivalent to the infinite-size/uniform tensor network state described in Refs.~\cite{McCulloch08,sZau18} when it reaches its fixed-point. 

We wish to deal with the above description of the laser dynamics, under the 
conditions discussed in 
the main text. As per requirements of Condition 1, 
we are approximating continuum-limit description of the output beam by a discrete model, with an arbitrarily short section of beam of length $\delta t$ having at most one photon occupation so being representable by a qubit. 
This is achieved  
by performing the following transformation:   
$\sqrt{\delta t}~\hat{b} \rightarrow \hat{\sigma}^-_{ \rm b}$, where 
$\hat{\sigma}^-_{ \rm b} =  \ket{0}_{{\rm b}}\bra{1}$.
(In fact, $\delta t$
translates to MPS lattice spacing in physical units -- more details below.)
Note that $\hat{\sigma}^{\{x,y,z,+,-\}}_{\rm s}$ are always operators on the sink (where 
$\hat{\sigma}^z_{\rm s}\equiv2\hat{n}_{\rm s}-1$). 
Taking  $\tilde{\frak{c}}(\omega)$ to have a maximum at $\omega\!=\!0$  as above, 
the coherence finds the following form:
\begin{equation}
\frak{C} = \sum_{q'=-\infty}^\infty \la \sigma^+_{ \rm b}(q+q') \sigma^-_{ \rm b}(q) \ra~,  
\label{eq:C-FinalDef}
\end{equation}
which is how we calculate the coherence numerically  (see also below).

Let us now use a prime to denote the operators after the application of the 
unitary, $\hat{U}_{\rm int}$, forwarding one time step for a pair of input qubits and the cavity. 
Condition 2  follows naturally in this MPS model by imposing 
conservation of energy.
 That is, we impose the constraint $\hat n_{\rm c} + \hat n_{\rm v} + \hat n_{\rm p} 
=  \hat n_{\rm c}' + (\hat n_{\rm v}' + \hat n_{\rm p}')$ on the unitary $\hat{U}_{\rm int}$. 
But, as illustrated in \fref{fig:model}, the term $ (\hat n_{\rm v}' + \hat n_{\rm p}')$ here can be equated with 
$\hat n_{\rm b} + \hat n_{\rm s}$, 
while 
the pump and vacuum are prepared in photon number eigenstates $0$ and $1$. Thus, we can equally well consider the isometry $\hat{V}$ with the constraint $1 + \la \hat n_{\rm c}\ra = \la\hat n_{\rm c}'+ \hat n_{\rm b} + \hat n_{\rm s}\ra $. 
Importantly, this means the $D \times D$ dimension $A$-matrices  are  highly sparse: 
\begin{equation}
	A^{[0]} = \left(
\begin{matrix}
0  & \cdots  & \cdots & \cdots & 0 \\
\bullet  & \ddots&& & \vdots \\
0  & \bullet & \ddots& &\vdots\\
\vdots  & \ddots & \ddots &\ddots & \vdots\\
0 & \cdots &  0 & \bullet & 0 \\
\end{matrix}
\right)
, \quad
A^{[1]}, A^{[2]} = \left(
\begin{matrix}
\bullet  & 0  & \cdots & \cdots & 0 \\
0  & \bullet& \ddots &  & \vdots \\
\vdots  & \ddots & \ddots &\ddots & \vdots\\
\vdots  &  & \ddots&\bullet & 0\\
0 & \cdots &  \cdots& 0 & \bullet\\
\end{matrix}
\right),
\quad
A^{[3]} =  \left(
\begin{matrix}
0  & \bullet  & 0 & \cdots & 0 \\
\vdots  & \ddots&\bullet& \ddots & \vdots \\
\vdots  &  & \ddots & \ddots & 0\\
\vdots  &  & &\ddots & \bullet\\
0 & \cdots &  \cdots& \cdots & 0 \\
\end{matrix}
\right)~,
\label{eq:A-forms}
\end{equation}
where bullets indicate the only elements allowed to be non-zero
\footnote{ \blu We note Eq.~(\ref{eq:A-forms}) tells us that MPS matrices of a steady-state laser are, indeed, analogous to those describing an eigenstate of an explicitly $U(1)$-symmetric physical Hamiltonian -- see also~\cite{sSaadatmand20}.\blk}. 
In other words, each $A$-matrix has at most a single non-zero diagonal, with $\Theta(D)$ free parameters.  
We will take those parameters to be real (and nonnegative), which is consistent with a spectral peak at $\omega=0$. 
It follows that the left-handed orthogonality (isometry) condition, \eref{eq:IsometryCond}, reduces to
\blu
\begin{align}
    \begin{cases}
    (A_{m+1,m}^{[0]})^2 + (A_{mm}^{[1]})^2 + (A_{mm}^{[2]})^2 = 1 \quad\text{for}~~m=0, \\
    (A_{m+1,m}^{[0]})^2 + (A_{mm}^{[1]})^2 + (A_{mm}^{[2]})^2 +
    (A_{m-1,m}^{[3]})^2 = 1 \quad\text{for}~~0<m<D-1, \\
    (A_{mm}^{[1]})^2 + (A_{mm}^{[2]})^2 + (A_{m-1,m}^{[3]})^2 = 1 \quad\text{for}~~m=D-1~,
    \end{cases}
\label{eq:ExplicitLeftOrth}
\end{align}
\blk which additionally implies that the absolute value of the $A$-matrices' elements are bounded by unity, $|A^{[j]}_{mn}| \leq 1$. 

\begin{figure}
  \begin{center}
    \includegraphics[width=0.85\linewidth]{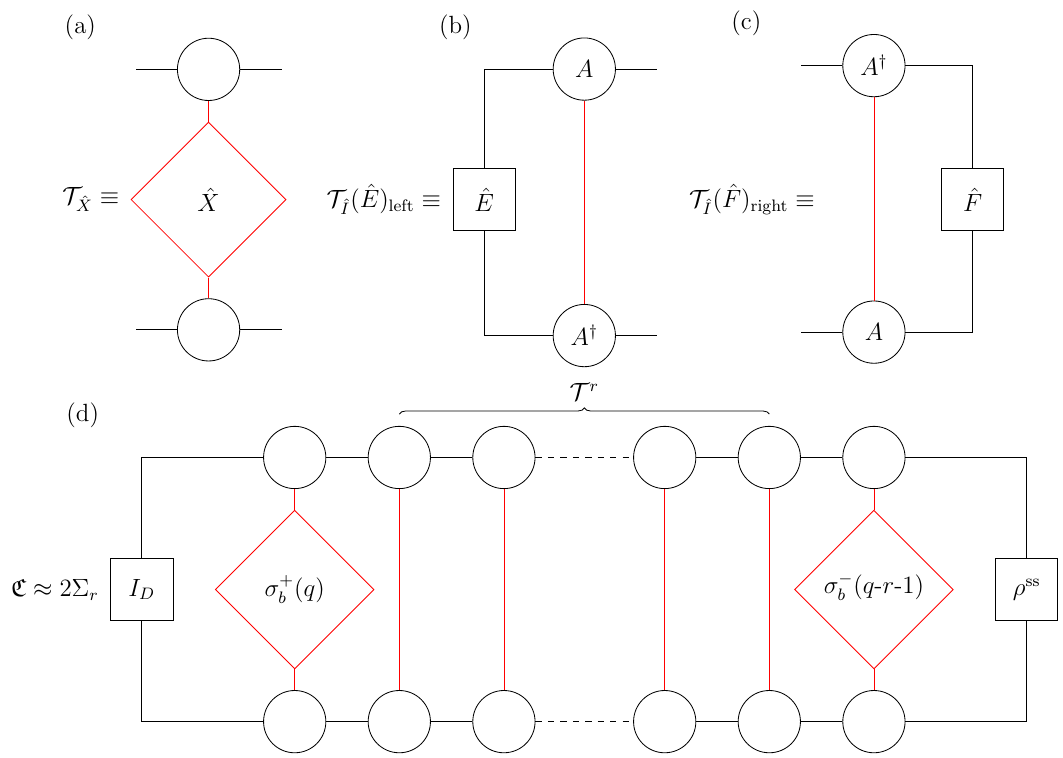}
    \caption{
    MPS diagrams for the definitions of the transfer 
    operators, \eref{eq:T-L&R-operators}, 
    and $\mathfrak{C}$. Here, circles
    display $A$-matrices  or their conjugates, diamonds represent physical local operators, 
     black lines  
    denote $D$-dimensional virtual bonds, and  red lines 
    show four-dimensional physical bonds in the \blu laser \blk system. 
    Displayed sub-figures demonstrate 
    the followings: (a) the diagrammatic definition of $\mathcal{T}_{\hat{X}}$,
    the transfer operator of a local operator $\hat{X}$ (for the identity transfer
    operator, $\mathcal{T}$, one only needs to set $\hat{X}=\hat{I}$). (b) The action of an operator 
    $\hat{E}$ on $\mathcal{T}_{\hat{I}}$ from the left, (c) the action of an operator 
    $\hat{F}$ on $\mathcal{T}_{\hat{I}}$ from the right, and (d) the diagrammatic definition of the discretized coherence, \eref{eq:C-FinalDef},  excluding the one-site 
    $\la \sigma^+_{ \rm b}(q) \sigma^-_{ \rm b}(q) \ra$-term\blk.
    \label{fig:T-def}}
  \end{center}
\end{figure}

We proved that in the model of \fref{fig:model} the output is equivalent to an iMPS; naturally, our preferred method to evaluate the steady-state  quantities of interest is the well-established approach of manipulating MPS transfer operators~\cite{sSaadatmand17_thesis,sMic10,sSchollwock11,McCulloch08,sOrus14,sZau18}, 
also known as transfer matrices. 
Using a standard MPS diagram, we present the definition for the general transfer operator, $\mathcal{T}_{\hat{X}}$, equipped with the local physical operator $\hat{X}$ in \fref{fig:T-def}(a).
Note that $\mathcal{T}$ matrices are superoperators acting on $D \times D$-size MPS operators. 
The most interesting superoperator in this class is the identity transfer operator, $\mathcal{T}_{\hat{I}} \equiv \mathcal{T}$, which is equipped with the physical identity operator, $\hat{I}_4\equiv\hat{I}$.
We present the actions of $\mathcal{T}$ on two left- and right-hand-side MPS operators
in \fref{fig:T-def}(b) and (c). Accordingly, the equation forms of 
these can be written as
\begin{align}
  \mathcal{T}(\hat{E})_{\rm left} &= \sum_j A^{[j]}{}^\dagger \hat{E} A^{[j]} \notag \\
  \mathcal{T}(\hat{F})_{\rm right} &= \sum_j A^{[j]} \hat{F} A^{[j]}{}^\dagger~.
\label{eq:T-L&R-operators0}
\end{align}
Equivalently, one can reshape any $D \times D$-size operator into a flattened $D^2 \times 1$-dimensional vector 
form as $\hat{E}_{m,n}\rightarrow ( E |_{(m,n)}$ or a $1 \times D^2$-dimensional vector form as $\hat{F}_{m,n}\rightarrow | F )_{(m,n)}$,
where $(m,n)$ stands for a collective index.
We refer to this space as the \emph{flattened space}, where the transfer-type operators become large $D^2 \times D^2$ matrices and MPS operators are represented by $D^2$-size vectors.
Furthermore, we use a bra- and ket-like notation to write the left-hand-side and right-hand-side acting vectors in the flattened space language, where \eref{eq:T-L&R-operators0} can be rewritten as
\begin{align}
  \big( ( E | \mathcal{T} \big)_{l l^\prime} &= \sum_{j,m,n} ( E |_{(m,n)} (A^{[j]}{}^\dagger)_{lm} A^{[j]}_{nl^\prime} \notag \\
  \big( \mathcal{T} | F ) \big)_{l l^\prime} &= \sum_{j,m,n} A^{[j]}_{lm} (A^{[j]}{}^\dagger)_{nl^\prime} | F )_{(m,n)}~.
\label{eq:T-L&R-operators}
\end{align}
 In some cases,  transfer operator approaches involve the calculations of leading eigenvalues 
of $\mathcal{T}$---in particular, the second largest eigenvalue, $\lambda_2$, which specifies the principal correlation length of the wave function; see \cite{sSaadatmand17_thesis} for some examples. 
However, unlike such approaches, for reasons that become clearer shortly, we formulate an 
analogous, but subtly different, transfer operator method --- more efficient for the laser system in \fref{fig:model} --- that does \emph{not}
require the direct calculation of the spectrum of $\mathcal{T}$. 
The numerical formalism below is partly equivalent to the 
more general transfer-matrix-based variational MPS optimizer introduced in Ref.~\cite{sZau18}, though independently developed. Note that, however, those authors were \emph{not} concerned with the iMPS description of a laser beam, and, naturally,
the essential connections between the steady state and $A$-matrices, and expressions for expectation
values of infinite-range operators, are unique to our formalism. Refer to~\cite{sSaadatmand20} for more detailed description and comparisons regarding the presented transfer matrix method. 

In its matrix form (in the flattened space) the transfer operator can be constructed as
$\mathcal{T} = \sum_j A^{[j]}{}^* \otimes A^{[j]}$.
The matrix $\mathcal{T}$, in this particular form, possesses a set of interesting properties: 
it has a spectral radius of $1$ and is generally non-Hermitian (a necessary feature of the map, 
since we can show that the symmetric transfer operators lead to trivial laser output states -- see also below). Moreover, its left leading
eigenvector (or eigenmatrix), $(1| \leftrightarrow \hat{I}_D$, is the identity operator of 
the Hilbert space $\mathcal{H}_{c}$, which is clear from \eref{eq:IsometryCond}; 
in other words, $(1| \mathcal{T} = (1| \lambda_1 = (1|$. 
We further suppose that the eigenvalues of $\mathcal{T}$ are arranged as 
$1=|\lambda_1| \geq  |\lambda_2| \geq \cdots \geq |\lambda_{D^2}|$.
The corresponding right eigenmatrix is then the familiar (right) reduced density matrix, ${ \rho}^{\rm ss}$, which, by construction, satisfies the following fixed-point (steady-state) equation
\begin{align}
  \sum_j \hat{A}^{[j]} \rho^{\rm ss} \hat{A}^{[j]}{}^\dagger = \rho^{\rm ss}~,
\label{eq:rho_ss}
\end{align}
or equivalently $\mathcal{T} |1) = |1)$ in the flattened space language. 
(As mentioned, ${ \rho}^{\rm ss}$ is always diagonal in 
the $\ket{m}_{ \rm c}$-basis.) Importantly, it is \emph{not} guaranteed for the map $\mathcal{T}$ to be injective (meaning that the dimension of the null space of $\mathcal{T}-I_{D^2}$  is  exactly $1$). For an MPS transfer matrix, the injectiveness assumption, alongside the  quantum version of the 
Perron-Frobenius  theorem~\cite{sAlb78,sFar96}, guarantees that $\lambda_1=1$ is unique with
a pair of nonnegative left and right eigenvectors,
i.e.~having a single fixed-point -- we iterate that a non-degenerate $\lambda_1$ also
guarantees that the boundary states are irrelevant. 
For the laser model in \fref{fig:model}, where no pulse or 
oscillatory lasing behaviors exist, injectivity  holds, so we shall assume it in what follows.   

It is worthwhile to look at the explicit matrix form of the transfer operator,
\setcounter{MaxMatrixCols}{15}
\begin{align}
  \mathcal{T} = \begin{pmatrix}     \bullet  & 0      & \cdots    & 0         &  \bullet     &     0      &     \cdots    &   \cdots      &       \cdots &    \cdots       &      0   \\   
                                   0      & \ddots  & \ddots         &     & \ddots          & \ddots     &    \ddots      &   &        &        &    \vdots     \\ 
                                   \vdots & \ddots      & \ddots   &    \ddots     &      & \ddots        & \ddots    &      \ddots   &      &     &      \vdots   \\
                                   0      &  & \ddots         & \ddots   &      \ddots      &           &    \ddots   &  \ddots &   \ddots   &             &      \vdots   \\
                                   \bullet  & \ddots      &     &       \ddots    & \ddots    &     \ddots      &          & \ddots   &  \ddots     &  \ddots &       \vdots   \\ 
                                   0       & \ddots  & \ddots         &           &       \ddots     & \ddots   &     \ddots     &       &  \ddots  &     \ddots     &       0    \\ 
                                   \vdots       &     \ddots   & \ddots   &      \ddots     &            &      \ddots     &  \ddots    &  \ddots &           &    \ddots       &     \bullet    \\ 
                                   \vdots       &        &      \ddots     & \ddots    &      \ddots      &           &    \ddots   &   \ddots  &   \ddots  &           &     0    \\
                                   \vdots       &        &           &       \ddots   & \ddots   &     \ddots      &           &  \ddots  &     \ddots     &   \ddots & \vdots                  \\
                                   \vdots       &        &           &          &   \ddots &     \ddots      &    \ddots       &        & \ddots   &  \ddots & 0   \\
                                   0       &  \cdots      &     \cdots      &     \cdots     & \cdots   &     0      &    \bullet      &   0   &  \cdots    &   0 & \bullet                 
                 \end{pmatrix}.
\label{eq:T-ExplicitMatrixForm}                                      
\end{align}
Here, bullets once again represent matrix elements which are allowed to be non-zero.
These three diagonals are non-sparse,  but  both the upper and lower bands have strictly $D-2$ evenly distributed  zero elements due to the sparsity of $A$-matrices \eref{eq:A-forms}.
This explicit form clearly indicates that $\mathcal{T}$ is a sparse, banded matrix with three non-zero diagonals.
To the best of our knowledge, there are no known connections   
between eigenvalue gaps and the sizes of  such matrices  
in the literature of linear algebra. 
Conventional linear algebra tools are, of course, capable of finding 
bounds on the eigenvalue gaps of square matrices having a higher degree of structure simplicity such as: being Hermitian, explicitly 
nonnegative and irreducible, or having Toeplitz forms. Importantly, $\mathcal{T}$ does not generally and should not 
be restricted to satisfy any such special criterion.
We reiterate that $\mathcal{T}$ is never irreducible, due to possessing some strictly zero  elements in off diagonals, and it is 
pointless  for it  to be set as symmetric; our thorough numerical investigations, 
similar to the ones presented in Fig.~1 of the main text, found that
a symmetric $\mathcal{T}$, having a flat photon number distribution,  $\rho^{\rm ss} = I_{D}/D$, 
simply recovers close to the SQL scaling, $\frak{C^{\rm ideal}_{\rm SQL}} =\Theta (D^2)$. Furthermore, 
the strict zeros in off diagonals of $\mathcal{T}$ also
forbid it to be approximated by any constant-diagonal matrices that possess simple and exact relations for the spectrum, as in Toeplitz matrices.

\subsection{iMPS numerical optimizations}
\label{sec:MPS-optimizations}

Due to the lack of analytical results on the spectral properties of $\mathcal{T}$-type matrices represented in Eq.~\eqref{eq:T-ExplicitMatrixForm}, we devise a cost-efficient numerical approach to calculate $\frak{C}$ based on a projected inverse-form \blu (see also~\cite{sSaadatmand20})\blk.
This allowed us to estimate the magnitude of the coherence still very precisely both numerically 
and semi-analytically. 
Using the definition in \fref{fig:T-def}(a), one can write \eref{eq:C-FinalDef} in terms of $\mathcal{T}$-operators,
as we clarified in \fref{fig:T-def}(d):
\begin{align}
  \frak{C} = (1| \mathcal{T}_{\sigma^+_{ \rm b}\sigma^-_{ \rm b}} |1) 
    + 2\sum_{r=0}^\infty (1| \mathcal{T}_{\sigma^+_{ \rm b}} \mathcal{T}^{r} 
    \mathcal{T}_{\sigma^-_{ \rm b}} |1)~.
\label{eq:C-in-Tmatrices}
\end{align}
It is easy to verify that every term in the sum is positive if the $A$-matrices have all nonnegative elements. 
We reiterate that this is the choice we make below, consistent with the numerics, and corresponds to the spectral peak being at $\omega=0$ as discussed 
in Sec.~\ref{sec:1Dfield}.  
Additionally, we note that the leftover $(1| \mathcal{T}_{\sigma^+_{ \rm b}\sigma^-_{ \rm b}} |1)$ 
single-site term is equivalent to the photon flux of the beam.
This term can be excluded from the above expression---another simplification that we do.
This is because the magnitude of this term is non-increasing with $D$ and, in fact, bounded from above by unity,
contributing only a negligible amount to the final coherence value.

Since the vector space of $\lambda_1=1$ has no contribution to the coherence (and also because we are interested in using
a geometric-series-type relation to simplify the above equation), we project out its space from the set of eigenvectors,  as prescribed in Ref.~\cite{sZau18}, 
by defining the following projector:
\begin{align}
  \mathbb{Q} = I_{D^2} - |1)(1|~,
\label{eq:Q-projector}
\end{align}
which implies $\mathcal{T} = \mathbb{Q} \mathcal{T} \mathbb{Q} + |1)(1|$.  
Replacing $\mathcal{T}$ with this expression in 
\eref{eq:C-in-Tmatrices} leads to

\begin{align}
  \frak{C} = 2\sum_{r=0}^\infty \big[ (1| \mathcal{T}_{\sigma^+_{ \rm b}} (\mathbb{Q}\mathcal{T}\mathbb{Q})^r   \mathcal{T}_{\sigma^-_{ \rm b}} |1)
             + (1| \mathcal{T}_{\sigma^+_{ \rm b}} |1) (1| \mathcal{T}_{\sigma^-_{ \rm b}} |1) \big]~.
\label{eq:C-progress}
\end{align}
Now, both expectation values in the second term of Eq.~\eqref{eq:C-progress} are strictly zero.  
More importantly, the superoperator $\mathbb{Q} \mathcal{T} \mathbb{Q}$ in the first term has no unity eigenvalue. 
Therefore, the inverse of the object $I_{D^2} - \mathbb{Q} \mathcal{T} \mathbb{Q}$ is now well-defined. Using geometric-series-type identities for 
the infinite sum appearing in the first term of Eq.~\eqref{eq:C-progress}, we find
\begin{align}
  \frak{C} &= 2(1| \mathcal{T}_{\sigma^+_{ \rm b}} \cdot \text{inv}(I_{D^2} - 
             \mathbb{Q}\mathcal{T}\mathbb{Q}) \cdot \mathcal{T}_{\sigma^-_{ \rm b}} |1) \notag \\
           &= 2(\sigma^+_{ \rm b}| \text{inv}(\bar{\mathcal{T}}) |\sigma^-_{ \rm b})~,
\label{eq:C-final}
\end{align}
where $\text{inv}$ denotes the 
matrix inverse, and we have we defined $\bar{\mathcal{T}} \coloneqq  I_{D^2} - \mathbb{Q}\mathcal{T}\mathbb{Q}$, $(\sigma^+_{ \rm b}| \coloneqq  (1| \mathcal{T}_{\sigma^+_{ \rm b}}$, and $|\sigma^-_{ \rm b}) \coloneqq  \mathcal{T}_{\sigma^-_{ \rm b}} |1)$. 
We note that $\bar{\mathcal{T}}$(-type) matrices are often ill-conditioned (almost singular);
however, even if the inverse is practically nonexistent, one can
estimate the above expression \blu using the
Moore-Penrose inverse. In practice, we calculate \eref{eq:C-final} by regularizing $\bar{\mathcal{T}}$ through the
addition of a very small term as $\epsilon I_{D^2}$ and then proceeding by a standard inversion, 
or by approximating Krylov-based iterative methods calculating the action of $\bar{\mathcal{T}}^{-1}$ on the 
vector in front of it. This \blk is why we used the notation $\text{inv}(\mathcal{X})$, and not $\mathcal{X}^{-1}$, to point to a generalized
approach of matrix inversion. 
Equation \eqref{eq:C-final} is our main recipe to calculate $\frak{C}$ numerically. 

In practice, accurately finding the optimum $A$-matrices and ${ \rho}^{\rm ss}$ (the $|1)$-vector) that maximizes $\frak{C}$ in \eref{eq:C-final}
for a given non-small $D$ is yet a relatively challenging task, even using numerical iteration. This would be due to the accumulated costs
of required computational steps
and instabilities caused by using a relatively large number of control quantities such as tolerances, maximum iteration numbers, and number of Krylov vectors contributed by such steps (see below). In particular, we soon realised that working with an approximately found vector $|1)$, ensuring that 
$\mathcal{T}$ is injective, and then estimating
the inverse of $\bar{\mathcal{T}}$ make the whole numerics 
unstable to changes in the set of tolerances and starting
vectors for the free parameters. 
Fortunately, this problem can be avoided; considering that we require $\mathcal{T}$ to be injective, the exact unique structure of ${ \rho}^{\rm ss}$ is analytically solvable.
From \eref{eq:IsometryCond} and \eref{eq:rho_ss}, we derived the recursive exact solution for the reduced density matrix as stated in the main text (for the laser's loss and gain operators -- see the connections below):
\beq \label{rhossAs}
{ \rho}^{\rm ss}_{m} = \left( \frac{A^{[0]}_{m,m-1}}{A^{[3]}_{m-1,m}} \right)^2 { \rho}^{\rm ss}_{m-1}~,~0< m < D. 
\eeq
This means that if the $A$-matrices are known, this equation will fully determine ${ \rho}^{\rm ss}$ (or equivalently $|1)$).
The first diagonal element of the reduced density matrix can be found using the fact that the diagonal elements of $\rho^{\rm ss}$ form a normalized probability distribution. 
Since a further $D-1$ free parameters are fixed by specifying $\rho^{\rm ss}$ (which ties $A^{[0]}$ and $A^{[3]}$ via the recurrence relation), the optimum $A$-matrices can be now found in quite an efficient manner.

Let us first detail the settings employed in optimization calculations, which were used to find the ans\"atze appearing in Eq.~(3) in the main text. 
We employed a constrained nonlinear optimizer, namely the highly-scalable interior-point method described in Refs.~(\cite{sWal06,sByr00,sByr99}), to find the maximum of the function given in Eq.~\eqref{eq:C-final}. 
In practice, the run-time complexity of this part scales at worst as $O(N_{\rm iter} N^3_{\rm param})$, where $N_{\rm param}$ denotes the number of free parameters to optimize for and $N_{\rm iter}$ 
is the number of iterations required
to reach a certain accuracy.  In practice, our stopping criteria was set simply for the step tolerance of the 
interior-point to reach a desired, very small, fixed value while satisfying all the constraints -- see below.  
Furthermore, we found the inverse in \eref{eq:C-final} (or in similar expressions) using the conventional memory-efficient method of 
banded LU-decomposition for $\bar{\mathcal{T}}$---an approach particularly optimized for sparse matrices---to form a linear system of equations at each iteration, which has the complexity of $O(N_{\rm param}^{2})$. 
Considering that $N_{\rm param}=O(D)$ for our generic no-ansatz calculations, the total computational cost scaled as $O(N_{\rm iter}D^{3})$, which was \blu quite \blk practical (at least for \blu $N_{\rm iter}\sim D$ and \blk dimensions up to $D=100$\blu---more details in \cite{sSaadatmand20}\blk). Finally, the memory cost of our optimization calculations \blu remained \blk
highly manageable, since we always exploited efficient methods to save and manipulate matrices as sparse-type inputs. 

\begin{figure}
  \begin{center}
    \includegraphics[width=\linewidth]{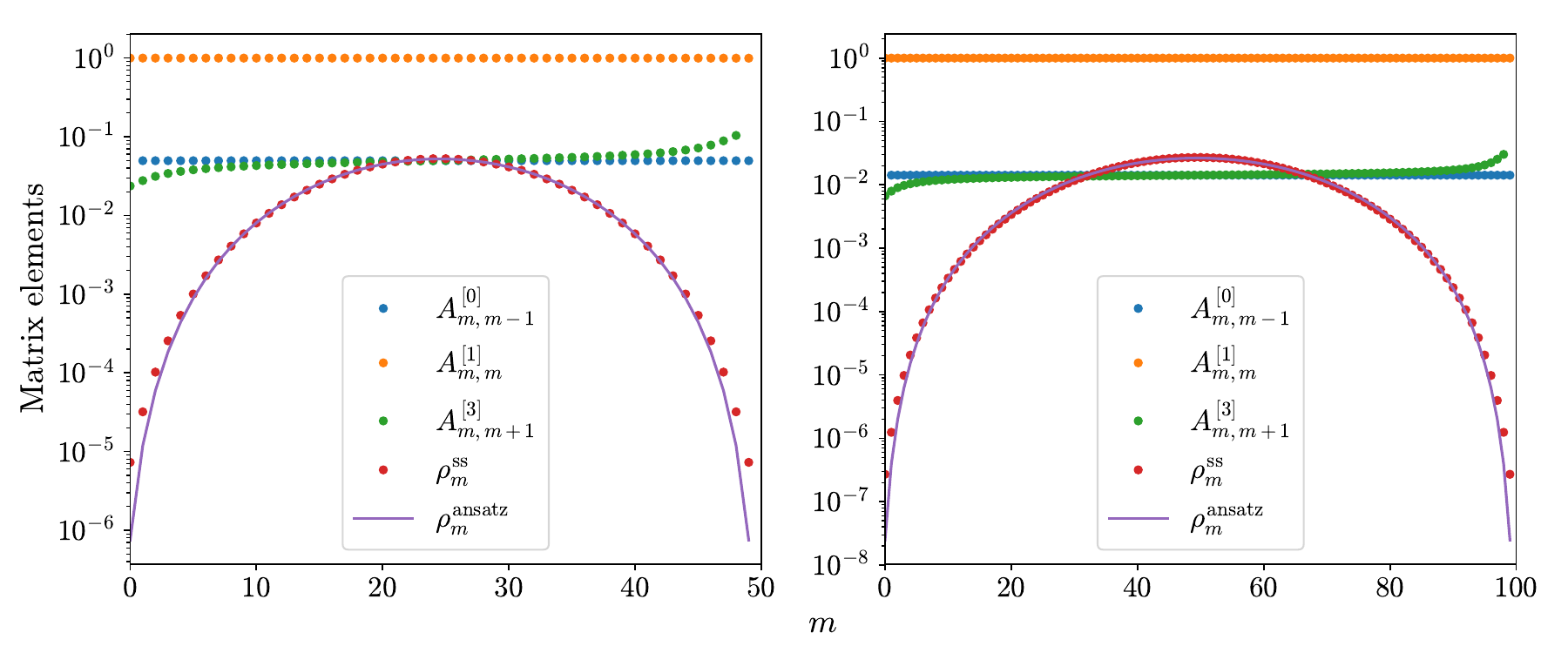}
    \caption{
Interior-point optimization results for the profile of non-zero diagonals of iMPS matrices, \eref{eq:A-forms}, when the limit of a maximum $\frak{C}$ is achieved.
We show the results for $D=50$ (left) and $D=100$ (right), when the ans\"{a}tze $A^{[2]}_{mn} = 0$ and \blu $A^{[0]}_{mn} = \text{const.}$ \blk (for their allowed non-zero elements) are placed as constraints. 
The solid lines display an ansatz for the steady-state presented in the main text, $\rho^{\rm ansatz}_m \propto \sin^4( \pi \frac{m+1}{D+1} )$.
    \label{fig:OptimizedMatrices-general}}
  \end{center}
\end{figure}
            
In the first series of our numerical calculations---the interior-point optimization part---we constrained the optimizations by Eqs.~\eqref{eq:ExplicitLeftOrth} and \eqref{rhossAs}, and forced the $A$-matrices to be of the forms prescribed in \eref{eq:A-forms}.
Moreover, we set $A^{[2]} = 0$ and all non-zero elements of $A^{[0]}$ equal to
some unique constant.
The justification behind setting $A^{[2]} = 0$ will be alluded to in Sec.~\ref{sec:MPS-interp}  below. 
Specifically, the $A^{[2]} = 0$ operator corresponds physically to transporting the uncorrelated pump qubit straight into the beam, and so should not play an important role in the creation of a coherent laser beam.
The reason behind setting the non-zero elements of $A^{[0]}$ to a constant was to ensure our model satisfied Condition 4 of the main text (see Sec.~\ref{sec:proving_all_conditions_hold} below for details).
We explored a number of possible models in our numerical optimizations, many of which attained Heisenberg limit scaling for $\coh$.
Importantly, the most general optimizations where no constraint was placed on $A^{[0]}$, did not result in a laser beam which was Glauber$^{(2)}$-ideal.
Setting the non-zero elements of $A^{[0]}$ to be equal to a constant rectified this problem.

Under these conditions, we found through numerous careful interior-point optimizations of the $A$-matrices for a number of states up to $D\!=\!100$, that when the maximum of $\frak{C}$ is achieved, these always possess the forms displayed in \fref{fig:OptimizedMatrices-general}.
There, we present exemplary cases $D=50$ (left) and $D=100$ (right).
We find that the ansatz
\beq \label{SMansatz1}
\rho^{\rm ansatz}_m \propto \sin^4\left( \pi \frac{m+1}{D+1} \right),
\eeq
\blu which was introduced in Eq.~(3) of the main text, \blk
works increasingly well in matching the numerically optimized  $\rho^{\rm ss}$-results as $D$ increases. 
For example, the fidelity, which for these diagonal states equals $\left( \sum_m \sqrt{ \rho^{\rm ss}_m } \sqrt{ \rho^{\text{ansatz}}_m }\right)^2$, evaluates to 
$90.76\%$ and $99.20\%$ for $D=50$ and $D=100$,  respectively. 

The reader may wonder about the significance of the relatively small size of the matrices $A^{[0]}$ and $A^{[3]}$ in \fref{fig:OptimizedMatrices-general}. 
While it is important that they are small (of the order of $10^{-1}$), there is no \emph{physical} significance in the particular value found by numerical optimization.
We have shown that the same Heisenberg scaling for $\coh$---to within numerical uncertainty---arises if the optimized  $A^{[0]}$ and $A^{[3]}$ matrices are manually scaled to be ever smaller, and $A^{[1]}$ increased even closer to the identity according to \eref{eq:ExplicitLeftOrth}. 
The state $\rho^{\rm ss}$ also remains unchanged, as per \eref{rhossAs}. 
The particular value obtained numerically was simply because the algorithm ceases to decrease the size of $A^{[0,3]}$-elements as soon as the interior-point optimization problem achieves the assigned tolerances explained above. 
This assured us that when $A$-matrices are optimized, we can introduce a small parameter $\gamma>0$ corresponding to the order of magnitude of the  elements of $A^{[0]}$ and $A^{[3]}$. 
Specifically, we will take $\gamma^2$ to be the number of photons emitted into the beam in a single time step.  
Since this number is small, we can regard the time step as infinitesimal, $\delta t$, so that we can fix its relation to the flux ${\cal N}$ by $\gamma^2=\mathcal{N}\delta t$. 
With this definition of $\delta t$, the physical and discrete time have the correspondence $t=q\delta t$. 

Now, assuming the form of the steady state in Eq.~\eqref{SMansatz1}, a constant gain operator $A^{[0]}_{m+1,m} = \gamma,~\forall m \neq D-1$, and  
$A^{[3]}$ and $A^{[1]}$ determined by $\rho^{\rm ss}$ and  the completeness relation, \eref{eq:IsometryCond}, we were able to perform numerical calculations of $\coh$ for larger values of $D$.
Thus, in the \emph{second} series of our calculations, 
the set of $A$-matrices are defined as follows:
\begin{align}
   \begin{cases}
     (1)~ \text{All the forms prescribed in \eref{eq:A-forms}}~, \\
     (2)~ A^{[2]} = 0~, \\
     (3)~ A^{[0]}_{m+1,m} = \gamma,~m<D-1~, \\     
     (4)~ A^{[3]}_{m-1,m} = \gamma \xfrac{\sin^2( \pi \frac{m}{D+1} )}{\sin^2( \pi \frac{m+1}{D+1} )},~m>0~, \\
     (5)~ A^{[1]} = \sqrt{I_D - A^{[0]\dagger} A^{[0]} - A^{[3]\dagger} A^{[3]}}~. \\ 
   \end{cases}
\label{eq:PoissonianCollectionAnsatz}
\end{align}
 Note that the expression for $A^{[3]}$ follows from the recurrence relation for $\rho^{\rm ss}$, which for the above choice of $A^{[0]}$ becomes very simple,  
$\rho_{m} = ( {\gamma}/{A^{[3]}_{m-1,m}})^2 \rho_{m-1},~\forall m\neq0$. 
We remind readers that $\gamma=\sqrt{\mathcal{N}\delta t}$ is an  arbitrarily small, but strictly nonzero, 
parameter. All physical observables of interest can be expressed independent of $\gamma$ 
(as expected from the final form of a discretized description of a continuum model), meaning that $D$ is the only relevant parameter.


Based on \eref{eq:PoissonianCollectionAnsatz}, we are now ready to derive virtually exact results for the coherence scaling in the limit of $\gamma \rightarrow 0^{+}$ and  $D\gg 1$, by fitting a power law to numerically calculated values of $\frak{C}$ for $D$-values as large as $1000$. 
To obtain results independent of $\gamma$, we define $B^{[i]} \coloneqq  A^{[i]}/\gamma $ 
for $i = 0$ and $3$; note that the elements of $B$-matrices are independent of $\gamma$. 
Then the transfer matrix  $\mathcal{T} = \sum_j A^{[j]}{}^* \otimes A^{[j]}$ can be  rewritten  as 
%
\begin{align}
   \mathcal{T} = I_{D^2} + { \gamma}^2 \mathcal{L}.
\label{eq:T&L-matrix}
\end{align}
%
Taking the limit $\gamma \rightarrow 0^{+}$,  the following  expression becomes exact
\begin{align}
   \mathcal{L} &= B^{[0]*} \otimes B^{[0]} + B^{[3]*} \otimes B^{[3]} - \frac{1}{2}(I_D \otimes L_0 + L_0 \otimes I_D)~,
\label{eq:L-matrix-explicit}
\end{align}
where we have  defined  a new diagonal MPS matrix as 
\begin{equation}
	L_0 \coloneqq  B^{[0]\dagger}B^{[0]} + B^{[3]\dagger}B^{[3]}.
\end{equation}
For our model, this matrix has elements
\begin{align}
(L_0)_{mm} \coloneqq  
 1  -\delta_{m,D-1}  + \frac{\sin^4( \pi \frac{m}{D+1} )}{\sin^4( \pi \frac{m+1}{D+1} )}~.
\end{align}
Substituting the expressions for the $B$-matrices from \eref{eq:PoissonianCollectionAnsatz} and the transfer matrix from \eref{eq:T&L-matrix}
into \eref{eq:C-final}, we find
\begin{align}
   \frak{C} =
    - 2 (1| ~ B^{[3]*} \otimes I_D \cdot \text{inv}(\mathbb{Q}\mathcal{L}\mathbb{Q}) \cdot I_D \otimes B^{[3]} ~ |1)~.
\label{eq:C-final-kappa0_indep}
\end{align}
The above result is how we evaluated $\frak{C}$ for the second series of our 
calculations, i.e.~for the results presented in Fig.~1 of
the main text. Note 
although there is no parameter optimization in this calculation, due to the inclusion of $D^2\times D^2$ matrices and an inverse operation, in practice, precisely evaluating coherences from \eref{eq:C-final-kappa0_indep} is still a \blu nontrivial \blk numerical task for large $D$.

\subsection{Physical interpretation and connections to the master equation}
\label{sec:MPS-interp}

In the main text, we presented the dynamics of our laser model in terms of the master equation $\dot{\rho} = \mathcal{L}\rho$.
There, the superoperator $\mathcal{L}$ was defined in terms of gain and loss operators $=$ and $\hat{L}$ as $\mathcal{L} = \mathcal{D}[\hat{G}] + \mathcal{D}[\hat{L}]$.
In fact, $\hat{G}$ and $\hat{L}$ correspond directly to $B^{[0]}$ and $B^{[3]}$ respectively, in taking the length of discrete time intervals to  be infinitesimal,  $\delta t \rightarrow 0^+$.

This correspondence becomes clear if we physically interpret the $A$-matrices for our iMPS model in terms of the processes involved in producing a beam from a laser cavity.
The operator $A^{[0]}$ relates to the amplitude of the process by which the cavity receives one uncorrelated photon from the input pump and no photon is emitted from the cavity.
The laser cavity is therefore excited by one level, meaning that $A^{[0]}$ can be interpreted as the operator describing the laser gain.
This gain process relates most obviously to that of a micromaser~\cite{sWalther06}, with $\ket{1}_{\rm p}$ being a single excited-state atom entering the cavity  and the sink being the (possibly de-excited) atom leaving the cavity. 
In a more typical laser~\cite{sLou73,sSarScuLam74,sCarmichael99_book}, with a gain medium comprising many atoms, the event analogous to the arrival of $\ket{1}_{\rm p}$ is the incoherent excitation of an individual atom to the upper level of the transition resonant with the cavity mode. In addition, the input $\ket{0}_{\rm v}$ can be thought of as the vacuum incident on the cavity output mirror, which becomes the output beam upon reflection~\cite{sCarmichael99_book}.

The operator $A^{[1]}$ relates to the amplitude of the process by which the cavity receives an uncorrelated photon from the pump and sends it directly to the sink, and the operator $A^{[2]}$ relates to the amplitude of the process by which the cavity receives one uncorrelated photon and sends it directly to the beam. 
Since such photons are uncorrelated with any other photons in the beam, they will add noise rather than contributing to the coherence of the beam. It is thus intuitive that a good choice would be to set $A^{[2]}=0$,  which is a choice we made in the simulations presented above. 
Finally, $A^{[3]}$ relates to the amplitude of the process  by which the cavity receives one uncorrelated photon, and emits two photons, one to the sink and one to the beam, so that it is de-excited by one level.  
Thus, $A^{[3]}$ can be thought of as describing the laser loss which creates the output beam.
 
In a time step $\delta t$, the state of the cavity evolves in our iMPS description according to
\begin{equation}
	\rho(t + \delta t) = \sum_j A^{[j]} \rho (t) A^{[j]}{}^\dagger.
\end{equation}
This is related to the master equation presented in the main text as follows.
Expanding the right hand side of the above equation in terms of the small parameter $\gamma=\sqrt{\mathcal{N}\delta t}=\sqrt{\delta t}$ (where we chose $\mathcal{N}=1$ as in the main text), we have
\begin{align}
	\dot{\rho}(t) &={} \lim\limits_{\delta t \rightarrow 0^+} \frac{\rho(t + \delta t) - \rho(t)}{\delta t} \\
	&={} B^{[0]} \rho(t) B^{[0]\dagger} + B^{[3]} \rho(t) B^{[3]\dagger} - \frac{1}{2}( L_0 \rho(t) + \rho(t) L_0) \\
	&={} \mathcal{D}[{B^{[0]}}]\rho(t) + \mathcal{D}[{B^{[3]}}]\rho(t),
\end{align}
where ${\cal D}[\hat c]\bullet \coloneqq \hat{c}\bullet\hat{c}\dg - \half\cu{\hat c\dg \hat c,\bullet}$, and $\cu{\bullet,\bullet}$ denotes the anti-commutator.
This is the master equation presented in the main text, with the gain and loss operators $\hat{G}$ and $\hat{L}$ replaced by the iMPS operators $B^{[0]}$ and $B^{[3]}$; in other words, $\hat{G} \equiv \hat{B}^{[0]}$ and 
$\hat{L} \equiv \hat{B}^{[3]}$, if we consider $B$-matrices as operators belonging to ${\cal B}({\cal H}_c)$.
 
\section{Proving our Laser Model satisfies Conditions 1--4}
\label{sec:proving_all_conditions_hold}

The laser model we have constructed above was derived in the framework of tensor networks in order to achieve $\coh = \Theta(\mu^4)$.
However, it is necessary that this Heisenberg limit for the coherence is saturated for the model \emph{while} satisfying Conditions 1--4.
In this final section of the SM, we demonstrate this claim.

Condition 1 (one-dimensional beam) follows simply from taking the continuum limit of the laser model, i.e. $\delta t\rightarrow 0^+$ as in Sec.~\ref{sec:MPS-interp}. 
In this limit, the lowering operator at site $q$ for the beam in the iMPS description $\hat{\sigma}^-_{ \rm b}(q)$,
satisfying $[\hat{\sigma}^-_{ \rm b}(q),\hat{\sigma}^+_{ \rm b}(r)]=-\delta_{qr}\hat{\sigma}^z_{ \rm b}(q)$, becomes the field operator $\hat{b}(t)$ satisfying $[\hat b(t),\hat b\dg(s)] = \delta(t-s)$, where $t$ is the time at which the infinitesimal part of the beam was created.
 Note that the operator $\hat{\sigma}^z_{ \rm b}(q)$, in this commutator, can, in this limit, be replaced by $-1$ because the local beam state is only infinitesimally different from the vacuum state, $\ket{0_q}_{\rm b}$, satisfying $\hat{\sigma}^z_{ \rm b}(q)\ket{0_q}_{\rm b}=-\ket{0_q}_{\rm b}$. 
Moreover, the relevant properties of the  iMPS  model which ensure Condition 3 (stationarity) is satisfied were alluded to in the main text.
The existence of a unique steady state for the cavity is ensured provided that all matrix elements $B^{[0]}_{m,m-1}$ and $B^{[3]}_{m-1,m}$ for $m=1,...,D-1$ are  nonzero. 
This is true for our laser model detailed in Eq.~\eqref{eq:PoissonianCollectionAnsatz}.

The demonstrations that Conditions 2 and 4 hold for our laser model, presented below, are slightly more involved.

\subsection{Condition 2: endogenous phase}
\label{sec:proving_endogenous_phase}

Condition 2 (endogenous phase) can be shown to hold in the continuum limit for our laser model, as a consequence of the following theorem.
The theorem requires considering the continuum laser field generated by the cavity in discrete time intervals \cite{sWisMil10},  where the state of the segment of the beam created in $[t,t+dt)$ belongs to a truncated Fock space ${\rm span}\cu{\ket{0},\ket{1}}$.
In this space, the beam operators are $\sqrt{\delta t}\hat{b}(t)=\ketbra{0}{1}$ and the photon number operator over this small beam segment becomes $dt\,\hat{b}\dg\hat{b}=\ketbra{1}$. 

\begin{theorem}[Sufficient conditions for endogenous phase] \label{theo:sufficient_cond2}
Suppose a laser model in which the evolution over small time $\delta t$ of the cavity, beam, and environment is governed by the unitary superoperator $\mathcal{U}^{t\rightarrow t+\delta t}_{\rm ce}: \mathcal{B}(\mathcal{H}_c \otimes \mathcal{H}_{\rm e}) \rightarrow \mathcal{B}(\mathcal{H}_{c} \otimes \mathcal{H}_{\rm b} \otimes \mathcal{H}_{\rm e'})$, defined through the relation 
\begin{align}
	\Tr_{\rm e'}\left[ \mathcal{U}^{t\rightarrow t+\delta t}_{\rm ce}(\rho_{\rm c}\blu \otimes \rho_{\rm e}\blk) \right] &={} \rho_{\rm c}\otimes\ketbra{0} + \sqrt{\delta t}\left( \hat{L}\rho_{\rm c}\otimes\ketbra{1}{0} + \rho_{\rm c}\hat{L}^\dagger \otimes\ketbra{0}{1}\right) \nonumber \\
	&{} \qquad+ \delta t \left( \hat{L} \rho_{\rm c} \hat{L}^\dagger \otimes \ketbra{1} - \frac{1}{2}\left\{ \hat{L}^\dagger\hat{L}, \rho_{\rm c}\right\}\otimes\ketbra{0} +  \mathcal{D}[\hat{G}](\rho_{\rm c})\otimes\ketbra{0}\right) + o(\delta t).
	\label{eq:theorem_cavity_beam}
\end{align} 
In such a model, if the two identities 
\begin{align}
	\mathcal{U}^\theta_{\rm c}( \mathcal{D}[\hat{G}](\bullet)) &={} \mathcal{D}[\hat{G}](\mathcal{U}^\theta_{\rm c}(\bullet))\label{eq:theorem2_phase1}\\
	\mathcal{U}^\theta_{\rm c}(\hat{L}) &={} \blu e^{-i\theta}\blk\hat{L}
	\label{eq:theorem2_phase2}
\end{align}
hold, for generic phase shift superoperators $\mathcal{U}^\theta_{\rm c} (\bullet) \coloneqq \blu e^{i\theta\hat{n}_{\rm c}} \bullet e^{-i\theta\hat{n}_{\rm c}}$, \blk then the endogenous phase condition holds.
\begin{proof}
The theorem follows from a straightforward calculation which shows that the endogenous phase condition holds, if Eqs.~\eqref{eq:theorem_cavity_beam} through \blu \eqref{eq:theorem2_phase2} \blk hold. 
We have
\begin{align}
	\Tr_{\rm e'}\left[ \mathcal{U}^\theta_{\rm cb} (\mathcal{U}^{t\rightarrow t+\delta t}_{\rm ce}(\rho_{\rm c} \blu \otimes \rho_{\rm e}\blk)) \right] &={} \mathcal{U}^\theta_{\rm c} (\rho_{\rm c})\otimes\mathcal{U}^\theta_{\rm b} (\ketbra{0}) + \sqrt{\delta t}\left( \mathcal{U}^\theta_{\rm c} (\hat{L}\rho_{\rm c})\otimes\mathcal{U}^\theta_{\rm b} (\ketbra{1}{0}) + \mathcal{U}^\theta_{\rm c} (\rho_{\rm c}\hat{L}^\dagger) \otimes\mathcal{U}^\theta_{\rm b} (\ketbra{0}{1})\right) \nonumber \\
	&{} \qquad+ \delta t \left( \mathcal{U}^\theta_{\rm c} (\hat{L} \rho_{\rm c} \hat{L}^\dagger) \otimes \mathcal{U}^\theta_{\rm b} (\ketbra{1}) - \frac{1}{2}\mathcal{U}^\theta_{\rm c} \left(\{ \hat{L}^\dagger\hat{L}, \rho_{\rm c}\}\right) \otimes \mathcal{U}^\theta_{\rm b} (\ketbra{0}) \right. \nonumber \\
	&{} \left. \qquad +\,\mathcal{U}^\theta_{\rm c} (\mathcal{D}[\hat{G}](\rho_{\rm c}))\otimes\mathcal{U}^\theta_{\rm b} (\ketbra{0}) \right) +  o(\delta t)  \\
	&={} \mathcal{U}^\theta_{\rm c} (\rho_{\rm c})\otimes\ketbra{0} + \sqrt{\delta t}\left( \blu e^{-i\theta} \blk \hat{L}\mathcal{U}^\theta_{\rm c} (\rho_{\rm c})\otimes \blu e^{i\theta}\blk\ketbra{1}{0} + \mathcal{U}^\theta_{\rm c} (\rho_{\rm c}) \hat{L}^\dagger \blu e^{i\theta}\blk \otimes \blu e^{-i\theta}\blk\ketbra{0}{1}\right) \nonumber \\
	&{} \qquad+ \delta t \left( \hat{L} \mathcal{U}^\theta_{\rm c} (\rho_{\rm c}) \hat{L}^\dagger \otimes \ketbra{1} - \frac{1}{2}\{ \hat{L}^\dagger\hat{L}, \mathcal{U}^\theta_{\rm c}(\rho_{\rm c})\} \otimes \ketbra{0} \right. \nonumber \\
	&{} \left. \qquad+\, \mathcal{D}[\hat{G}]\mathcal{U}^\theta_{\rm c} (\rho_{\rm c})\otimes \ketbra{0} \right) +  o(\delta t) \\
	&\equiv{} \Tr_{\rm e'}\left[ \mathcal{U}^{t\rightarrow t+\delta t}_{\rm ce}(\mathcal{U}^\theta_{\rm c} (\rho_{\rm c})) \right].
\end{align}
\end{proof}
\end{theorem}
 Since there is no evolution of the system ${\rm b}_t$ beyond $t+\delta t$, the above result generalizes for a finite evolution $\mathcal{U}^{t\rightarrow t+\tau}_{\rm ce}$. 
To prove our laser model satisfies Condition 2 (endogenous phase) in the continuum limit, it remains to show that it meets all conditions stated in Theorem \ref{theo:sufficient_cond2}, Eqs.~\eqref{eq:theorem_cavity_beam} through \eqref{eq:theorem2_phase2}.
Recall the unitary interaction operator which describes time evolution in our tensor network model, defined below Eq.~\eqref{eq:interaction_isometry}.
It maps the system of the $D$-dimensional cavity, pump and vacuum qubits into the cavity plus beam and sink qubits in a single time step.
Thus, the environment state space $\mathcal{H}_{\rm e}$ in Theorem \ref{theo:sufficient_cond2} corresponds to the two-qubit Hilbert space $\mathcal{H}^2_{\rm p}\otimes\mathcal{H}^2_{\rm v}$, which become the beam and sink output spaces $\mathcal{H}^2_{\rm b}\otimes\mathcal{H}^2_{\rm s}$ in the proceeding time step.
In other words, $\rho_{\rm e}\equiv \ket{1}_p\bra{1} \otimes\ket{0}_v\bra{0}$ for our laser model.
This means that the unitary superoperator $\mathcal{U}^{t\rightarrow t+\delta t}_{\rm ce}(\bullet)$ in Eq.~\eqref{eq:theorem_cavity_beam} which acts at time $t$, is equivalent to the isometry acting at time $q$ as $\hat{V}_q (\bullet) \hat{V}_q^\dagger$ defined in Eq.~\eqref{eq:interaction_isometry}.
 Recall from Eq.~\eqref{eq:interaction_isometry} each discrete time step in our laser model, the cavity and beam evolves as
\begin{align}
	\Tr_{\rm s}\left[ \hat{V}\rho_c\hat{V}\dg \right] &={} \Tr_{\rm s}\left[\sum_j A^{[j]}\rho_c A^{[j]\dagger} \otimes\ket{j}_{ \rm o}\bra{j}\right] \\
	&={} \rho_{\rm c}\otimes\ketbra{0} + \gamma\left( \hat{B}^{[3]}\rho_{\rm c}\otimes\ketbra{1}{0} + \rho_{\rm c}\hat{B}^{[3]\dagger} \otimes\ketbra{0}{1}\right) \nonumber \\
	&{} \qquad+ \gamma^2 \left( \hat{B}^{[3]} \rho_{\rm c} \hat{B}^{[3]\dagger} \otimes \ketbra{1} - \frac{1}{2}\left\{ \hat{B}^{[3]\dagger}\hat{B}^{[3]}, \rho_{\rm c}\right\}\otimes\ketbra{0} +  \mathcal{D}[\hat{B}^{[0]}](\rho_{\rm c})\otimes\ketbra{0}\right) + o(\gamma^2).
	\label{eq:final_theorem_MPS_expansion}
\end{align}
Now, in \sref{sec:MPS-interp} we saw that in the limit $\gamma^2=\delta t\rightarrow0^+$, we can consider $B$-matrices as operators belonging to ${\cal B}({\cal H}_c)$, and thus can make the association $\hat{G} \equiv \hat{B}^{[0]}$ and $\hat{L} \equiv \hat{B}^{[3]}$ in that limit.
From these arguments, Eq.~\eqref{eq:final_theorem_MPS_expansion} is equivalent to  Eq.~\eqref{eq:theorem_cavity_beam} in the continuum limit. 

It remains to check that Eqs.~\eqref{eq:theorem2_phase1} and \eqref{eq:theorem2_phase2} are also satisfied in our laser model.
This follows directly from the forms of the $A$-matrices in the cavity number basis (see Eq.~\eqref{eq:A-forms}), and exchanging $B^{[0]}$ and $B^{[3]}$ with $\hat{G}$ and $\hat{L}$, respectively.
Hence, all conditions of Theorem \ref{theo:sufficient_cond2} are satisfied by our laser model, and therefore taking $\sqrt{\delta t}\hat{b}(t) = \sigma^-_{\rm b}(q)$ as the beam operator of beam $\hat{b}(t)$ produced at time $t$ to recast our model in the continuum limit, Condition 2 (endogenous phase) is satisfied.

\subsection{Condition 4:  ideal Glauber-$^{(1),}$$^{(2)}$ coherence }
\label{sec:G1ideality}

Condition 4 in the main text requires that the first- and second-order Glauber coherence functions for our laser model ``well approximate'' those for the standard model of a single mode laser beam. \blu In Sec.~\ref{sec:glauber_ideality} we quantified these requirements by \eqref{firstwellapp} and \eqref{secondwellapp}.  
In this section, we show how these inequalities arise, as the sufficient conditions of the Glauber coherence functions to allow Theorem~\ref{theo:main} to be derived. 
We then provide simpler conditions on $g^{(1)}$ and $g^{(2)}$, and show numerically that they are indeed satisfied by our model. 

\blu 
In order for our model to sufficiently accurately approximate the ideal laser beam, the asymptotic bound on the MSE for measuring the phase in Eq.~\eqref{eq:asymptotic_msse} should still hold, which implies that the upper bound on the coherence of approximately $3\mu^4$ still holds.
To estimate the difference between the MSE for the model and the MSE for the ideal laser, note that
in Eq.~\eqref{eq:g12var} the first term comes from $g^{(1)}$ and the second term comes from $g^{(2)}$.
If the model does not have ideal coherence, then using Eq.~\eqref{eq:SSG2} and \eqref{eq:SSG2_oneterm}, 
the difference in the MSE from that for ideal coherence is 
\begin{align}
&\left|\frac{1}{{\cal N}\tau^3} \int_0^{\tau} dt \int_0^{\tau} ds \, \delta g_{\rm model - ideal}^{(1)}(s,t) + \frac{1}{\tau^4}\left[\frac{1}{2} \int_0^{\tau} ds \int_{-\tau}^0 ds' \, \int_0^{\tau} dt' \int_{-\tau}^0 dt \,\delta g^{(2)}_{\rm model - ideal}(s,s',t',t) \right.\right.\nonumber\\
	&{} \qquad \qquad \ \left. - \frac{1}{2}\left. \int_0^{\tau} ds \int_0^{\tau} ds' \,\int_{-\tau}^0 dt' \int_{-\tau}^0 dt \, { \delta} g^{(2)}_{\rm model - ideal}(s,s',t',t) \right]\right| . \blu \label{eq:gdif} 
\end{align}
From this we can see that the {\em relative} difference in the MSE from that for ideal coherence, Eq.~(\ref{eq:asymptotic_msse}), ignoring multiplicative constants, 
is bounded above by 
\begin{align}
&\max_{s,t\in[0,\tau]} \left|\delta g_{\rm model - ideal}^{(1)}(s,t)\right| + {\mathcal{N}\tau}  \max_{s,s',t',t\in[-\tau,\tau]} \left|\delta g_{\rm model - ideal}^{(2)}(s,s',t',t)\right|.\label{eq:gbnd}
\end{align}

\begin{figure}
\begin{center}
\includegraphics[width=0.73\linewidth]{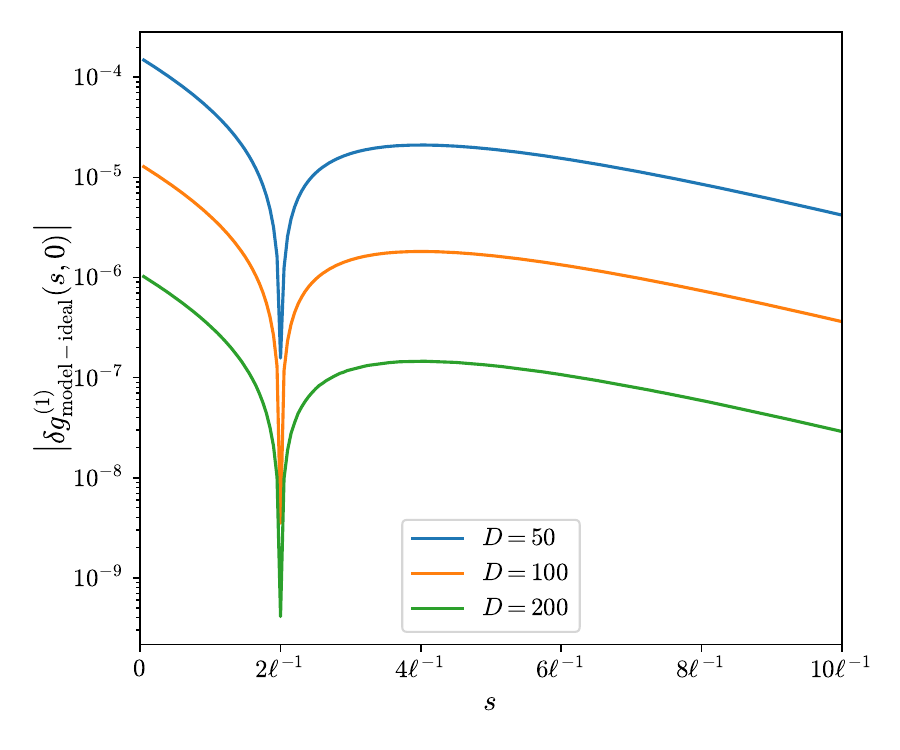}
\caption{Deviations from the ideal laser model  (a coherent state with its phase undergoing pure diffusion)  of the first-order Glauber coherence function for our laser model. 
For $\mathcal{N}=1$, the magnitude of the quantity $\delta g_{\rm model - ideal}^{(1)}(s,t)$ (with $t=0$ without loss of generality) is shown over ten coherence times.
For increasing $D$, we see this difference is converging toward zero like a power law, which implies $G^{(1)}$-ideality is satisfied for our laser model.
}
\label{fig:g1corrections}
\end{center}
\end{figure}

First, consider the first term in \eqref{eq:gbnd}. This will be negligible if 
\beq \label{eq:g1bnd}
\max_{s,t\in[0,\tau]} |\delta g_{\rm model - ideal}^{(1)}(s,t)| =o(1),
\eeq 
as stated in \eqref{secondwellapp}, without the constraints on the arguments.  
We now show numerically that the magnitude of $\delta g_{\rm model - ideal}^{(1)}$ \blu is indeed small, 
for a range of times far larger than needed for  \eqref{eq:g1bnd}. \blk
Since the beam is translationally invariant, we take $t=0$ and plot $|\delta g_{\rm model - ideal}^{(1)}(s,0)|$ against $s$ 
 (in units of $\ell^{-1}$) in Fig.~\ref{fig:g1corrections}. \blu
For these calculations we have used \blu ${\cal N}=1$ (as in the main text) so that \blk $g_{\rm model}^{(1)}(s,0) = (\sigma^+| \mathbb{L}_{\rm exp}(s) |\sigma^-)$, 
where we have defined $\mathbb{L}_{\rm exp}(t) \coloneqq  \exp(\mathcal{N}t\mathcal{L}) = \exp(t\mathcal{L})$
and used the transformation $\sqrt{\delta t}~\hat{b} \rightarrow \hat{\sigma}^-_{ \rm b}$. 
The time interval in this plot, up to $10\ell^{-1}$, corresponds to that in the inset in Fig.~1 in the main text,  \blu
and is much longer than $\tau=\ell^{-1}\sqrt{6/\coh}$, since we are interested in the limit $\coh \gg 1$.  
Moreover, from the values for different $D$, we see that the value of $|\delta g_{\rm model - ideal}^{(1)}(s,t)|$ decreases inversely with $\coh=\Theta(D^4)$. 

\begin{figure}
  \begin{center}
    \includegraphics[width=0.85\linewidth]{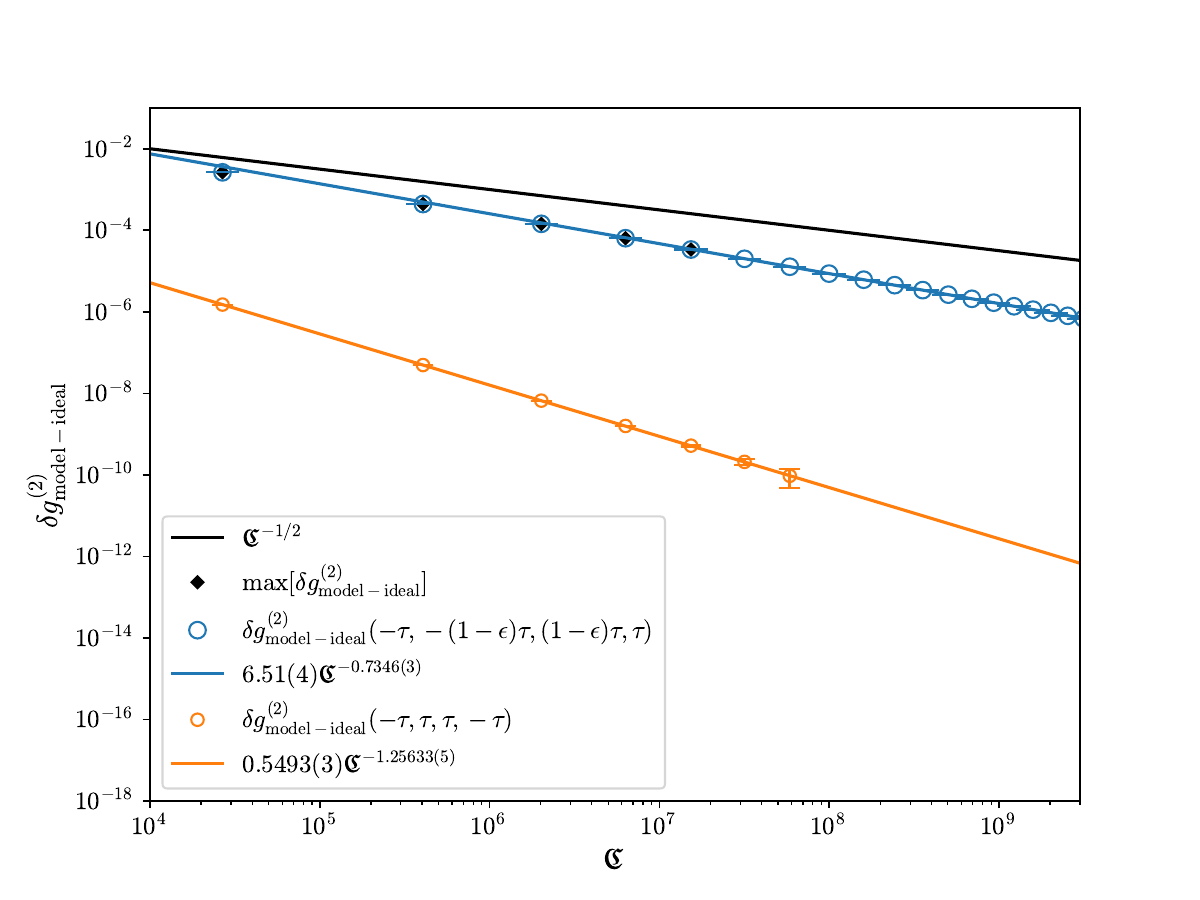}
    \caption{
    The global maxima of \blu$\delta g_{\rm ideal}^{(2)}(-\tau,s',t',t)$ \blk versus the coherence (black diamonds) 
    calculated for \blu $\{s',t',t\} \in [-\tau, \tau]$ \blk employing interior-point optimizations of 
    the iMPS forms  for bond dimensions up to $250$.  Some examples for $\delta g_{\rm ideal}^{(2)}$ are also presented for 
    comparison for bond dimensions up to $1000$, where the iMPS forms of Glauber
    coherence functions can be found in \eref{eq:exampleG2s}.
    Error bars are smaller than  the  symbol size for for the  black diamonds. \blu Coloured \blk lines are power-law fits to large-$D$ points of 
    these examples \blu and the black line is $\coh^{-1/2}$ for comparison purposes\blk.
    \label{fig:g4corrections}}
  \end{center}
\end{figure}


Next consider the second term, arising from $g^{(2)}$. 
By comparing Eq.~\eqref{eq:gdif} to Eq.~\eqref{eq:dgvar} we see that a sufficient condition is that $|\delta g_{\rm model - ideal}^{(2)}(s,s',t',t)| \ll |\delta g_{\rm ideal-1}^{(2)}(s,s',t',t)|$, as given in \eqref{firstwellapp}.  
Alternatively, from Eq.~\eqref{eq:gbnd} a simpler sufficient condition is that $|\delta g_{\rm model - ideal}^{(2)}(s,s',t',t)|\ll 1/({\cal N}\tau)$. 
With the choice of $\tau$ in Eq.~\eqref{eq:tauchoice}, this is equivalent to 
\begin{equation}
	\max\limits_{s,s',t',t \in [-\tau,\tau]} \left|\delta g_{\rm model - ideal}^{(2)}(s,s',t',t)\right| =  o \left( \frak{C}^{-1/2} \right).
	\label{eq:this_is_wellapproximate}
\end{equation}
This statement is the one given as Eq.~(11) in the Methods of the main text. (Recall that we use there ${\cal N}=1$, which means $G^{(n)} \equiv g^{(n)}$.) 
We now demonstrate that this condition holds for 
our model. 

To begin, we can take the first time argument in the $g^{(2)}$ function to be \blu $-\tau$ without loss of generality. 
This is because of the symmetries of $g^{(2)}$.
Because the annihilation operators commute, $g^{(2)}(s,s',t',t)=g^{(2)}(s,s',t,t')$, and because the creation operators commute, $g^{(2)}(s,s',t',t)=g^{(2)}(s',s,t',t)$.
We can also take the complex conjugate and find $[g^{(2)}(s,s',t',t)]^*=g^{(2)}(t,t',s',s)$.
For the ideal case $g^{(2)}$ is real, so $g^{(2)}_{\rm ideal-1}(s,s',t',t)=g^{(2)}_{\rm ideal-1}(t,t',s',s)$, and $|g^{(2)}_{\rm model-ideal}(s,s',t',t)|=|g^{(2)}_{\rm model-ideal}(t,t',s',s)|$.
Therefore, for the purpose of finding the maximum deviation of the model from the ideal case, for any set of times we can change the ordering so $s$ is earliest, without changing the deviation from the ideal coherence.
Then we can use time invariance to shift the times so that $s=-\tau$ and $s',t',t\in[-\tau,\tau]$. 
Thus, we need only calculate $\max_{s',t',t \in [-\tau,\tau]}[\delta g_{\rm ideal}^{(2)}(-\tau,s',t',t)]$.  

\blk Next, we display the flattened-space iMPS form for a representative example of time ordering, $-\tau<s'<t'<t$:  
\begin{align}
   \blu g\blk^{(2)}(\blu-\tau\blk,s',t',t)\blu|_{-\tau<s'<t'<t} \blk&= \blk (\sigma^+| \mathbb{L}_{\rm exp}(s'\blu + \tau\blk) (B^{[3]*}\otimes I_D)
     \mathbb{L}_{\rm exp}(t'-s') (I_D\otimes B^{[3]}) \mathbb{L}_{\rm exp}(t-t') |\sigma^-).
\label{eq:exampleG2s}
\end{align}
Here we again take $\delta t \rightarrow 0^+$ and use \blk
the transformation $\sqrt{\delta t}~\hat{b} \rightarrow \hat{\sigma}^-_{ \rm b}$ 
and commutator relations $[\hat{b}(t),\hat{b}(t')]\equiv 0$ and $[\hat{b}^\dagger(t),\hat{b}(t')]=0$ for $t\neq t'$.
\blu While \eref{eq:exampleG2s} is for the specific ordering $-\tau<s'<t'<t$, for the numerical calculations below we consider all nontrivial permutations of the bosonic operators, for which the correlators can be computed from similar expressions. \blk

The main numerical challenge for calculating such second-order (and even first-order) Glauber coherence functions is to exponentiate $D^2 \times D^2$ matrix $\mathcal{L}$ for large bond dimension $D$; this is equivalent to precisely finding a large enough number of its eigenvalues. 
Direct scaling and squaring algorithms (directly estimating matrix exponential based on Pad\'{e} approximants) provide high precision, but are memory extensive with computational time scaling as $O(D^6)$. 
Luckily, the $\mathcal{L}$ superoperator is highly sparse (in fact, as mentioned, we manipulate all the operators in our codes as sparse inputs), 
and Krylov subspace  projection techniques can be used to efficiently 
find the result of the action of $\mathbb{L}_{\rm exp}(t)$
on a vector. Here, we employed the `matrix-free' Krylov method introduced in \cite{sSid98}. As such,
we constructed the iMPS forms for all necessary permutations of arguments in 
$\delta g_{\rm ideal}^{(2)}(\blu -\tau\blk,s',t',t) = g^{(2)}(\blu -\tau\blk,s',t',t) - g_{\rm ideal}^{(2)}(\blu -\tau\blk,s',t',t)$
as above. Then, we employed the highly-scalable optimization method of 
interior-point, discussed in \sref{sec:MPS-methods}, to find the global maximum of all 
such functions for $\{s',t',t\} \in \blu [-\tau, \tau]\blk$ for bond dimensions up to $250$. 

In \fref{fig:g4corrections}, 
we present the results of these interior-point optimizations and additionally some 
representative sub-maximal $\delta g_{\rm ideal}^{(2)}$ deviations. 
\blu As the plot shows, for $D$ up to $250$, \blk $\max_{s',t',t \in [\blu-\tau,\tau]\blk}[\delta g_{\rm ideal}^{(2)}(\blu -\tau\blk,s',t',t)]$ almost exactly coincides
with easier-to-calculate $\delta g^{(2)}_{\rm ideal}(\blu -\tau,-(1-\epsilon)\tau, (1-\epsilon)\tau, \tau\blk)$, 
where $\epsilon$ is an arbitrarily small but strictly positive number. \blu We calculate the latter for far larger $D$, from which we extract \blk 
a power-law fit 
of $\Theta( \frak{C}^{-0.7346(3)})$, a scaling \blu certainly strictly smaller \blk than $\frak{C}^{-1/2}$. 
Eq.~\eqref{eq:this_is_wellapproximate} therefore holds, which implies that Condition 4 of the main text is indeed satisfied for our laser model.

\blu

\section{Engineering a beam performing at the Heisenberg Limit}
\label{sec:pathway}

Thus far, we have proven theoretically that the Heisenberg limit for the coherence of a close-to-ideal laser beam is $\coh =\Theta(\mu^4)$. Part 
of this was our numerical proof, using the MPS framework in \sref{sec:MPS-methods} that a model exists which achieves this scaling. 
However, it is a different matter to ask whether our model, which involves couplings very different from those of a standard laser, 
can be realised with current, or foreseeable, technology. In this section we answer that question in the affirmative,  
by introducing an engineered and controlled system that could, in principle, demonstrate the required dynamics. 
The system we propose operates in the microwave domain, 
but the fundamental physics is independent of the frequency band. 
Historically, the production of coherent optical fields at microwave frequencies (masers) \cite{sGor55} preceded the production of such fields at optical frequencies (lasers)~\cite{sSchTow58, sMai60}, and we expect this will be repeated for the development of devices 
surpassing the standard quantum limit.

\begin{figure}
\begin{center}
\includegraphics[width=\linewidth]{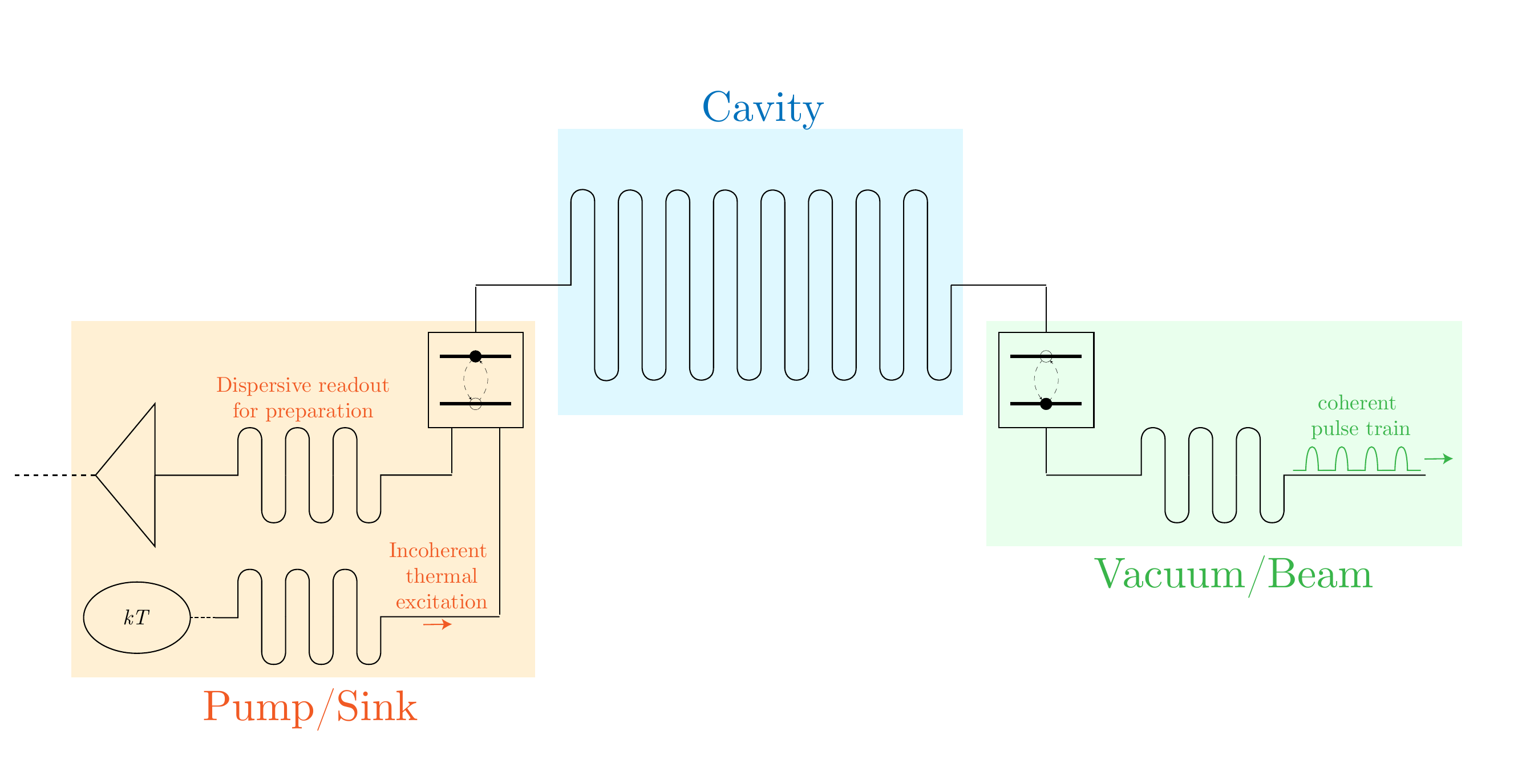}
\caption{\blu Proposal for a controllable physical system to produce a coherent field attaining the Heisenberg limit.
The setup implements the cavity, input, and output qubits as a system of two superconducting qubits capacitively coupled to a large resonator cavity 
with dynamically controllable detunings.  
Also required are a thermal source for creating an excitation in the pump/sink qubit, an amplifier for detecting the presence (or absence) of such an excitation, an output transmission line for the vacuum/beam qubit, and secondary cavities for mediating the couplings of these elements. 
See the text for details.}
\label{fig:superconducting_maser}
\end{center}
\end{figure}

Motivated by the current technologies available from the field of circuit quantum electrodynamics (CirQED), 
we consider the superconducting system illustrated in Fig.~\ref{fig:superconducting_maser}.
The cavity (blue) is a superconducting coplanar waveguide resonator, which stores a mean number $\mu$ of excitations at microwave frequency $\omega_c$. 
This resonator is capacitively coupled to two superconducting qubits (SQs), both with the ground-to-excited frequency separations which 
can be tuned by relatively small amounts away from $\omega_c$~\cite{sKrastanov2015}. Each of these plays two roles in terms of the laser model in Fig.~\ref{fig:model}: as the pump/sink and vacuum/beam respectively, with the solidus separating the role at the beginning of a control cycle from that at its end.  
We denote the Hilbert spaces of the left and right qubits, and the cavity, as $\mathcal{H}_\text{l}$, $\mathcal{H}_\text{r}$ and $\mathcal{H}_\text{c}$, respectively. 
In the description below, the labels for the non-cavity components are omitted, and the interaction unitary of our model corresponds to a superoperator which maps bounded operators onto their original Hilbert spaces, ~$\mathcal{U}^{t\rightarrow t+\delta t}_{\rm clr}: \mathcal{B}(\mathcal{H}_c \otimes \mathcal{H}_{\rm l} \otimes \mathcal{H}_{\rm r}) \rightarrow \mathcal{B}(\mathcal{H}_{c} \otimes \mathcal{H}_{\rm l} \otimes \mathcal{H}_{\rm r})$.
The laser (or, if one prefers, maser) operates by repeating the following gain and loss processes.

Gain is achieved using the left SQ in Fig.~\ref{fig:superconducting_maser}. 
It is driven by a thermal reservoir (bottom rail), which pumps the SQ with incoherent excitations. This is mediated by a secondary resonator, enabling the 
coupling to the reservoir to be dynamically controlled by detuning.  
The SQ is monitored (represented by an amplifier) via another off-resonant resonator, which permits dispersive readout (i.e.~by measuring a shift in phase of the coupled resonator) of the excitation, $\s{z}$ \cite{sWal04, sLup04}. 
Such monitoring allows, via multiple trials, deterministic preparation of the SQ in the excited state. 
When this excitation is prepared, the detuning of this SQ is changed again to bring it closer to resonance with the 
main resonator (the cavity), and the detuning is dynamically controlled to transfer the excitation into the cavity mode, 
with some probability, in the desired way, as will be explained below. The right SQ then becomes a 'sink'---the information it 
contains is discarded when it is allowed to thermalize again.   
During this step the right SQ is kept decoupled from the cavity. 

Loss, generating the output beam, is achieved by the right SQ in Fig.~\ref{fig:superconducting_maser}. This SQ, initially in the ground state (vacuum), 
is brought close to resonance with the cavity, and its detuning is dynamically controlled to extract an excitation from the cavity mode,
with some probability, in the desired way, as will be explained below. 
The SQ is then coupled to the secondary resonator on the right, which itself is coupled to a transmission line, 
so that the excitation in the SQ leaves a well-defined pulse in the transmission line. This leaves the SQ in the ground state 
as required for repeating the cycle.  
Repeating this loss sequence, intercalated with gain, generates the beam as a coherent train of pulses. 
While not being precisely a continuous-wave (CW) beam, the statistical properties of the beam on a time scale 
large compared to the pulse separation are identical to that of a CW beam.  

The remainder of this section is to show that the required gain and loss unitaries can be engineered via the usual  
dipole coupling (in the rotating-wave approximation) between a fixed harmonic mode (the cavity) and a SQ of approximately the same frequency, 
with the only control parameter being the detuning for the SQ \cite{sBla04}. 
Importantly, we do not assume the ability to coherently drive the cavity or qubit at the microwave frequency of the output beam, as this would require a coherent source at that frequency, which would violate the fundamental idea (see main text) of the laser as a device which creates a coherent output from inputs that have negligible coherence.  

\subsection{Controllability}

Our proposed setup for achieving the Heisenberg limit relies on the ability to faithfully implement the dynamics of the family of laser models presented in the main text.
That is, it hinges on the question of \emph{sufficient controllability}: is the desired interaction unitary 
contained within the set of evolutions obtainable from the allowed fixed and controllable couplings.  
We begin by reviewing some fundamental results from the theory of quantum control, which we introduce along with some useful notations from Ref.~\cite{sdAl07}.

Consider a system governed by a Hamiltonian written in terms of non-controlled and controlled parts,
\begin{equation}
	\hat{H}(\mathbf{u}(t)) = \hat{H}_0 + \sum_{s=1}^{S} u_s (t) \hat{H}_s,
	\label{eq:general_hamiltoniancontrolled}
\end{equation}
where $\mathbf{u}(t)$ is the vector of $S$ independent control functions $(u_1(t),u_2(t), \dots ,u_S(t))$, each of which can be taken to be a bounded, piece-wise constant function of time.
The evolution of the system governed by this Hamiltonian is described by the unitary operator $X(t)$ 
which 
is a solution to 
\begin{equation}
	\dot{X}(t) = -iH(\mathbf{u}(t)) X(t), \qquad X(0) = I.
	\label{eq:schrodinger_operator}
\end{equation}
If the set of operators which are obtainable solutions to Eq.~\eqref{eq:schrodinger_operator} is the entire set of unitary operators, the system is said to be \emph{completely} controllable. 
However, for most systems---including the maser system we propose---this is not the case.
For a system governed by Eq.~\eqref{eq:general_hamiltoniancontrolled}, the set of allowable unitaries are precisely those contained in the Lie group associated with the Lie algebra over the field of real numbers, generated by the set $\mathbb{H}\coloneqq \{ i\hat{H}_j\}_{j=0}^{S}$.
That is, all achievable unitaries belong to $e^\mathfrak{L}$, where $\mathfrak{L}$ is the Lie algebra generated by $\mathbb{H}$ (see e.g. \cite[Theorem 3.2.1]{sdAl07}). 
This means the span, over the real numbers, of the repeated Lie brackets (here, commutators) of the operators in $\mathbb{H}$, of all depths from 0 to $\infty$. 
Here, a repeated Lie bracket of depth $p-1$ is the set of all operators of the form $[\hat{O}_1,[\hat{O}_2,[\dots[\hat{O}_{p-1},\hat{O}_p]]]]$, with $\hat{O}_1, \dots, \hat{O}_p \in \mathbb{H}$. 
The repeated Lie brackets of depth 0 are, by convention, just the elements of $\mathbb{H}$.
The Lie algebra $\mathfrak{L}$ is often referred to as the \emph{dynamical Lie algebra} of the system.

A method for generating a basis for the dynamical Lie algebra is as follows \cite{sdAl07}.
First, one starts with $\mathbb{H}$, the set of operators with depth 0.
By definition, $\mathfrak{L}$ is completely determined by all possible repeated Lie brackets generated from elements of this set.
This means that, in order to find a basis for $\mathfrak{L}$, one can first calculate the operators of depth 1, and discard any which can be written as a linear combination of those in $\mathbb{H}$.
Those which are linearly independent are appended to this set, as new elements of the basis.
One then repeats this computation of repeated Lie brackets, increasing the depth by one each time, and keeping only those which are linearly independent with respect to the elements generated at lower depth.
This procedure is repeated until no new basis elements are generated, or until sufficient basis elements are generated for the Lie group to contain the desired unitary.

\subsection{Heisenberg Limit dynamics can be implemented}

We now return to Eq.~(5) of the main text,
\begin{equation}
	\hat H(t) = \sum_{s={\rm l,r}} \left[ g_s\left( \hat a\dg \hat  \sigma_s^- + \hat a  \hat \sigma_s^+ \right)  + \Delta_s(t) \hat \sigma^z_s \right],
	\label{eq:maser_couple}
\end{equation} 
where $\omega_s(t) = \omega_{\rm c} + \Delta_s(t)$, and we have set $\hbar=1$.
Here we are working in a frame rotating at the cavity frequency $\omega_{\rm c}$, and 
$\hat{H}_0=\sum_{s={\rm l,r}}  g_s\left( \hat a\dg \hat  \sigma_s^- + \hat a  \hat \sigma_s^+ \right)$ describes, in the rotating-wave approximation, the exchange of excitations between each of the left (l) and right (r) qubits with the cavity, where $g_s$ is the coupling strength. 
The two controllable \blu terms $\hat{H}_{\rm s}=\hat \sigma^z_s$ with corresponding controls $\Delta_s(t)$ (where $s={\rm l}, {\rm r}$) describe relatively small changes in the qubit frequency, i.e.~detuning. 
Note that $\hat H(t)$ is excitation preserving, since it commutes with the total number operator, $\hat{a}^\dagger \hat{a} + \sum_{s={\rm l,r}}\sigma^+_s \sigma^-_s$. 
We show that the unitary dynamics required by the laser model presented in the main text can be implemented, by showing it is contained within the set of allowable evolutions permitted by the setup in Fig.~\ref{fig:superconducting_maser}. 

The dynamics of this setup are governed by a unitary superoperator $\mathcal{U}_{\rm clr}$ acting on the spaces of the cavity and the left and right qubits, 
which are prepared in the states $\ket{1}_{\rm l}\bra{1}$ and $\ket{0}_{\rm r}\bra{0}$ respectively. That is, the left (right) qubit here corresponds to the pump (vacuum) 
qubit for the model discussed in \sref{sec:MPS-description}, and $\mathcal{U}_{\rm clr}$ is simply the 
superoperator version of the interaction unitary $\hat{U}_{\rm int}$ defined in \eref{eq:iMPS_interaction_unitary}.
Moreover, in terms of the notation used in \eref{eq:theorem_cavity_beam} describing endogenous phase, the left and right qubits play the role of the relevant state of the environment, $\rho_{\rm e}\equiv\ket{1}_{\rm l}\bra{1} \otimes \ket{0}_{\rm r}\bra{0}$.
Formally,  
\begin{equation}
	\Tr_{\rm l} \left[\mathcal{U}_{\rm clr} (\rho_c\otimes  \ket{1}_{\rm l}\bra{1} \otimes \ket{0}_{\rm r}\bra{0}) \right] \equiv 
	\Tr_{\rm e'}\left[ \mathcal{U}^{t\rightarrow t+\delta t}_{\rm ce}(\rho_{\rm c} \otimes \rho_{\rm e}) \right], \label{clrtheorel}
\end{equation}
and the right qubit following the action of $\mathcal{U}_{\rm clr}$ on the system is precisely the segment of the beam generated at that time, ${\rm b}_t$.

Now, recall that, in the description of the maser operation above, the processes of gain and loss occur while only one of the qubits is coupled to the cavity. 
This works because the interaction unitary 
can be written as a composition of consecutive gain and loss unitaries acting on the cavity and one of the qubits: 
\begin{equation}
	\hat{U}_{\rm clr} \coloneqq (\hat{U}_{\rm cr}\otimes \hat{I}_{\rm l})(\hat{U}_{\rm cl}\otimes \hat{I}_{\rm r}),
\end{equation}
where
\begin{align}
	U_{\rm cl} &\coloneqq \exp(\sqrt{\delta t}(\hat{G}\hat{\sigma}^-_{\rm l} - \hat{G}\dg\hat{\sigma}^+_{\rm l})), \label{eq:UCL}\\
	U_{\rm cr} &\coloneqq \exp(\sqrt{\delta t}(\hat{L}\hat{\sigma}^+_{\rm r} - \hat{L}\dg\hat{\sigma}^-_{\rm r})). \label{eq:UCR}
\end{align}
It is straightforward to check that this does indeed generate \eref{eq:theorem_cavity_beam}, via the relation (\ref{clrtheorel}).  


Now, in order for these two unitaries $\hat{U}_{\rm cl}$ and $\hat{U}_{\rm cr}$ to be permissible evolutions for the system, we require that \blu $\hat{G}\hat{\sigma}^-_{\rm l} - \hat{G}\dg\hat{\sigma}^+_{\rm l}$ and $\hat{L}\hat{\sigma}^+_{\rm r} - \hat{L}\dg\hat{\sigma}^-_{\rm r}$ are contained within the dynamical Lie algebra of the system.
We remind the reader that the only non-zero elements of $\hat{G}$ and $\hat{L}$ are $G_n = \bra{n}\hat G \ket{n-1}$ and $L_n = \bra{n-1}\hat L \ket{n}$, for $0<n<D$. 
Now both $\hat{U}_{\rm cl}$ and $\hat{U}_{\rm cr}$ individually act on only one of the qubits (l and r, respectively) and the cavity 
in a given time interval. The other qubit (r and l, respectively) can be decoupled from the evolution by shifting its frequency far off-resonance ($\Delta$ large 
compared to the other inverse time-scales), or by some more sophisticated dynamical decoupling  \cite{sVio99} if needed. Thus, for the purpose of generating 
each unitary, the Lie algebra we are interested in is that of the cavity and \emph{one} qubit.
Dropping the label for the qubit, this is the Lie algebra $\mathfrak{L}_\text{cq}$ generated by $\mathbb{H}_{\mathrm{cq}}\coloneqq \{ i \hat{H}^\text{cq}_0, i \hat{H}^\text{cq}_1\}$, where $\hat{H}^\text{cq}_0 \coloneqq  g(\hat a\dg \hat  \sigma^- + \hat a\hat \sigma^+ )$ and $\hat{H}^\text{cq}_1 \coloneqq \hat{\sigma}^z$. 

Generating a basis for $\mathfrak{L}_\text{cq}$ first requires computation of the repeated Lie brackets of the form \blu
\begin{equation}
	[i\hat{H}^\text{cq}_0,[i\hat{H}^\text{cq}_0,[\dots[i\hat{H}^\text{cq}_0,i\hat{H}^\text{cq}_1]]]],
	\label{eq:pure_H0_brackets}
\end{equation}
From these, using the identities $\commutator{\hat\sigma^\pm}{\hat\sigma^z} = \pm 2\sigma^\pm$ and $\commutator{\hat\sigma^\pm\hat\sigma^\mp}{\hat\sigma^z} = 0$, one can then calculate the repeated Lie brackets  
\begin{equation}
	[i\hat{H}^\text{cq}_1,[i\hat{H}^\text{cq}_0,[i\hat{H}^\text{cq}_0,[\dots[i\hat{H}^\text{cq}_0,i\hat{H}^\text{cq}_1]]]]]
	 \label{eq:impure_H0_brackets}
\end{equation}
to arbitrary depths. 
It can be shown by induction that the above two types of Lie brackets give the full Lie algebra. Explicitly, the Lie algebra is  
\begin{equation}
	 \mathfrak{L}_\text{cq} =\mathrm{span}\left( \bigcup\limits_{\xi=1}^3\mathcal{S}_\xi \right),
\end{equation}
where we have defined the following sets of skew-Hermitian operators,
\begin{align}
	\mathcal{S}_1 &\coloneqq \left\{ i(a^\dagger a)^m \sigma^-\sigma^+ - i(aa^\dagger)^m\sigma^+\sigma^- \right\}_{m=0}^\infty, \\
	\mathcal{S}_2 &\coloneqq \left\{ (a\dg a)^m a\dg \sigma^- - a (a^\dagger a)^m\sigma^+  \right\}_{m=0}^\infty, \\
	\mathcal{S}_3 &\coloneqq \left\{ i(a\dg a)^m a\dg \sigma^- + i a (a^\dagger a)^m\sigma^+ \right\}_{m=0}^\infty.
\end{align}
The sets $\mathcal{S}_1$ and $\mathcal{S}_2$ arise from the nested commutators in \eref{eq:pure_H0_brackets}, while $\mathcal{S}_3$ comes from those in \eref{eq:impure_H0_brackets}.

First, we show that \blu the term $\hat{G}\hat{\sigma}^-_{\rm l} - \hat{G}\dg\hat{\sigma}^+_{\rm l}$ from \eref{eq:UCL} is contained in this dynamical Lie algebra, by expanding it as a linear combination of the operators in $\mathcal{S}_2$.
We are required to show the existence of a real vector $\bm{v}^\mathrm{G} = (v^\mathrm{G}_1,\dots,v^\mathrm{G}_{D-1})$, the elements of which provide a valid solution to the equation 
\begin{equation}
	 \hat{G}\hat{\sigma}^-_{\rm l} - \hat{G}\dg\hat{\sigma}^+_{\rm l} = \sum_{m=1}^{D-1} v^\mathrm{G}_m \left( (a\dg a)^m a\dg \sigma^-_{\rm l} - a (a^\dagger a)^m\sigma^+_{\rm l} \right). \label{eq:gain_generator_maser}
\end{equation}
This matrix equation can be converted into systems of equations linear in the components of $\bm{v}^\mathrm{G}$ by considering matrix elements.
Introducing indices $n,n'=1,\dots,D-1$, we arrive at the system of equations
\begin{align}
	\bra{n}\hat{G} \ket{n'} = \bra{n} \sum_{m=1}^{D-1} v^\mathrm{G}_m (a^\dagger a)^m a\dg \ket{n'}.
\end{align} 
For $n'\neq n-1$ we have that $\bra{n}\hat{G} \ket{n'}=0$, which is consistent with the requirement that the only non-zero elements of $\hat{G}$ are $G_n = \bra{n}\hat G \ket{n-1}$.
For $n'= n-1$, one finds the system of $D-1$ linear equations
\begin{equation}
	 G_n = \sum_{m=1}^{D-1} F_{nm} v^\mathrm{G}_m,
	 \label{eq:gain_LP}
\end{equation}
where the $(D-1)\times(D-1)$ square matrix $F$ has elements defined by $F_{nm}\coloneqq n^{m+1/2}$ for $n,m=1,\dots,D-1$.
For this system of equations to be solvable, we require $F$ to be invertible.
This can be seen from the following result on the determinants of generalized Vandermonde matrices; see \cite[Theorem and Remark]{sRob00}, and also \cite[Theorem 1]{sYan01}.

\begin{lemma}[Generalized Vandermonde determinant]
\label{lem:determinant}
Given $N\geq 1$ and real vectors $\bm{c}\coloneqq(c_1,c_2,c_3\dots,c_N)$, $\bm{x}\coloneqq(x_1,x_2,x_3\dots,x_N)^\mathrm{T}$, define the $N\times N$ matrix
\begin{equation}
W(\bm{c},\bm{x})\coloneqq
\left(
\renewcommand\arraystretch{1.5}
\begin{matrix}
x_1^{c_1}  & x_1^{c_2}  & x_1^{c_3} & \cdots & x_1^{c_N} \\
x_2^{c_1}  & x_2^{c_2}  & x_2^{c_3} & \cdots & x_2^{c_N} \\
x_3^{c_1}  & x_3^{c_2}  & x_3^{c_3} & \cdots & x_3^{c_N}\\
\vdots  & \vdots & \vdots &\ddots & \vdots\\
x_N^{c_1}  & x_N^{c_2}  & x_N^{c_3} & \cdots & x_N^{c_N}\\
\end{matrix}
\right).
\label{eq:generalized_vandermonde}
\end{equation}
If $c_N>c_{N-1}>\dots>c_2>c_1$, and $x_N>x_{N-1}>\dots>x_2>x_1>0$, then the determinant of $W$ is positive, $\det(W(\bm{c},\bm{x}))>0$.
\end{lemma}
It is straightforward to verify that $F$ is equivalent to the matrix in \eref{eq:generalized_vandermonde}, if one takes $N=D-1$, $c_j=j+1/2$ and $x_j=j$ for $j=1,\dots,D-1$.
Lemma~\ref{lem:determinant} therefore shows that $F$ has non-zero determinant for any $D$, which implies that it is invertible.
In turn, this means that the system of linear equations in \eref{eq:gain_LP} can be solved by matrix inversion, with the vector of coefficients  given by $\bm{v}^\mathrm{G} = F^{-1}\hat{G}$.
Thus, $\bm{v}^\mathrm{G}$ exists.

The calculation showing the loss term belongs to the Lie algebra is similar.
We are required to show the existence a real vector $\bm{v}^\mathrm{L} = (v^\mathrm{L}_1,\dots,v^\mathrm{L}_{D-1})$
for which 
\begin{equation}
	 \hat{L}\hat{\sigma}^+_{\rm r} - \hat{L}\dg\hat{\sigma}^-_{\rm r} = \sum_{m=1}^{D-1} v^\mathrm{L}_m \left( (a\dg a)^m a\dg \sigma^-_{\rm r} - a (a^\dagger a)^m\sigma^+_{\rm r} \right). \label{eq:loss_generator_maser}
\end{equation}
The matrix elements of this equation give the system
\begin{equation}
	\bra{n}\hat{L}\ket{n'} = \bra{n} \sum_{m=1}^{D-1} v^\mathrm{L}_m a(a^\dagger a)^m \ket{n'}.
\end{equation}
For $n\neq n'-1$ we find $\bra{n}\hat{L} \ket{n'}=0$, which is consistent with the only non-zero elements of $\hat{L}$ being $L_n = \bra{n-1}\hat L \ket{n}$.
The other $D-1$ equations for  $n=n'-1$ reduces to solving the system of linear equations
\begin{equation}
	L_n = \sum_{m=1}^{D-1} F_{nm}  v^\mathrm{L}_m,
\end{equation}
where $F$ is the matrix defined below \eref{eq:gain_LP}.
Therefore, $\bm{v}^\mathrm{L}$ exists, and is given by $\bm{v}^\mathrm{L} = F^{-1}\hat{L}$.

In summary, the proposed system can be controlled, without any external 
microwave drive (thus ensuring it has an {\bf endogenous phase}), 
so as to produce, in {\bf stationary} operation, 
a {\bf one-dimensional beam}, which, allowing for its spatial modulation (pulses), 
has {\bf ideal Glauber coherence}, and which attains the {\bf Heisenberg limit} for $\coh$.
\vspace{30pt}

\blk


\cleardoublepage


\end{document}